\input harvmac
\input epsf

\def\bfone{\relax{\rm 1\kern-.35em 1}}
\def\cicy#1(#2|#3)#4{\left(\matrix{#2}\right|\!\!
                     \left|\matrix{#3}\right)^{{#4}}_{#1}}

\def\inbar{\vrule height1.5ex width.4pt depth0pt}
\def\IC{\relax\,\hbox{$\inbar\kern-.3em{C}$}}
\def\ID{\relax{\rm I\kern-.18em D}}
\def\IF{\relax{\rm I\kern-.18em F}}
\def\IH{\relax{\rm I\kern-.18em H}}
\def\II{\relax{\rm I\kern-.17em I}}
\def\IN{\relax{\rm I\kern-.18em N}}
\def\IP{\relax{\rm I\kern-.18em P}}
\def\IQ{\relax\,\hbox{$\inbar\kern-.3em{\rm Q}$}}
\def\us#1{\underline{#1}}
\def\IR{\relax{\rm I\kern-.18em R}}
\font\cmss=cmss10 \font\cmsss=cmss10 at 7pt
\def\ZZ{\relax\ifmmode\mathchoice
{\hbox{\cmss Z\kern-.4em Z}}{\hbox{\cmss Z\kern-.4em Z}}
{\lower.9pt\hbox{\cmsss Z\kern-.4em Z}}
{\lower1.2pt\hbox{\cmsss Z\kern-.4em Z}}\else{\cmss Z\kern-.4em
Z}\fi}
\def\nup#1({Nucl.\ Phys.\ $\us {B#1}$\ (}
\def\plt#1({Phys.\ Lett.\ $\us  {#1}$\ (}
\def\cmp#1({Comm.\ Math.\ Phys.\ $\us  {#1}$\ (}
\def\prp#1({Phys.\ Rep.\ $\us  {#1}$\ (}
\def\prl#1({Phys.\ Rev.\ Lett.\ $\us  {#1}$\ (}
\def\prv#1({Phys.\ Rev.\ $\us  {#1}$\ (}
\def\mpl#1({Mod.\ Phys.\ Let.\ $\us  {A#1}$\ (}
\def\ijmp#1({Int.\ J.\ Mod.\ Phys.\ $\us{A#1}$\ (}
\def\tit#1|{{\it #1},\ }

\def\Coe#1.#2.{{#1\over #2}}

\def\coe#1.#2.{\relax{\textstyle {#1 \over #2}}\displaystyle}

\def\cod{{\cal O }_D}
\def\t#1{{\theta_#1}}

\def\ra{{\rightarrow}} 

\def\npb#1(#2)#3{{ Nucl. Phys. }{B#1} (#2) #3}
\def\plb#1(#2)#3{{ Phys. Lett. }{#1B} (#2) #3}
\def\pla#1(#2)#3{{ Phys. Lett. }{#1A} (#2) #3}
\def\prl#1(#2)#3{{ Phys. Rev. Lett. }{#1} (#2) #3}
\def\mpla#1(#2)#3{{ Mod. Phys. Lett. }{A#1} (#2) #3}
\def\ijmpa#1(#2)#3{{ Int. J. Mod. Phys. }{A#1} (#2) #3}
\def\cmp#1(#2)#3{{ Comm. Math. Phys. }{#1} (#2) #3}
\def\cqg#1(#2)#3{{ Class. Quantum Grav. }{#1} (#2) #3}
\def\jmp#1(#2)#3{{ J. Math. Phys. }{#1} (#2) #3}
\def\anp#1(#2)#3{{ Ann. Phys. }{#1} (#2) #3}
\def\prd#1(#2)#3{{ Phys. Rev.} {D\bf{#1}} (#2) #3}

\def\fivepoint{\def\rm{\fam0\fiverm}
\textfont0=\fiverm \scriptfont0=\fiverm \scriptscriptfont0=\fiverm
\textfont1=\fivei \scriptfont1=\fivei \scriptscriptfont1=\fivei
\textfont2=\fivesy \scriptfont2=\fivesy \scriptscriptfont2=\fivesy
\textfont\itfam=\fivei \def\it{\fam\itfam\fiveit}\def\sl{\fam\slfam\fivesl}%
\textfont\bffam=\fivebf \def\bf{\fam\bffam\fivebf}\rm}
\def\phm{\phantom {-}}
\def\bra{\langle}
\def\ket{\rangle}

\def\C{{\bf C}}

\def\P{{\bf P}}

\def\cA{{\cal A}}

\def\cI{{\cal I}}

\def\cO{{\cal O}}
\def\cM{{\cal M}}
\def\cB{{\cal B}}
\def\cC{{\cal C}}
\def\cD{{\cal D}}
\def\cE{{\cal E}}
\def\cI{{\cal I}}

\def\cL{{\cal L}}

\def\cS{{\cal S}}
\def\cT{{\cal T}}

\newbox\strutfbox
\setbox\strutfbox=\hbox{\vrule height6.5pt depth1.5pt width 0pt}
\def\strutf{\relax\ifmmode\copy\strutfbox\else\unhcopy\strutfbox\fi} 
\def\tabsp{&&&&&&&&&&&&&&&&&\cr}
\def\tabspI{height8pt&\omit&&\omit&\omit&\omit&
\omit&\omit&&\omit&\omit&\omit&\omit&\omit
&\omit&&\omit&\cr }
\def\tm{\rlap*}

\def\cicy#1(#2|#3)#4{\left(\matrix{#2}\right|\!\!
                     \left|\matrix{#3}\right)^{{#4}}_{#1}}

\def\ra{\rightarrow}

\def\bs{\bigskip}

\def\Box{{\,\lower0.9pt\vbox{\hrule 
\hbox{\vrule height 0.2 cm \hskip 0.2 cm  
\vrule height 0.2 cm}\hrule}\,}}
\Title{ \vbox{\baselineskip12pt\hbox{hep-th/9609239}
\hbox{EFI-97-01}}}
{\vbox{
\centerline{Calabi-Yau fourfolds for M- and  
F-Theory compactifications}}}
\centerline{A. Klemm$^1$, B. Lian$^2$, S-S. Roan$^3$  and S-T. 
Yau$^4$}
 \medskip
\centerline{\sl $^1$ Enrico Fermi Institute, University of Chicago}
\centerline{5640 S. Ellis Ave.,  Chicago, IL 60637, USA}
\medskip
\centerline{\sl $^2$ Department of Mathematics, Brandeis University}
\centerline{Waltham, MA 02154 , USA}
\medskip
\centerline{\sl $^3$ Institute of Mathematics, Academia Sinica Taipei, Taiwan}
\medskip
\centerline{\sl $^4$ Department of Mathematics, Harvard University}
\centerline{\sl Cambridge, MA 02138, USA}
\medskip
{\bf Abstract}
We investigate topological properties of Calabi-Yau fourfolds and 
consider a wide class of explicit constructions in weighted 
projective spaces and, more generally, toric varieties. Divisors 
which lead to a non-perturbative superpotential in the effective 
theory have a very simple description in the toric construction. 
Relevant properties of them follow just by counting lattice points 
and can be also used to construct examples with negative Euler number. 
We study nets of transitions between cases with generically smooth 
elliptic fibres and cases with ADE gauge symmetries in the N=1 theory due 
to degenerations of the fibre over codimension one loci in the base. 
Finally we investigate the quantum cohomology ring of this fourfolds using
Frobenius algebras.

\centerline{\tt 
aklemm@maxwell.uchicago.edu,
maroan@ccvax.sinica.edu.tw,}
\centerline{\tt 
 lian@max.math.brandeis.edu,
yau@math.harvard.edu  }
\vskip 1cm
\noindent
\overfullrule=0pt
\Date{December 1996}

\newsec{Introduction}

Geometrizing the expected symmetries in the moduli space of 
supersymmetric theories has proven to be a simple and successful 
tool in the investigation of their non perturbative behaviour.
Especially the geometric interpretation  of the non perturbative 
$SL(2,\ZZ)$ of type $II_B$ string as coming {\sl really } from a 
two torus of an (elliptically fibred) compactification of $F$-theory 
has helped to uncover many non perturbative properties of string 
compactifications to dimensions greater 
then five \ref\ftheory{C. Vafa, {\it Evidence 
for F-Theory}, hep-th/9602022}\ref\mvI{D.\ Morrison, C.\ Vafa, 
hep-th/9602114}\ref\mvII{D. Morrison and C. Vafa, hep-th/9603161}
\ref\bikmsv{M.\ Bershadsky,
K.\ Intriligator, S.\ Kachru, D.\ R.\ Morrison, V.\ Sadov and 
C.\ Vafa, hep-th/9605200}\ref\bj{M.\ Bershadsky and 
A.\ Johansen, hep-th/961011}.   

Elliptically fibred complex four dimensional K\"ahler manifolds 
$X$ with $SU(4)$ holomony, Calabi-Yau fourfolds for short, are the 
geometry relevant for $N=1$ compactifications of $F$-theory to 
four dimensions \ftheory . Orbifold constructions
\ref\orbi{E.\ Witten, \npb 463 (1996) 506, 
P. \ Horava and E.\ Witten,\npb460 (1996) 506,
K.\ Dasgupta  and S.\ Mukhi, \npb465 (1996) 399}
\ref\sen{A.\ Sen, \mpla 11 (1996) 1339, A.\ Sen, 
{\sl F-theory and Orientifolds}, hep-th/9605150}
\ref\kr{A.\ Kumar and K.\ Ray, hep-th/9602144, hep-th/9604133}
\ref\bz{J.\ Blum and A.\ Zaffaroni, hep-th/9607019, J.\ Blum, hep-th/9608053}
of $M$ and $F$ theory are particular useful to 
get a fast view on the spectrum and  the symmetries. Using results 
of \ref\borcea{C.\ Borcea, {\sl K3 Surfaces with and Mirror Pairs of 
Calabi-Yau Manifolds}, in Mirror Symmetry II, Ed. 
B. Greene and S.T. Yau, International Press (1991)} they were considered 
to compactify $F$ ($M$, type II) theory to four (three, two) 
dimensions  in~\ref\fourorb{R.\ Gopakumar and S.\ Mukhi, {\sl Orbifold 
and Orientifold Compactifications of F-Theory and M-theory to six 
and four Dimensions}, hep-th/9607057}.
However in order to study the moduli space and in particular transitions, 
one wishes to have a deformation family {\sl and} the knowledge about 
the enhanced symmetry points\foot{There are elegant ways of finding
such symmetric configurations in the deformation families, see e.g.
\ref\hw{F.\ Hirzebruch and J. Werner, {\sl Some examples of Threefolds
with trivial canonical bundle}, Preprint Max-Planck-Institut Bonn
MPI/85-58}.}. To get some overview of the possible 
deformation families of Calabi-Yau fourfolds, especially the 
elliptically fibred ones, is the first objective of this paper.
We will therefore consider a rich class of hypersurfaces and 
complete intersections in weighted projective spaces and toric 
varieties. For the hypersurfaces
we obtain a large scan $(104\ 021$ configurations) over possible 
Hodge numbers by classifying {\sl all} Fermat type configurations and 
generic hypersurfaces up to degree 400. The Euler number ranges between
$-240\le \chi\le 1\, 820\, 448$. Examples with negative Euler number 
could eventually lead to supersymmetry breaking in three 
dimensions by anti-branes,  which have to be included to cancel the 
tadpoles if $\chi<0$ and the fourform background 
flux vanishes~\ref\SVW{S. Sethi, C. Vafa and E. Witten, Constraints 
on low-dimensional string compactifications, 
hep-th 9606122}\ref\wittenII{E.\ 
Witten, {\sl  On Flux Quantization in $M$-theory and the effective Action},
hep-th/9609122}. A hyperk\"ahler fourfold with $\chi<0$ was 
constructed in~\SVW~and and K\"ahler examples appear in\borcea, 
both orbifold constructions. 
Here we find the first K\"ahler manifolds with $\chi<0$ realized as 
deformation families.

The properties of elliptically fibred fourfolds can 
be understood from properties of the bases and the degeneration 
of the fibres. The most basic properties, like the triviality 
of the canonical bundle, are decided from the degeneration over 
codimension one. For instance if the singularity here is as mild 
as possible ($I_1$ fibres only, compare A.1) one can express the 
Euler number for the fourfold $X$ by formulas, which refer only to 
properties of the bases and the generic type of fibre. 
Likewise physically the most basic properties like the unbroken gauge 
group\foot{More exotic theories could also arise from degeneration 
on codimension one.} are decided from the degeneration on codimension 
one. We therefore aim for examples in which we can control the 
degeneration at codimension one in a simple way.

We will first study examples, which are simplest in two respects, 
namely the fibres degenerates homogeneously on a subspace $B'$ 
of codimension one in the base to an $ADE$ singularity and it does 
so for generic values of the moduli. In this situation we find 
formulas for the Euler number, which depend on the cohomology of 
$B'$ and the invariants of the gauge group. 
The manifolds provide a realizations of $N=1$ gauge theories, discussed 
recently in \ref\kvII{S.\ Katz and  C.\ Vafa, 
{\sl Geometrical Engineering of $N=1$ Quantum Field Theories}, 
hep-th/96011091} \ref\bjpsv{M.\ Berschadsky, A.\ Johanson, 
T.\ Pantev, V.\ Sadov and C.\ Vafa,{\sl F-theory, Geometrical 
Engineering and $N=1$ dualities\ }, hep-th/9612052}. 

$F$-theory on the fourfolds has beside the complex and the 
K\"ahler moduli of the manifold, also the moduli associated with 
three branes, which live in space-time and intersect the base in points, 
as well as a choice of discrete back ground fluxes which take (half)integer 
values in the unimodular selfdual lattice $H^4(X,\ZZ)$, which is even if 
$\chi=0\ {\rm mod} \ 24$. About the global moduli space of the first 
two types of moduli, we can learn by Kodaira \& Spencer deformation 
theory and mirror symmetry 
(see e.g. \ref\Hayakawa{Y.\ Hayakawa, {\sl Degeneration of 
Calabi-Yau Manifold with Weil-Petersen Metric}, alg-geom/9507016} for 
recent results on dimension$>3$). On the moduli space of the
three branes one can learn locally in non generic situations with 
orbifold symmetries~\sen\ref\bds{T.\ Banks, M.\ Douglas, 
N.\ Seiberg, {\sl Probing F-theory with Branes}, hep-th/9605199} or 
more generically in situations as above~\bjpsv~and at the transitions 
points which connect the $F$-theory vacua. 
Using Batyrev and Watanabes classification
of toric Fano threefolds we can construct systematically a rather 
dense net of such fourfold transitions (fig 1), which are 
again very simple in that they keep the elliptic fibre structure 
{\sl and} the generic degeneration type\foot{The geometrically 
interesting fourfold transitions considered in~\ref\bls{I. Brunner, 
M. Lynker and R. Schimmrigk, hep-th/9610195.} are not particularly 
useful for F-theories, because they behave randomly w.r.t. to the
fibre structure (if any).}. These extremal transitions correspond 
to shrinking $E_8$, $E_7$, $E_6$, $D_5$ Del Pezzo surfaces along one 
dimensional (T-stable) subsets in the base or generalized elliptic 
threefolds to (T-fixed) points in the base.     

Of course physically one would like to understand perturbative or 
non-perturbative enhancements of the gauge symmetries, which correspond 
to codimension one degenerations, which occur only for specific values 
of the moduli and the meaning of the codimension two (and three) 
degenerations. A good guidance to these more complicated situations can be 
obtained by considering those three dimensional elliptic fibrations 
over Hirzebruch surfaces $F_n$, for which the degeneration on 
codimension one and two has been studied in \mvI\mvII\bikmsv\bj and 
\ref\bkkm{P.\ Berglund, S.\ Katz, 
A.\ Klemm and  P.\ Mayr, {\sl New Higgs Transitions between Dual N=2 Models}, 
hep-th/9605154} and replacing the base $\IP^1$ of $F_n$ by a rational 
surface. In easy cases this can be done so that part of the 
singularity structure at codimension one and two essentially carries over 
to fourfolds.  Here we can also obtain systematically chains
of now more complicated extremal fourfold transitions, which keep the 
elliptic fibre structure, but frequently violates the evenness of 
$H^4(X,\ZZ)$.

\subsec{Divisors which lead to a non-perturbative superpotential in three
dimensions.}
Some aspects of the four dimensional theory can be
investigated more easily by compactifying first $M$-theory or 
type IIB on $X$ to three dimensions 
or two dimensions and considering decompactification limits to learn
about four dimensions.
Eleven dimensional M-theory compactifications, on not necessarily 
elliptically fibred, Calabi-Yau fourfolds $X$, leads to $N=2$ supersymmetric 
theories in three dimensions~\ref\witten{E.\ Witten,
{\sl Non-Perturbative Superpotentials in String Theory}, hep-th9604030}
\ref\bb{K.\ Becker, M.\ Becker, {{\cal M}-Theory on Eight-Manifolds},
hep-th/9605053} .
There is a general mechanism to generate a non-perturbative 
superpotential in the three-dimensional theory 
from supersymmetric instantons, which arise from wrapping the 
$5$-branes of the M-theory around complex divisors $D$ of $X$.

\vskip 5 pt
\noindent
i.) Under the assumption that $D$ is smooth, 
the following necessary condition 
on the arithmetic genus of $D$
for the occurrence of 
instanton induced terms in the superpotential was derived 
from the anomaly vanishing requirement in \witten:
\eqn\nec{\chi(D,{\cal O}_D)=\sum_{n=0}^3 h^n(\cod)=1} 

\vskip 5 pt
\noindent
ii.) If $h^0(\cod)=1$ and $h^1(\cod)=h^2(\cod)=h^3(\cod)=0$ a 
non-perturbative contribution of the form
\eqn\superpotential{\int d\theta e^{-(V_D+ i \phi_D)} T(m_i)}
must be generated in the superpotential, as no cancellation from extra 
fermionic zero modes can occur. 
Here $V_D$ is the volume of 
$D$ measured in units of the 5-brane tension, $(V_D+i \phi_D)$ are 
real and complex moduli components of a chiral superfield and $T(m_i)$ is
a non-vanishing section of a holomorphic line bundle 
over the moduli space of the theory on $X$.

Using the fact that $h^n(\cod)$ describes the dimension of the 
deformation space of $D$ it was shown in \witten~that divisors 
given by a polynomial constraints in a Calabi-Yau fourfolds 
defined as hypersurfaces or complete intersections in (products) 
of ordinary projective spaces have $\chi(D,\cod)<1$. The reason is 
basically that such polynomials have too many possible deformations.  
These divisors will therefore not lead to nontrivial contributions 
to the superpotentials.  Using the Hirzebruch-Riemann-Roch index 
formula \ref\hirzebruch{F. Hirzebruch, {\sl Topological Methods
in Algebraic Geometry}, Springer-Verlag (1966), Berlin, Heidelberg, 
New York}
\eqn\index{\chi(D,\cod )=\int (1- e^{-[D]}) td(X),}
the explicit expansion of the Todd polynomials $T_0=1$, 
$T_1={1\over 2} c_1(X)$ $T_2={1\over 12} (c_2(X)+c_1(X)^2)$ 
and the fact that $c_1(X)=0$ for manifolds of $SU(4)$ holonomy we can
rewrite~\nec~in the  more useful form 
\eqn\newform{[D]^4+c_2(X) [D]^2=-24.}
With this topological formula the above statement follows  
from the fact that all intersection numbers on the left of~\newform~come from 
semi ample divisors in projective spaces and are hence positive. 
On the other hand the fact that the left hand side of~\newform~has to 
be negative suggests that $D$'s with the desired properties occurs 
preferably as exceptional divisors or in situations where the 
deformation space is for some reasons small. For instance because 
we make an orbifoldisation and thereby killing most of the deformation 
space or we work with weighted projective spaces, where the possible 
deformations are restricted by the weights. This hints that
weighted projective space and more generally toric varieties
will lead to interesting configurations of such divisors.
In fact we will see that the intersection of the $T$-invariant 
orbits of the toric ambient space with the Calabi-Yau fourfold will
lead under very simple combinatorial conditions, which are explained 
in section 4, to such divisors.  A special situation where one can 
construct infinitly many divisors, which contribute to the 
superpotential, was described in~\ref\DGW{R.\ Donagi, A.\ Grassi and 
E.\ Witten, a non-perturbative superpotential  with $E_8$-symmetry, 
hep-th 9607091}. 

\subsec{Preferred physical situations, additional 
geometrical data and dualities}

If the Calabi-Yau manifold\foot{Other interesting 
compactifications are on manifolds with
$Spin(7)$ holonomy, the so-called Joyce manifolds. 
They lead to $N=1$ supersymmetry for $M$-theory 
compactifications to three dimensions (see \ftheory
\ref\acharya{B.\ S.\  Acharya, {\sl N=1 M-theory Heterotic duality in 
Three-Dimensions and Joyce Manifolds}, hep-th/9604133; 
{\sl M-theory Compactification and Two Brane/Five Brane Duality}, 
hep-th/9605047}).}
 $X$ admits an elliptic fibration
\eqn\ellX{\cE \longrightarrow X 
{{\pi \atop \longrightarrow}\atop \phantom{.XS}} B}
then a compactification of $M$ theory on $X$ is equivalent to
$F$-theory~\ftheory~on $X\times S^1$, which in turn is
equivalent to Type IIB on $B\times S^1$. If $\varepsilon$ is the 
area of $\cE$ one can use for $\varepsilon\rightarrow 0$ the fiberwise 
equivalence of M theory compactification on $R^9\times T^2$ with Type IIB on 
$R^8\times S^1$. This means that $M$ theory compactification to
three dimensions on $X$ has the same moduli as Type IIB compactified 
to three dimensions on $B\times S^1$. Denoting the radius of 
the $S^1$ by $R$ one has $\varepsilon \propto 1/R$ such that 
the $\varepsilon\rightarrow 0$ limit is the decompactification 
limit for the type IIB theory.

W.r.t. this limit $\varepsilon\rightarrow 0$ one has two principally 
different situations for the location of the divisor $D$ on $X$ to 
distinguish 

\noindent
a.) $\pi(D_a)=B$, i.e. $D_a$ is a section or multisection. $D_a$ is 
called {\sl horizontal}.

\noindent
b.) $D_b=\pi^{-1}(B')$ with $B'$ a divisor in $B$. $D_b$ is called 
{\sl  vertical}.

As was explained in \witten~for {\sl generic fixed} geometry of the base
non perturbative superpotentials in the four dimensional Type IIB theory 
will only occur in case b.). The reason is that the action of the non 
perturbative configuration in $F$-theory units is proportional to the 
volume of the divisors, which for the two types of divisors goes like 
$D_a\sim {1/\epsilon} D_b$ in the limit\foot{Of course one can 
enhance the contribution of the $D_a$ divisors by going to a 
singular point in $B$.} $\epsilon\rightarrow 0$. 

For phenomenology it might be more useful to think about
the situation in terms of the heterotic $N=1$ string.
This is possible if $B$ admits a holomorphic $\IP^1$ fibration
\eqn\ruled{\IP^1 \longrightarrow B 
{{\pi' \atop \longrightarrow}\atop \phantom{.}} B' }
then one can consider an elliptic fibration 
\eqn\ellZ{\cE' \longrightarrow Z  
{{\pi^{\prime \prime} \atop \longrightarrow}\atop \phantom{.}} B' }
over $B'$ and get, by fiberwise application of type IIB/heterotic 
string duality, a description of the heterotic string on the 
Calabi-Yau threefold $Z$. The effect of a divisor of type b.) can be 
interpreted in the heterotic string theory description \witten~as 
worldsheet or as spacetime instanton effect 
depending of whether $D_b$ maps in $Z$ to a 
{\sl vertical} or {\sl horizontal} divisor 
w.r.t. $\pi''$. Both types can occur as $T$-invariant toric divisors
as discussed in section 5 and 6.

The organization of the material is as follows. In section two we
will summarize the basic topological properties of Calabi-Yau
fourfolds. Then we give in section three some overview of the 
class of complete intersections in weighted projective spaces. 
In section four we explain the toric construction of elliptically 
fibred toric fourfolds. 
We extend the formulas of Batyrev and give a characterisation of
the divisors on $X$, which come from the divisors of the ambient space, 
which are invariant under the torus action. This gives a very
easy criterium, when such a divisor contributes to the superpotential. 
Section five contains a complete list of elliptically 
fibred Calabi-Yau manifolds over 
toric Fano bases and the transitions among them. 
In section six we also  discuss degenerations of the
fibre, which lead to gauge symmetry in four dimensions.
Sections seven and eight contains proofs for the formulas of the 
Euler number of the fourfolds in terms of the topological properties of the 
base and the the type of the fibre. Some cases have been already discussed 
in \SVW . In section nine we discuss the quantum cohomology of fourfolds
using Frobenius algebras. Especially we give the generalization of the 
formulas for quantum cohomology ring obtained for threefolds in 
\ref\hktyII{S.\ Hosono, A.\ Klemm, S.\ Theisen, S.\ Yau, 
\npb 433 (1985) 501}
\ref\hly{S. Hosono, B. Lian and S. T. Yau, 
alg-geom-9511001, to appear in CMP.}  
~to the n-fold case. In section (9.6) we discuss
in some details examples which are connected by transitions. 

\noindent{\bf Acknowledgements.} We would like to thank P. Candelas, 
X. de la Ossa, S. Katz and C. Vafa for very helpful discussions.
We also like to thank S. Hosono for his help and frequent 
correspondences.

\newsec{General topological properties of Calabi-Yau fourfolds} 
We will first employ Hirzebruch-Riemann-Roch index theorems to 
derive some relations and general divisiblity conditions
among the topological invariants of Calabi-Yau fourfolds.
If $W$ is a vector bundle over $X$, 
$\chi(X,W)=\sum_{i=0}^n (-1)^i {\rm dim} H^i(X,W)$ and
$c_0[X],\ldots,c_n[X]$, Chern classes of $X$ and $d_0[W],
\ldots,d_r[W]$ Chern classes of $W$ one has \hirzebruch
\eqn\hirz{
\chi(X,W)=\kappa_n\left[\sum_{i=1}^q e^{\delta_i} 
\prod_{i=1}^n {\gamma_i\over 1- e^{-\gamma_i}}\right],}
where $\kappa_n[ ]$ means taking the coefficient of the n'th 
homogeneous form degree, the $\gamma_i$ and $\delta_i$ are 
the formal roots of 
the total Chern classes: $\sum_{i=0}^n c_i[X]=\prod_{i=1}^n (1-\gamma_i)$ 
and $\sum_{i=0}^q d_i[X]=\prod_{i=1}^q(1-\delta_i)$. We want
to use the index formula to compute
the arithmetic genera $\chi_q=\sum_p (-1)^p 
{\rm dim}H^p (X,\Omega^q)$. First we will evaluate~\hirz~for $W=T_X$, the
tangent bundle of $X$. One way of to do so is
to express the formal roots, via symmetric polynomial, in terms 
of the Chern classes $c_i$. This yields for the two,three and four 
dimensional cases the following formulas for $\chi_q=
\sum_{p=1}^{{\rm dim}(X)}(-1)^p h^{p,q}$:
\eqn\chII{{\rm dim}(X)=2:\qquad \chi_0={1\over 12} \int_X(c_1^2+c_2),}
\eqn\chiIII{{\rm dim}(X)=3:\qquad\eqalign{
a.)& \quad \chi_0={1\over 24} \int_X(c_1 c_2)\cr
b.)& \quad \chi_1={1\over 24}\int_X(c_1 c_2-12 c_3), }}
\eqn\ch{{\rm dim}(X)=4:\qquad\eqalign{
a.)& \quad \chi_0={1\over 720}\int_X(-c_4 + c_1 c_3+3 c_2^2 +4 c_1^2 c_2 -c_1^4)\cr
b.)& \quad \chi_1={1\over 180}\int_X(-31 c_4-14 c_1 c_3+3c_2^2+4c_1^2 c_2-c_1^4)\cr 
c.)& \quad \chi_2={1\over 120}\int_X( 79 c_4- 19 c_1 c_3+ 3 c_2^2+4 c_1^2 c_2-c_1^4)
}}

We are mainly interested in  K\"ahler fourfolds 
with $c_1[X]=0$. This is equivalent to the statement that a
Ricci flat K\"ahler metric exists and the manifold has 
holonomy inside $SU(4)$. In the following, by a Calabi-Yau manifold, we mean
a manifold for which the holonomy is strictly\foot{Which
excludes 4-tori $T^8$, $K3\times T^4$, $T^2\times CY$-3-folds etc. 
Note however that there are Hyperk\"ahler 
fourfolds with $h^{2,0}\neq 0$, which are not of of this simple product type.} 
$SU(4)$. In this case there is a unique holomorphic four-form and 
no continuous isomorphisms, i.e.  $h_{0,0}=1, h_{1,0}= h_{2,0}=h_{3,0}=0, 
h_{4,0}=1$. Hodge $*$-duality and complex conjugation reduces the
independent Hodge numbers in the Hodge square  
$$\matrix{1&      0&      0&      0&      1\cr
          0&h^{3,1}&h^{3,2}&h^{3,3}&0\cr    
          0&h^{2,1}&h^{2,2}&h^{2,3}&0\cr
          0&h^{1,1}&h^{1,2}&h^{1,3}&0\cr
          1&      0&      0&      0&      1}$$
to four, say $h^{1,1}=h^{3,3}$, $h^{3,1}=h^{1,3}$, $h^{2,1}=h^{3,2}$ 
and $h^{2,2}$. For Calabi-Yau manifolds in the sense above 
we have $c_1=0$, $\chi_0=2$. Using this in \ch\ implies\foot{Beside this
it implies $\int_X c_2^2$ is even. It also seems
that $c_2^2\geq0$, indicating that $\chi\ge -1440$.} a further 
relation among the Hodge numbers say 
\eqn\rel{h^{2,2}=2(22+2 h^{1,1}+2 h^{3,1}- h^{2,1}).}
The Euler number can thus be written as
\eqn\eul{\chi(X)=6(8+h^{1,1}+h^{3,1}-h^{2,1}).}  

The middle cohomology splits into a selfdual ($*\omega =\omega$) 
$B_+(X)$ subspace and an anti-selfdual ($*\omega =-\omega$) 
subspace $B_-(X)$
$$H^4(X)=B_+(X)\oplus B_-(X),$$
whose dimensions are determined by the Hirzebruch signature as
\eqn\sig{\eqalign{\tau(X)&={\rm dim} B_+(X)-{\rm dim} B_-(X)=
                   \int_X L_2={1\over 45} \int_X (7 p_2-p_1^2)\cr
                  &={\chi \over 3}+32.}}
The symmetric inner product 
$(\omega_1,\omega_2)=\int_X \omega_1\wedge * \omega_2$ 
is positive definite on $H^4(X)$ and  $H^4(X,\ZZ)$ is by Poincare 
duality unimodular. The symmetric quadratic form $Q(\omega_1,\omega_2)=
\int_X \omega_1 \wedge \omega_2$ is positive definite on $B_+(X)$ and 
negative on $B_-(X)$. Beside this we expect a split of 
$H^4(X,\ZZ)$ from mirror symmetry, see section 4 .

We note furthermore that from the definition of the Pontryagin 
classes $p_i\in H^{4i}(X,\ZZ)$ in terms of the Chern classes 
\eqn\pontryagin{\eqalign{
p&=\sum_{i=0}^{[{\rm dim}(X)/2]}(-1)^i p_i=\sum_{i,j} (-1)^i c_i\wedge c_j,
\qquad {\rm hence}\ \cr
p_1&=c_1^2-2 c_2,\qquad p_2=c_2^2-2 c_1 c_3 +2 c_4,\ldots}}
one has, using the Gauss-Bonnet formula, for Calabi-Yau fourfolds\foot{
The later condition was found in \ref\ip{ C.\ J.\ Isham, C.\ N.\ Pope, 
{Class. Quant. Grav. {\bf 5} (1988) 257};  C.\ J.\ Isham, C.\ N.\ Pope
and N.\ P.\ Warner {Class. Quant. Grav. {\bf 5} (1988) 257}} requiring
the existence of a nowhere vanishing eight dimensional Majorana-Weyl spinor
in the $8_c$ representation of $SO(8)$.}  
always $\chi={1\over 8}\int_X (4 p_2-p_1^2).$

It was shown in \SVW~that $I(R)=-\int_X X_8(R)=
\int_X (4 p_2-p_1^2)/192=\chi/24 \neq 0$
gives rise to a non vanishing contribution one-point function for 
the two,  three or four form in $II_A$, M-  or F-theory compactification 
on $X$. Assuming that there are no further non integral contributions 
to the one-point functions it was argued in~\SVW~that these 
compactification are unstable if the one-point functions cannot be canceled 
by introducing integer quanta of string, twobrane or threebrane charge 
in these theories, that is \foot{We have collected in table B.2 a couple 
of non-trivial examples, in which $\chi$ { \sl is} actually zero.} 
$\chi=0\ {\rm mod}\ 24.$ 

In~\wittenII~it was argued that there is a flux quantization 
$\left[G \right]-{p_1\over 4}\in H^4(X,\ZZ)$, where $G$ is the 
four form field strength to which the twobrane of $M$ theory couples \bb . 
If $G$ is zero that means that $p_1/4$ {\sl has} to be  an integral class
($c_2=2 y$ with $y\in H^4(X,\ZZ)$) and as explained in~\wittenII~this implies 
by the formula of Wu $x^2=0\ {\rm mod }\ 2$ for any $x\in H^4(X,\ZZ)$.
That means especially by~\sig~that $H^4(X,\ZZ)$ is an even selfdual 
lattice with signature\foot{Note that $\tau=0\ {\rm mod}\ 8$ as it must be
for even selfdual lattices if$\chi= 0\ {\rm mod } \ 24$.} $\tau$ and
implies by \ch~a.) again that $\chi=0\ {\rm mod}\ 24$. 
On the other hand if $p_1/4$ is half integral then  $[G]$ has to be half 
integral and potentially non-integral contributions to the one point 
function  $I(R,G)=-\int_X X_8-{1\over 8 \pi^2 } \int_X G^2=
{1\over 8}\int_X c_2^2-{1\over 2}\int_X G^2-60$ can be canceled also
for Calabi-Yau's for which $\chi \neq 0 {\rm mod } \ 24$. 

We will find in chapter three and six various chains of 
geometrically possible transition between elliptically fibred 
fourfolds in which the Euler number {\sl is} divisible by $24$ 
in an element of the chain, while it is {\sl not} divisible after 
the transition (see especially table 6.5). This is somewhat 
disturbing as the flux $G$ would have to jump by one half unit if 
one tries to follow this transition in $M$- or $F$-theory, suggesting
that these transitions are impossible in these theories.

Beside the three brane source terms there are contributions from the 
fivebranes~\ref\pema{P. Mayr, hep-th/9610162}~which can cancel 
$I(R,G)$. In~\SVW~it has been also suggested to calculate the 
Euler number of an 
elliptic fibration by counting locally the three-brane charge which 
is induced from the seven branes whose world volume $W$ is the discriminant 
locus $\tilde \Delta$ of the projection map $\pi:X\rightarrow B$ times 
the uncompactified space-time. This three brane charge is 
$Q={1 \over 48}\int_W p_1(W)={1\over 48}\int_{\tilde \Delta}
p_1(\tilde \Delta)$. It might be that such induced three brane charges 
can explain the occurrence of three brane charge quanta in $\ZZ/4$ 
if one tries to 
follow  the transition.

\newsec{Constructions of Calabi-Yau fourfolds}
The classification of Calabi-Yau manifolds with dimension $d\ge 3$ is an 
open problem\foot{It was shown in~\ref\mg{M.\ Gross, Duke Math. Jour. 
{\bf 74} (1994) 271}~that there are, up to birational equivalence, 
only a finite number of families of {\sl elliptically fibred  } 
Calabi-Yau threefolds.}. The purpose of this section is to get a 
preliminary overview over Calabi-Yau fourfolds by investigating simple 
classes: namely hypersurfaces in weighted projective spaces, 
Landau-Ginzburg models and some complete intersections in toric varieties.
Some examples of Calabi-Yau fourfolds appear in \borcea (orbifolds) 
\witten\bls\ref\bs{I. Brunner, R. Schimmrigk, hep-th/9606148} 
(hypersurfaces and
complete intersections)~\pema~ (toric hypersurfaces). 
 
\subsec{Hypersurfaces in weighted projective spaces}
 
There is well studied connection between $N=2$ (gauged) 
Landau-Ginzburg theories and conformal $\sigma$-models 
on Calabi-Yau complete intersections in weighted projective spaces
\ref\vafalg{C. Vafa, \mpl4 (1989) 1169 , \mpl4 (1989) 1615}, 
\ref\wittenlg{E. Witten,
\npb 403 (1993) 159}. 
For example consider a Landau-Ginzburg models which flows in 
the infrared to a conformal theory with $c=3\cdot d$.
If such a model has a transversal quasi-homogeneous superpotential 
of degree $m$, and $r=d+2$ chiral super-fields with positive 
charges (w.r.t. the $U(1)$ of the $N=2$ algebra) $q_i=w_i/m$ 
subject to the constraint 
\eqn\cI{\sum_{i=1}^r(1-2 q_i)=d} 
then it corresponds to a $\sigma$-models on the Calabi-Yau 
hypersurfaces $X_m(w_1,\ldots, w_r)$ of 
degree $m$ in a 
weighted projective space $\IP^{r-1}(w_1,\ldots,w_r)$. 
Due to fixed sets of the $\IC^*$-action of the weighted projective 
space, the Calabi-Yau hypersurface $X_{m}(w_1,\ldots,w_r)$ is in 
general singular. 
The Hodge numbers of the resolved Calabi-Yau hypersurface can be 
obtained from the Landau-Ginzburg model formula for the 
Poincar\`e polynomial of 
the canonical twisted LG model \vafalg, i.e.
\eqn\pp{ {\rm {\bf tr}}\ t^{m J_0}{\bar t}^{m {\bar J}_0}
= \sum_{l=0}^{m-1} 
\prod_{l {w_i\over m} {\ \rm mod\ }\ZZ=0}{1- (t\bar t)^{m-w_i}\over 1-(t\bar t)^{w_i}}
\prod_{l {w_i\over d} {\ \rm mod\ }\ZZ \neq 0} (t\bar t)^{m/2-w_i} 
\left(t\over \bar t\right)^{m ( l {w_i \over d} {\ \rm mod}\  \ZZ-
{1\over 2})}.}
Here the Hodge numbers $h^{p,q}$ are given simply by the degeneracy of
states with $(J_0,\bar J_0)$-charges $(d-p,q)$. 
For $d\leq 3$ there is always a geometrical 
desingularization of theses singularities \ref\roan{S.-S. Roan, Int. J. 
Math. {\bf 2} (1991) 439}.
For $d\ge 4$ there need not be such a geometrical resolution. 
However we note that for
all Landau-Ginzburg models described in the following the relation 
derived from the index theorem \rel\eul~holds, {\sl independent} of whether 
a  geometrical resolution exist or not. This is a hint that the index 
theorem (and many other apparently geometrical aspects relevant to 
$M$ and $F$-theory compactifications) could be stated in terms of an 
internal $N=2$ topological field theory.

To get some overview of this class of Calabi-Yau fourfolds we 
classify first the Fermat type constraints. In these cases, all weights 
divide the degree. It is easy to see that the  maximal allowed 
degree of these configurations growth with 
$m_d=m_{d-1}(m_{d-1}+1)$ ($m=6$ for tori, $m=42$ for 
$K_3$ etc. ) much faster then factorial in the dimension. 
In fact the maximal configuration in dimension $d$ is a fibration
with maximal number of branch points over $\IP^1$ as base, 
whose fibre is in turn the maximal configuration in dimension $d-1$. 
The extreme Calabi-Yau fourfold\foot{Let us use the
notation $X_{m}(w_1,\ldots, w_r)^{h^{1,1},h^{d-1,1}}_{h^{2,1}}$ to 
summarize the three independent Hodge numbers of a fourfold} with
degree $m=326548$ is hence the top of the following vertical chain of 
self mirrors\foot{For the
$K3$ case in the chain the statement is that the Picard lattice of $X_{42}$ can
be identified with the Picard lattice of the mirror. Especially half
of the Picard-lattice has to be invariant under automorphismus by which
the mirror $K_3$ is constructed see e.g. \ref\Kondo{S. Kondo, J. Math. Soc. 
Japan {\bf 44} (1992) 75}.} ($h^{d-1,1}=h^{1,1}$) 
in $d=1,\ldots,4$ 
\eqn\chain{\fivepoint{\eqalign{
X_{3265248}(1,1806,75894,466206,108714,1631721&)^{151700,151700}_{0} \qquad
X_{3612}(1,1,84,516,1204,1806)^{252, 303148}_{0}\cr
&\downarrow \qquad\qquad\qquad\qquad\quad\swarrow\cr
X_{1806}(1,&42,258,602,903)^{251,251}\qquad X_{84}(1,1,12,28,42)^{11,491}\cr  
&\downarrow \qquad\qquad\qquad\qquad\quad
\swarrow\cr
&X_{42}(1,6,14,21)\qquad\quad X_{12}(1,1,4,6)\cr
&\downarrow \qquad\qquad\qquad\swarrow \cr
X_{6}&(1,2,3)^{1,1},}}}
and has $\chi=1\, 820\, 448=24\cdot 75852$. It is the Calabi-Yau fourfold with 
the highest Euler number in this class. There are in total 
$3462$ Fermat type fourfolds\foot{$N=2$ Landau-Ginzburg models with $c=3\cdot d$ 
can have maximally $3\cdot d$ nontrivial ($q_i<{1\over 2}$) fields.
For $d=4$ one has $157,43,14,10,2,1$ Fermat examples for $r=7,\ldots,12$.} 
(to be compared with $147,14,3$ in $d=3,2,1$).  The bounds on the topological
numbers for the Fermat Calabi-Yau fourfold hypersurfaces are
$$\eqalign{
288\le & \chi\le 1\, 820\, 448,\qquad
1\le  h^{1,1}\le 151\, 700,\qquad 
0\le  h^{2,1}\le 1008\cr 
284\le & h^{2,2}\le 1\, 213\, 644\qquad 
60\le  h^{3,1}\le 303\, 148.}$$
Note that all upper bounds up to the last one are saturated by the 
$X_{3265248}$ case, while the configuration with maximal $h^{n-1,1}$  
$$X_{3612}(1,1,84,516,1204,1806)^{252, 303148}_{0},$$
is constructed by taking the minimal number of branch points over $\IP^1$ 
for the top fibration in \chain. Configurations for which 
$\delta=h^{n-1,1}-h^{1,1}$ is maximal\foot{For $K^3$ the statement
is that the invariant part of the Picard Lattice under the mirror automorphism 
is maximal \Kondo.} fit as branches in the chain of $d$-fold fibrations over $\IP^1$ as 
indicated in \chain . Among the $3462$ Fermat cases there are $59$($7$) 
for which 
the Euler number is not divisible by $24$($12$).

Using the transversality conditions \ref\fletcher{A. R. Fletcher,
{\sl Working with Weighted complete Intersections}, Preprint MPI/89-34,
Max-Planck-Institut f. Math. Bonn (1989)}\ref\ks{A. Klemm, R. Schimmrigk,
\npb(411 (1994) 559, M. Kreuzer, H. Skarke \npb388, (1993) 113}~one
can show similarly as in \ks~that the number of all quasi homegenous
hypersurface fourfolds is finite. It is straightforward but very time 
consuming to enumerate all of them. To get an estimate on the number 
of these configurations we note that there are $100\ 559$ configurations 
with\foot{Nine examples appear in \bs . 
Some Other examples of complete intersections 
in products of projective spaces are considered in \bls .} 
$m\le 400$, which exhibits topological numbers  in the range 
$$\eqalign{
-240&\le \chi\le  239\, 232,\qquad 1   \le h^{1,1}\le 173\qquad
0   \le h^{2,1}\le 716,\cr 
82  \le &h^{2,2}\le 159\, 506,\qquad 
6   \le h^{3,1} \le 39\, 840.}$$

Among them there are $21641$ ($9654$) cases for which the Euler number is not 
divisible by $24$ ($12$).      
 
Furthermore a small fraction of it, 138 cases, are  examples of 
Calabi-Yau fourfolds with negative Euler number, which  give 
the possibility to break supersymmetry at least for the $M$-theory compactification 
to three dimensions. E.g. the hypersurface 
$$X_{180}(10,17,36,36,36,45)^{30,36}_{78}$$
has Euler number $\chi=-24$. 
Because of \eul~for $\chi$ to be negative $h^{2,1}$ has to be large.
Elements in $H^{2,1}$, or by the Hodge $*$ and Poincare duality 
we may actually count elements of $H_{3,2}$, are generally generated 
if we have a singular curve $C$ 
of genus $g$ in the unresolved space $X_{sing}$. 
In this example we have a genus $6$ singular 
curve $X_5(1,1,1)$ living in the $x_3,x_4,x_5$ stratum of the 
weighted projective space with a $\ZZ_{36}$ action on its 
transversal space in $X_{sing}$. 
Putting the curve in the origin the singularity 
in the transverse direction is a 
$\IC^3/\ZZ_{36}$, where the $\ZZ_{36}$ acts by phase multiplication by 
$\exp(2\pi i {10\over 36}),\exp(2 \pi i {17\over 36}),
\exp({2 \pi i 45\over 36})$ on the $\IC^3$ coordinates. 
The resolution of this singularity can be described easily 
torically (see section below). It gives rise to a 2 dimensional 
toric variety $E$ whose fan  $\Sigma_E$ is spanned by $\nu_1^*=(-1;-1,-1)$ 
$\nu_2^*=(-1;1,-1)$, $\nu_3^*(-1;11,19)$ from the orign. 
The triangle in the $(-1;0,0)$ plane contains $13$ points in the
interior which correspond to rational surfaces with an intersection form
which will depend on the triangulation of $\Sigma_E$.   
Thus $X$ contains a divisor which has the
fibre structure of a fibre bundle  $E\rightarrow Y \rightarrow C$ 
which contributes $13\cdot 6$ independent $H_{2,3}$-cycles, all of 
them made up  from a $(1,0)$-cycles of the base and the $(2,2)$-cycles 
of the rational surfaces in the fibre. This reasoning will be 
generalized in the toric description to yield formula (4.9). 
Further examples with $\chi \le 0$ appear in table B.2. For many 
cases constructed in the literature as orbifold examples we obtain 
candidates of deformation families\foot{A list containing admissible 
weights and the dimensions of $H^{*,*}$ is available on request.}. 
For example the Hodge numbers of the model discussed in \fourorb \ref\gj{E. Gimeon and C. Johnson, {\sl Multiple Realizations 
of $N=1$ Vacua in Six-Dimensions} hep-th/9606176} coincide with the
deformation family $X_{47}(3,5,7,8,11,13)^{100,4}_0$.

\subsec{Elliptic fibrations with sections and multisections as complete 
intersection CY}

We will describe  here a method for constructing elliptic 
fibred Calabi-Yau spaces as hypersurfaces in 
weighted projective spaces. The starting point
are the elliptic curves  
\eqn\ellc{\eqalign{
 E_6\ :& \quad X_{3}(1,1,1)=\{
x^3+y^3+z^3-s x y z=0\ |\ (x,y,z)\subset \IP^2(1,1,1)\}\cr
 E_7\ :& \quad X_{4}(1,1,2)=\{
x^4+y^4+z^2-s x y z=0\ |\ (x,y,z) \subset \IP^2(1,1,2)\}\cr
 E_8\ :& \quad X_{6}(1,2,3)=\{
x^6+y^3+z^2-s x y z=0\ |\ (x,y,z) \subset \IP^2(1,2,3)\}\cr 
 D_5\ :& \quad X_{2,2}(1,1,1,1)=\left\{ \matrix{&x^2+y^2-szw=0\cr  
 &z^2+w^2-sxy=0}\ \bigg|\ (x,y,z,w)\subset \IP^3(1,1,1,1)\right\},}}
which will appear as the generic fibers. Here we included the complete
intersection case $D_5$. We will focus in the following 
mainly on the first three cases. 

In the third case there are birational equivalent
representations, which give rise to additional
possibilities to construct the fibration space. 
To find them, consider the $\IC^*$-action 
$\sigma:(x\rightarrow \rho x,y\rightarrow \rho^2 y,x\rightarrow \rho^3 x)$,
with $\rho^6=1$ and construct the possible fractional transformations,
which are well defined under this action. There are two series
of fractional transformations, 
\eqn\ab{
(1) :\quad \matrix{
x&=\xi^{{2\over 3}+k}\ \cr
y&=\xi^{{1\over 3}} \eta\cr
z&=\zeta\phantom{\kappa^{{1\over 3}}}},\qquad\qquad
(2) :\quad \matrix{
x&=\xi^{{1\over 2}+k}\ \cr
y&=\eta\phantom{\kappa^{{1\over 2}}}\cr
z&=\xi^{1\over 2}\zeta}
}
which identify $X_{6}(1,2,3)$ with the following representations
\eqn\ar{\eqalign{
E_8' :\qquad & X_{4+6k}(1,1+2k,2+3k)=\cr & 
\qquad \{
\xi^{4+6k}+\xi \eta^3+\zeta^2-s \xi^k \eta \zeta=0| 
(\xi,\eta,\zeta)\subset \IP^2(1,1+2k,2+3k)\}
\cr
E_8'' : \qquad & X_{3+6k}(1,1+2k,1+3k)=\cr &
\qquad \{\xi^{3+6k}+\eta^3+\xi\zeta^2-s \xi^k \eta \zeta=0| 
(\xi,\eta,\zeta)\subset \IP^2(1,1+2k,1+3k)\}.}}

Our construction of elliptic fibred Calabi-Yau hypersurfaces (complete
intersections) will proceed by the following general process 
\eqn\rec{X^{(0)}_{d_1,\ldots,d_k}(w_1^{(0)},
w^{(0)}_2,\ldots w^{(0)}_{r^{(0)}})\rightarrow 
X^{(1)}_{pd_1,\ldots,pd_k}(w_1^{(1)},w_2^{(1)},\ldots,w_{r^{(1)}}^{(1)},p w^{(0)}_2,\ldots, 
pw^{(0)}_{r^{(0)}})}
with $\sum_{i=1}^{r^{(1)}}w_i^{(1)}+p\sum_{i=2}w^{(0)}_i=p 
\sum_{i=1}^k d_i$.
 In this cases the base is given by

This construction is a simple generalisation of the one used in 
\ref\klm{A. Klemm, W. Lerche and P. Mayr,\plb 357 (1995) 313} 
to get threefolds with $K_3$ fibre. It was used in 
\ref\hlyII{S. Hosono, B. Lian S.-T. Yau,alg-geom/9603020} to produce 
more such examples and in \bs~to get some fourfold configurations. 
Iteration of this process, with say $ r^{(i)}=2$ $i>0$, lead to 
sequences, e.g. for 
the $X_3$ case,
$$X_3(1,1,1)\ \rightarrow\  
\matrix {
X_6(1,1,2,2)&\rightarrow \cr
X_9(1,2,3,3)&\hfil           \cr
X_{12}(1,3,4,4)&\hfil\cr
X_{15}(2,3,5,5)&\hfil }
\ \ 
\matrix { X_{12}(1,1,2,4,4)&\rightarrow \cr
 X_{18}(1,2,3,6,6)&\cr
 X_{24}(1,3,4,8,8)&\cr
 X_{30}(1,4,6,6,6)&\cr
   \vdots\quad & \cr
\phantom{X_9(1,2,3,3)}&\cr
\phantom{X_{12}(1,3,4,4)}&\cr
\phantom{X_{15}(2,3,5,5)}}
\ \
\matrix {X_{24}(1,1,2,4,8,8)&\rightarrow \ldots \cr
X_{24}(1,2,3,6,12,12)& \cr
   \vdots\quad & \cr
\phantom{X_{18}(1,2,3,6,6)}&\cr
\phantom{X_{24}(1,3,4,8,8)}&\cr
\phantom{X_{30}(1,4,6,6,6)}&\cr
\phantom{\vdots\quad} & \cr
\phantom{X_9(1,2,3,3)}&\cr
\phantom{X_{12}(1,3,4,4)}&\cr
\phantom{X_{15}(2,3,5,5)}}
$$  
in which fiber of the threefold is itself 
an elliptic fibered $K_3$ and so on.  The birational equivalent cases
\ar~can be treated similarly. The table B.3  contains a complete 
list of all $K_3$ hypersurfaces which are obtained in the first 
step from this process.

Let us investigate some general properties of these types of fibrations. 
The condition for triviality of the 
canonical bundle of $X$ follows from the analysis in 
\ref\can{Y.\  Kawamata, J. Fac. Sci. Univ. Tokyo Sec. IA 
{\bf 30} (1983) 1; T. Fujita, J. Math. Soc. Japan {\bf 38} 20; 
N.\ Nakayama, in Algebraic Geometry and Commutative Algebra vol. II,
Kinokuniya, Tokyo (1988) 405, A.\ Grassi, Math. Ann. {\bf 290} (1991) 287}.
As summarized in \mvI~ one 
can choose a birational model to get a Calabi-Yau with $K_X=0$ if
\eqn\can{K_B=-\sum a_i [B_i'],}
where $B_i'$ is a divisor in the base $B$ and $a_i$ follows from
the type of singularity of the fibre over $B_i'$ according to 
Kodaira's list of singular fibres for Weierstrass models in table A.1. 

Our first aim is to relate the Euler number of the total 
space to topological data of the base. In the following we first 
concentrate on cases which have a section 
(or multisection) and for which the fibre degenerates no worse 
than with the $I_1$ fibre over codimension one in the base.
That means that the discriminant $\Delta$ of the normal form 
of the elliptic fibre vanishes with ${\rm ord}~ \Delta=1$, 
while the coefficient functions $e,f,g$ are generic (see section 5.1). 
Proofs of the formulas for the Euler numbers can be found in section 7,8. 
The $d=4$ $X_6$ case was first treated in \SVW .
 
If the dimension of the total space $X$ is $d=3$ we have the following formula
\eqn\eIII{\chi(X)=-2\cdot C_{(G)} \cdot \int_{B} c^2_1(B),}
where $\int_B c_1^2(B)$ is the integral of the square of the first 
Chern class over the base and $C_{(G)}$ is the dual Coxeter number of the 
group associated with the elliptic fibre \ellc, 
$$C_{(E_8)}=30,\quad C_{(E_7)}=18, \quad C_{(E_6)}=12,\quad C_{(D_5)}=8.$$ 
Using \rec~, with $r^{(1)}=3$, we can provide examples with $B=\IP^2$ 
for these cases 
$$\eqalign{
&X_{18}(1,1,1,6,9)^{2(0),272},
\quad X_{12}(1,1,1,3,6)^{3(1),165},
\quad X_{9}(1,1,1,3,3)^{4(2),112},\cr
&X_{6,6}(1,1,1,3,3,3)^{5(3),77.}}$$
{}From the index theorem (2.2) and $\chi(\IP^2)=c_2^B=3$ we conclude 
$\int_B c_1^2(B)=9$ and application of \chiIII~gives 
$\chi=2(h^{1,1}-h^{2,1})$.
We can represent these manifolds torically as described in the
next section. $\IP^2$ is then encoded in the fan spanned by 
$(1,0),(0,1),(-1,-1)$ and the blow up  can be represented torically by
adding the successively the vectors $(-1,0),(0,-1)$ and $(1,1)$ to the
$\IP^2$ fan. This enhances $h^{1,1}(B)$ and hence the Euler number of 
the bases by $1$, but does not introduce singularities of the fibre 
in higher codimension therefore $h^{1,1}(X)\rightarrow h^{1,1}(X)+1$ and 
by \eIII~we get chains of models with 
$(h_{(i+1)}^{1,1}(X),h_{(i+1)}^{2,1}(X))$=
$(h_{(i)}^{1,1}(X)+1,h_{(i)}^{2,1}(X)-C_{(G)}+1)$. Transitions
of this type involve the vanishing of real 2 (d-1)-cycles and for 
$d=3$ they have been analysed in \mvII\ref\kvm{A.\ Klemm, 
P.\ Mayr and C.\ Vafa, {\sl BPZ States of Exceptional Non-Critical 
Strings}, hep-th/9607139}\ref\lsty{J.\ Louis, J.\ Sonnenschein, 
S.\ Theisen and S.\ Yankielowicz, {\sl Non-perturbative properties
of heterotic String Vacua conpactified on $K3\times T^2$}, 
hep-th/9606049} and we generalise this situation to $d=4$ in section 5.

For the general dimension $d$ of $X$ we show that  
\eqn\eulel{\chi(X)= a \sum_{r=1}^{d-1} (-1)^{r-1} b^r\
\int_B c_1^r(B) c_{d-r-1}(B)}
with $a=2,3,4$, $b=6,4,3$ for the $E_8$, $E_7$, $E_6$ fibre respectively.
For $D_5$ the Euler number likewise  only depends on the 
Chern classes of the base. Let us summarize the formulas for 
$d=4$  
\eqn\eulfoursmooth{\eqalign{
&E_8:\  
\chi(X)=12 \int_B c_1 c_2+360\int_B c_1^3,\qquad 
E_7:\
\chi(X)=12 \int_B c_1 c_2+144 \int_B c_1^3,\cr
&E_6:\  
\chi(X)=12 \int_B c_1 c_2+72 \int_B c_1^3, \phantom{0}\qquad 
D_6:\
\chi(X)=12 \int_B c_1 c_2+36 \int_B c_1^3.}}

The study of examples with low Picard numbers has helped a lot to
establish the $N=2$ Type II/hetetoric duality in four dimensions.
Fourfold cases with low Picard numbers are expected to play a role in 
the investigation of the dynamics of $M$ theory compactifications 
to three dimensions and  $N=1$  $F$-theory/heterotic duality in four
dimensions. For the general LG-models we found respectively 
$31,108,255,411,508,800$ 
configurations with $h^{1,1}=1,2,3,4,5,6$. The ones which have an 
elliptic fibration of type $E_6, E_7, E_8, E_8',E_8''$, which is apparent 
in the patches of the weighted projective space are collected in 
table B.4. 

It is clear from table B.4 and \eulfoursmooth~ that the cases in 
which the fibre degenerates only to $I_1$ are very rare. 
Such cases are for instance (5,9,27), where the base is 
$\IP^3$ with $\int_{\IP^3} c_1^3=64$. 
Let us check for these manifolds \can~and the fact that the fibre 
degenerates with $I_1$ over a generic point of the codimension one 
locus. $c_1(\IP^n)=n [H]$, where $[H]$ is the hyperplane class. 
So $K_B=-n [H]$ and from section (5.1) we see that the discriminant
$\tilde \Delta=0$ is a singular degree $12 n$ polynomial in 
$\IP^n$, i.e. $[\tilde \Delta]=-12 K_B$. However $f,g,h$ are are generic
such that over codimension one the fibre degenerates to $I_1$. As 
$d\tilde \Delta=3 f df + 2 g dg $ (e.t.c)  $\tilde \Delta $ will degenerate
in codimension two  at $f=g=0$ to a cusp, but this does not contribute 
to \can . So $a=12$ and hence $[\tilde \Delta]= - K_B$.  Similar 
cases are (23,41,79) where the base is  a $\IP(\cO_{\IP} \otimes 
\cO_{\IP}(3))\rightarrow B\rightarrow \IP^2$ bundle with 
$\int_B c_1^3=72$ and case (109) where the base has a 
$\IP(\cO_{\IP} \otimes \cO_{\IP}(4))\rightarrow B\rightarrow 
\IP^2$ structure with $\int_B c_1^3=86$ etc.

The $E_8'$, $E_8''$ cases are very interesting because the 
Weierstrass form degenerates for them over codimension 
one in the base. For example for the $E_8'$ case (60) in table B.4 
the Weierstrass form degenerates to a conic bundle for 
$x_4=0$, which splits over codimension two in the base into 
pairs of lines. In this respect it is very similar to the case 
$X_{20}(1,1,2,6,10)$ described in~\bkkm . Similar as in \bkkm~it is
part of a chain of transitions $(110)\rightarrow (60)\rightarrow (23)$, 
which is analogous to the $X_{18}(1,1,2,6,8)\rightarrow X_{20}(1,1,2,6,10)
\rightarrow X_{24}(1,1,2,8,12)$ transitions. Note that the Euler number 
of (110) and (23) is divisible by $24$ while the one of (60) only by six.
We will discuss such chains further in the toric setup in section 6.   

Most of the time the models of table B.4 have a much more 
intricate singularity structure over the base. As these
give rise to gauge groups, matter spectrum and more exotic physics 
in the low energy field theory, it is very important to investigate 
these cases. It turns out however that the realisation of simple 
generalisations e.g. to gauge groups without matter are easier to 
engineer in the toric framework, which we will do in the 
next section.

\newsec{Toric construction and mirror symmetry for
Calabi-Yau Fourfolds}

Next we consider a generalization of the previous construction 
namely a $d$-dimensional hypersurfaces $X$ in a compact toric variety
\foot{See e.g. \ref\fulton{W. Fulton, 
{\sl Introduction to Toric Varieties}
Princeton University Press, Princeton 1993} and section 2.6 for the 
construction of $\IP_{\Delta^*}$.}
$\IP_{\Delta^*}$. This hypersurface is defined by the zero 
set of the Laurent polynomial~\ref\batyrevI{V.\ Batyrev, {\sl Dual Polyhedra 
and Mirror Symmetry for Calabi-Yau Hypersurfaces in Toric Varieties}, 
J. Alg. Geom. 3 (1994) 493} 
\eqn\htoric{P=\sum_{\nu^{(i)}} a_i U_i=0,\ {\rm  where}\  
U_i=\prod_{k=1}^{d+1} X^{\nu_k^{(i)}}} 
and $\nu^{(i)}$ are the integral points in $M\sim \ZZ^{d+1}$, 
whose convex hull defines the polyhedron $\Delta$. 
The hypersurface \htoric~defines a Calabi-Yau space if 
$\Delta$ contains the origin as the only interior point \batyrevI .
The polar polyhedron 
$\Delta^*=\{y\in M^* |\langle x,y\rangle =-1,\ \forall x\in \Delta\}$ 
is likewise the convex hull of integral points $\nu^{*(i)}\in M^*$
with this property. Such a pair of polyhedra $(\Delta,\Delta^*)$
is called reflexive pair. Note that $(\Delta^*)^*=\Delta$.

In \batyrevI~Batyrev has given the following combinatorial formulas for 
$h^{1,1}$ and $h^{d-1,1}$ in terms of the numbers of 
points in $(\Delta,\Delta^*)$:

\eqn\hp{\eqalign{h^{1,1}(X_\Delta)&=h^{d-1,1}(X_{\Delta^*})\cr
  &=l(\Delta^*)-(d+2)-\sum_{{\rm dim}\Theta^*=d} l'(\Theta^*)
    +\sum_{{\rm codim}\Theta_i^*=2} l'(\Theta_i^*) l'(\Theta_i),\qquad a.)
\cr
h^{d-1,1}(X_\Delta)&=h^{1,1}(X_{\Delta^*})\cr
  &=l(\Delta)-(d+2)-\sum_{{\rm dim}\Theta=d} l'(\Theta)
    +\sum_{{\rm codim}\Theta_i=2} l'(\Theta_i) l'(\Theta_i^*), \qquad b.)}}
where $\Theta$ ($\Theta^*$) denotes faces of $\Delta$ ($\Delta^*$),
$l(\Theta)$ is the number of all points of a face $\Theta$ and 
$l'(\Theta)$ is the number of points inside that face. In the last
term the sum is over dual pairs $(\Theta_i,\Theta^*_i)$ of faces.
The fan $\Sigma(\Delta^*)$ {\sl over} $\Delta^*$ defines in the standard 
way \fulton~a toric variety $\IP_{\Delta^*}(\Sigma(\Delta^*))=
\IP_{\Delta^*}$ in which $X$ is embedded. 

The following facts are relevant for the discussion of the divisors

\noindent
{\bf i.)}  {\sl Divisors and sub-manifolds in} $\IP_{\Delta^*}$: 
Every ray $\tau_k$ through a  point $P_k$ in $\Delta^*$ (or more generally
a cone in $\Sigma(\Delta^*))$ defines a $\IQ$-Cartier divisor 
(or more generally a sub-manifold) in $\IP_{\Delta^*}$, 
denoted $D'_k:=V(\tau_k)$, which by itself has a very simple 
toric description. Take all cones $\cS_k=\{\sigma_{k_i}\}$ for which 
$\tau_k$ is a face and consider the image of $\cS_k$ in 
$M^*(\tau)=M^*/M^*_{\tau_k}$, 
where $M^*_{\tau_k}$ is the sub-lattice of $M^*$ generated by vectors in 
$\tau_k$. This image is called ${\rm star}(\tau_k)$ and can be visualized
as the projection of the $\cS_k$ along $\tau_k$ 
on the hyperplane perpendicular to $\tau_k$. Now $V(\tau_k)$ is the toric 
variety constructed from the fan over ${\rm star}(\tau_k)$. Especially 
all these divisors in $\IP_{\Delta^*}$ have $h^{0,0}=h^{d,d}=1$ and $h^{i,j}=0$ for $i\neq j$ and
one can construct $l(\Delta^*)-(d+2)$ independent divisors classes 
which are a basis for $H^d(\IP_{\Delta^*})$.

\noindent
{\bf ii.)} {\sl Divisors and sub-manifolds in} $X$: The intersections
$D_K=D_K'\cap X$ leads to divisors in $X$. In fact the divisors 
classes $[D_K]$ obtained this way generate $H^{d-1}(X)$.  
The manifold $X$ can be thought as being constructed from a singular 
variety $X_{sing}$ with quotient singularities along subsets $R_K$ 
of $codim>1$, which are induced from quotient singularities of the
ambient space $\IP_{\Delta^*}$. The divisors $D_K$ will therefore be 
bundles of exceptional components $E_k$ coming from the desingularisation 
of the ambient space over the regular component $R_k$. The dimension of the 
regular and singular components depend simply on the dimension of the 
face of  $\Delta^*$ on which the point $P_i$ lies. 
The real dimension $d_{\Theta^*_k}$ of the face $\Theta^*_k$  is 
the complex dimension $d_{E_k}$ of the exceptional component of $D_k$, while
the  complex dimension of $R_k$ is $d_{R_k}=d-1-d_{E_k}$. In fact 
$E_k$ and $R_k$ have a very simple toric description.
If $\Theta^*_k$ is a face of the $d+1$ dimensional polyhedron 
$\Delta^*_i$ then the dual face 
$\Theta_k$, of dimension ${\rm dim}(\Theta)=d-{\rm dim}(\Theta^*_k)$,
is defined as 
\eqn\df{\Theta_k= 
\{ u\in \Delta|\langle u,v\rangle=-1, \forall v\in \Theta^*_k\}.}
The sets $R_i$ can be viewed as  $D'_i\cap X_{sing}$ and are constructed
as follows. Remember that the coordinate ring
of the singular ambient space  is generated by the corners $E_i$  of 
$\Delta^*$, especially $X_{sing}$ is given in this coordinates by the 
vanishing of 
\eqn\sing{p=\sum_{i=1}^{l(\Delta)} a_i \prod_{j=1}^{\# E}
x_j^{\langle \nu^i , E_j\rangle}.}
Now from the construction of $D'_i$ as above it is clear that 
$D_k \cap X_{sing}$ is given by the vanishing of
\eqn\intersection{p_k=\sum_{i=1}^{l(\Theta_k)}a_{k_i} \prod_{j=1}^{
\# E(\Theta^*_k)}
x_j^{\langle \nu^i , E_j(\Theta^*_k)\rangle},}
where $E_j(\Theta^*_k)$ are the corners of the face $\Theta^*_k$.
The structure of the exceptional component of $D_k$ is given by
the toric variety constructed from ${\rm star}' (\tau_k)$;  
the projection of $\cS_k$ on $\Theta^*_k$ along $\tau_k$. This implies
especially that $h^{0,0}=h^{ d_{\theta^*_k}, d_{\theta^*_k}} =1$ and
$h_{i,j}=0$ if $j\neq j$ \fulton . 
Particularly useful is the fact that number of parameters by which we 
can move $R_k$ in $X$ namely $l(\Theta_K)$ is also the dimension of 
$H^{d_{R_k},0}(R_k)$, i.e. 
\eqn\usefull{h^{d_{R_k},0}(R_k)=l(\Theta_k).}

This structure gives a useful classification of the
divisors in $X$ in types (a-d) below just according to the
dimension of the face on which $\tau_k$ lies. 

\noindent
{\bf o.)} $d_{\Theta_k^*}=d$, then $\Theta_k$ is a point
and $R_k=\{p_k=0\}=\emptyset$. Therefore divisors associated with
these points have no intersection with $X$ and the corresponding points 
are therefore  subtracted in the third term in \hp~ a).   

\noindent
{\bf a.)} $d_{\Theta_k}^*=d-1$, then $\Theta^k$ is one dimensional 
and $R_k=\{Q_i|i=1,\ldots,deg(p_k)\}$ are points  in $X$ whose number
is given by the the degree of $p_k$ or equivalently by $l(\Theta_k)+1$. 
The fact that one has $l(\theta_k)+1$ divisor components $D^i_k$ of the
type $p_i\times E_k$ explains addition of the fourth term in \hp~a). 
That $E_k$ is toric variety implies  $h^{i,j}(D^i_k)=0$ 
if $i\neq j$ and in particular $\chi(D^i_k,\cO_{D^i_k})=1$.
So this case leads to divisors for which a  
non-perturbative superpotential due to five fivebrane wrappings 
is generated. 

\noindent
{\bf b.)} $d_{\Theta_k^*}=2$, $E_k$ are rational surfaces, while $R_k$ are 
Riemann surfaces whose genus $g$ is by \usefull~the number of points
inside $\Theta_k$ i.e. $l(\Theta_k)$. In this case
we get $l(\Theta^*_k)\cdot l(\Theta_k)$ $(3,2)$-forms from the 
pairing of the $(1,0)$-forms on $R_k$ with the $(2,2)$-forms of 
the $E_k$, which leads to the generalization of \hp~given below.
Especially we have for the irreducible component of the divisor
$h^{0,0}(D_k)=1$, $h^{1,0}(D_k)=l(\Theta_k)$, $h^{2,0}(D_k)=0$, 
$h^{3,0}(D_k)=0$. 

\noindent
{\bf c.)} $d_{\Theta_k^*}=1$ $E_k$ is a $\IP^1$ (in  general in a 
Hirzebruch Sphere three) and $R_k$ is a hypersurface in a 
three dimensional toric variety with
$h^{2,0}(R_k)=l(\theta_k)$, moreover we can use the Lefschetz theorem
to conclude $h^{1,0}(R_k)=0$. In this case we get additional $(3,1)$ forms
form the pairing of $(2,0)$-forms of $R_k$ with the $(1,1)$-forms 
of $E_k$, which gives rise to the fourth term in \hp~b). A 
superpotential is  generated if $l(\Theta_k)=0$.

\noindent
{\bf d.)} $d_{\Theta_k^*}=0$ in this case $D_k=R_k$. Similar as in \witten~
one can argue with the Lefschetz theorem that $h^1(D)=h^2(D)$ is zero,
so that $\chi(D_k,\cO_{D_k})=1-l(\Theta_k)$. Usually $h^3(D_k)$ is expected
to be very positive so that $D$ is movable and $\chi(D_k,\cO_{D_k})\le 1$.
However in toric varieties due to conditions imposed by the weights 
this deformation space can be actually very restricted
so that 
one can easily construct cases in which $h^3(D)=0$ for divisors of 
type d.), i.e. this divisors can lead to a non-perturbative 
superpotential. To summarize we have 
\eqn\chioD{\chi(D_k,\cO_{D_k})=1-(-1)^{dim(\Theta_k)} l(\Theta_k).} 

It should be clear by the above that $\chi(D,\cO_{D})=1$ 
divisors classes can be made abundant in the toric constructions 
of CY-manifolds.  To illustrate this point take the  mirror of any 
fourfold with small Picard number, e.g. the mirror of  
the sixtic in $\IP^5$. $\Delta^*$ is
now the  Newton polyhedron of the sixtic which has $6,75,200,150,30,1$ 
points on dimension $0,1,2,3,4,5$ faces, which lead,  as the $\Delta$ has 
only 6 corners and the inner point such that $l(\Theta_k)=0$, 
all to $\chi(D,\cO_D)=1$ divisors.  Some examples of this type of divisors 
have been considered in \witten\bls\pema .  
Very frequently one encounters the
situation were $l(\Theta_k)=1$, which means $\Theta_K$ is
a reflexive polyhedron of lower dimension and $c_1(R_k)=0$. 
The compactification of the fivebrane on such a divisor could 
lead to a sub-sector in the $N=1$ theory with enhanced supersymmetry.

Mirror symmetry implies for the Hodge diamonds of a mirror pair
$X,X^*$ that 
\eqn\mirprop{h^{p,q}(X)=h^{d-p,q}(X^*).} 
For threefolds this property follows from \hp~as $h^{2,1}(X)$ and 
$h^{1,1}(X)$
are the only independent Hodge numbers, if we construct 
$X^*=X_{\Delta^*}$ from $\Delta^*$ in the same way as $X=X_{\Delta}$ 
is constructed from  $\Delta$.

For fourfolds we have from \hp~$h^{3,1}(X)=h^{1,1}(X^*)$ but since we
have one more independent Hodge number we also have to establish 
$h^{2,1}(X)=h^{2,1}(X^*)$. This follows from the discussion of c.) above, 
which gives the formula
\eqn\othercohom{h^{d-r,1}(X)=h^{r,1}(X^*)=
\sum_{{\rm codim}\ \Theta_i=r+1} 
l'(\Theta_i) \cdot l'(\Theta_i^*), \  {\rm for}\ d-1> r>1.} 
Together with \rel~it shows for four-folds that $X$, $X^*$ as constructed 
from $\Delta,\Delta^*$ have indeed the mirror Hodge diamond.

It is somewhat more complicated to obtain $h^{2,2}(X)=h^{2,2}(X^*)$ 
directly from the polyhedron. If mirror symmetry is true however, 
then one expects to have very good control over $H^{2,2}(X)$ as
\eqn\split{H^{2,2}(X)=H^{2,2}_{prim}(X)\oplus H^{2,2}_{prim}(X^*),} 
were $H^{2,2}_{prim}(.)$ denotes the primitive part of the cohomology. 
This gives of course also a way of counting 
$h^{2,2}$ directly\foot{E.g. for the sixtic in $\IP^5$ 
($h^{1,1}=1,h^{2,1}=0,h^{3,1}=426$) it is easy to see that 
$h^{2,2}(X)=h^{2,2}_{prim}(X)+h^{2,2}_{prim}(X^*)=1+{\rm dim}
\left(\IC[x_1,\ldots,x_6]/\partial P|_{{\rm deg}=12}\right)=1752$, as
it also follows from the index theorem. Here $P$ is a degree $
6$ polynomial in $x_1,\ldots,x_6$. }.

To every quasi homogeneous polynomial $p$ in $d+2$ variables, like the
one discussed in the last section, we can associate a 
Newton polyhedron $\Delta_p$ by considering the $(d+2)$-tuples of 
the exponents of the monomials of $p$ as coordinates of points 
in $\IR^{d+2}$ and building the convex hull of them. 
Quasi homogeneity of $p$ implies that $\Delta_p$ lies in a hyperplane 
in $\IR^{d+2}$, while \cI~implies that $(1,\ldots,1)$ is always 
an interior point of $\Delta_p$, which we shift in the origin of 
$\IR^{d+1}$. For $d\le 3$ transversality of $p$ implies reflexivity 
of $\Delta_p$. That was actually shown by construction 
\ref\cok{P. Candelas, X. d. Ossa, S. Katz, \npb 372 (1995) 127} (see also 
\ref\hktyI{S. Hosono, A. Klemm, S.Theisen, S. T. Yau, \cmp 167 (1995) 301}). 
For $d\geq 4$ this property does not hold. A simple counter example 
is the manifold $X_7(1,1,1,1,1,2)$.

\newsec{Toric four-folds over Fano Bases.} 

Fano varieties of dimension two, so called del Pezzo surfaces, are
$\IP^2$, $\IP^1\times \IP^1$ and $\IP^2$ blown up in up to eight 
points. There are five toric del Pezzo surfaces classified in 
\ref\batyrev{V. V. Batyrev, {\sl Toroidal Fano 3-folds}, Math. 
USSR-Izv {\bf 19} (1982), 13-25, Izv. Akad. Nauk SSSR, Ser. Mat. 
{\bf 45} (1981), 704-717}. $\IP^2$, $\IP^1\times\IP^1$,
the Hirzebruch surface $F_1$, the equivariant blow up of $\IP^2$ 
at two points $B_2$, and the equivariant blow up of $\IP^2$ at
three points $B_3$.

There are 84 Fano varieties of dimension three which were
classified by Iskovskih and Mori-Mukai \ref\mm{S. Mori and S. Mukai,
{\sl On Fano 3-folds with $B_2\ge 2$}, in Algebraic Varieties and 
Analytic Varieties, Ed. S. Iitaka, Adv. Studies in Pure Math. 1 
(1983) 101}. From these we will consider the 18 which can be 
represented in toric varieties (see \ref\oda{T. Oda, {\sl Convex Bodies
and Algebraic Geometry}, Springer Verlag, Berlin Heidelberg (1988)} for
a review). From \batyrev \ref\watanabe{K.\ Watanabe, M.\  Watanabe,
{\sl The classification of Fano 3-folds with torus embeddings}, 
Tokyo J. Math. {\bf 5} (1982) 37-48 }~we have

\noindent 
(1) $\IP^3$

\noindent 
(2) $\IP^1\times \IP^2$

\noindent 
(3) The $\IP^1$-bundle $\IP({\cal O}_{B'}\oplus {\cal O}(1)_{B'})$ 
    over $B'=\IP^2$

\noindent 
(4) The $\IP^1$-bundle $\IP({\cal O}_{B'}\oplus {\cal O}(2)_{B'})$ 
    over $B'=\IP^2$

\noindent 
(5) The $\IP^2$-bundle $\IP({\cal O}_{B'}
    \oplus{\cal O}_{B'}\oplus {\cal O}(1)_{B'})$ 
    over $B'=\IP^1$

\noindent 
(6) $(\IP^1)^3$

\noindent 

\noindent 
(7) The $\IP^1$-bundle $\IP({\cal O}_{B'}\otimes {\cal O}_{B'}(f_1+f_2))$
over ${B'}=(\IP^1)^2$, where $f_1$ and $f_2$ are fibres of the two 
projections from ${B'}$ to $\IP^1$

\noindent 
(8) The $\IP({\cal O}_{B'}\otimes {\cal O}_{B'}(f_1-f_2))$ bundle over 
$\IP^1\times \IP^1$.

\noindent 
(9) $\IP^1\times F_1$ where $F_1$ is the Hirzebruch surface.

\noindent 
(10) The $\IP^1$-bundle $\IP({\cal O}_{B'}\otimes {\cal O}(s+f))$ over $F_1$, 
where $f$ is the fibre from $F_1$ to $\IP^1$, while s is the minimal
cross section for the projection with $-1$ as self-intersection number.

\noindent 
(13) $\IP^1\times B_2$ with $B_2$ as above

\noindent 
(17) $\IP^1\times B_3$ with $B_3$ as above
 
The other cases are equivariant blow ups of the ones mentioned. This
can be seen from the concrete fans below and is depicted in figure 1. 
Let us denote by $e_1=(1,0,0)$, $e_2=(0,1,0)$, $e_3=(0,0,1)$ unit
vectors which span a rectangular lattice in $\IR^3$.
Then we can represent the toric varieties by the complete
fans spanned by the following vectors 

\def\smi{{\fivepoint -}}
\noindent
(1)  $(e_1,e_2,e_3,\smi e_1\smi e_2\smi e_3)$, \quad
(2)  $(e_1,e_2,e_3,\smi e_1\smi e_2,\smi e_3)$, 

\noindent
(3)  $(e_1,e_2,e_3,\smi e_2,\smi e_1\smi e_2\smi e_3)$, \quad
(4)  $(e_1,e_2,e_3,\smi e_2,\smi e_1\smi 2 e_2\smi e_3)$,

\noindent
(5)  $(e_1,e_2,e_3,\smi e_1\smi e_2\smi e_3,\smi e_1\smi e_3)$ \quad
(6)  $(e_1,e_2,e_3,\smi e_1,\smi e_2,\smi e_3)$,

\noindent
(7)  $(e_1,e_2,e_3,\smi e_1\smi e_3,\smi e_2\smi e_3,\smi e_3)$, \quad
(8)  $(e_1,e_2,e_3,\smi e_1\smi e_3,e_3\smi e_2,\smi e_3)$,

\noindent
(9)  $(e_1,e_2,e_3,\smi e_2,e_2\smi e_1,\smi e_3)$, \quad
(10) $(e_1,e_2,e_3,\smi e_1\smi e_3,e_1\smi e_2,\smi e_3)$,

\noindent
(11) $(e_1,e_2,e_3,e_3\smi e_2,\smi e_2,\smi e_1\smi e_2\smi e_3)$, \quad
(12) $(e_1,e_2,e_3,e_3\smi e_2 ,\smi e_1\smi e_3,\smi e_2)$,

\noindent
(13) $(e_1,e_2,e_3,e_2\smi e_1,\smi e_2,e_1\smi e_2,\smi e_3)$, \ \ 
(14) $(e_1,e_2,e_3,e_2\smi e_1,\smi e_2,e_1\smi e_2,e_1\smi e_2\smi e_3)$

\noindent
(15) $(e_1,e_2,e_3,e_2\smi e_1,\smi e_2,e_1\smi e_2,\smi e_2\smi e_3)$
,\ \ 
(16) $(e_1,e_2,e_3,e_2\smi e_1,\smi e_2,e_1\smi e_2,e_1\smi e_3)$,
 
\noindent
(17) $(e_1,e_2,e_3,\smi e_1,\smi e_2,\smi e_3,e_1\smi e_2,e_2\smi e_1)$
(18) $(e_1,e_2,e_3,e_2\smi e_1,\smi e_1,\smi e_2,e_1\smi e_2,\smi e_1\smi e_3)$

 \goodbreak\midinsert
 \centerline{\epsfxsize 3.5truein\epsfbox{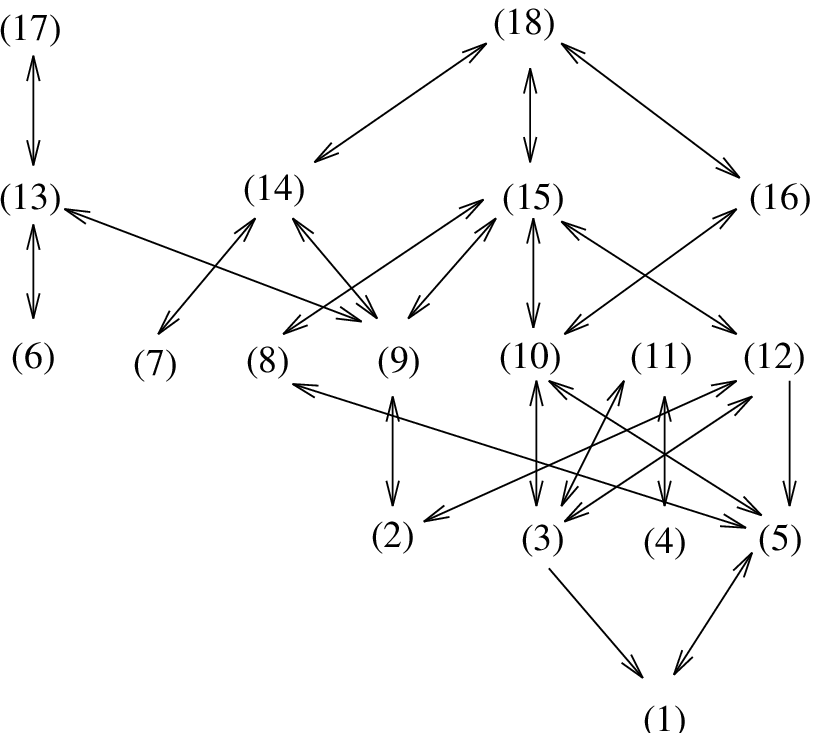}}\leftskip 1pc
 \rightskip 1pc\noindent{\ninepoint\sl \baselineskip=8pt {\bf Fig.1}:

The net of equivariant blow ups (downs) among the 
fano bases as in \watanabe. The blow ups are either at points 
$\downarrow$ or along one dimensional closed irreducible subvarities 
$\updownarrow$, which are stable under the torus action. 
By the construction below, they will be promoted to transitions 
between elliptically fibred CY fourfolds.}\endinsert
To construct $d$-dimensional elliptic fibration Calabi-Yau 
manifolds $X$ over this base spaces $B$ we consider polyhedra 
which are obtained from the toric description of the base spaces
as follows. We define the vectors in the rectangular $\ZZ^{d+1}$
latticed in $\IR^{d+1}$ 
$$\eqalign{
v_A&=(\underbrace{0,\ldots,0}, 2,3), \ \ 
v_B=(\underbrace{0,\ldots,0},1,2), \ \ 
v_C=(\underbrace{0,\ldots,0},1,1)\cr 
&\phantom{=(0.)}^{d-1},\phantom{(0,2,3),\ \ v_B=()}^{d-1}
\phantom{(0,2,3),\ \ v_B=(0.)}^{d+1}\cr
&{\rm as\  well\  as\ } 
e_{d}=(\underbrace{0,\ldots ,0},1,0)\ {\rm and\ }\ 
e_{d+1}=(\underbrace{0,\ldots ,0},0,1).\cr
&\phantom{{\rm as\  well\  as\ }e_d=(0)}^{d-1}\phantom{...,1,0)\ {\rm and} 
\ e_{d-1}=(0.)}
^{d+1}}$$ 
Let $\nu^{(i)}$ $i=1,\ldots,r$ 
be the vectors of the complete fan of the 
base space embedded in the 
$1,\ldots,d-1$-plane in $\IR^{d+1}$, and 
$\nu^{(r+1)}=(0,\ldots,0)$ the origin. 
Then we can define for any given base space $B$ (1)-(18) three reflexive
polyhedra $\Delta_I^*$, $I=A,B,C$ with vertices 
\eqn\dualpoints{\{\nu^{(i)*}=\nu^{(i)}+v_I\cdot( \sum_{j}
\nu^{(i)}_j-1),i=1,\ldots,r+1;e_d, e_{d+1}\}.} 
The hypersurfaces as defined by \htoric~in $\IP^{d+1}_{\Delta^*}$
correspond to elliptic fibrations over the base space $\Sigma$
with generic fibre of the type $X_6(1,2,3)$, $X_4(1,1,2)$ 
and $X_3(1,1,1)$. The topological data of these manifolds are summarized
in table 6.1:

\goodbreak\midinsert
$${\fivepoint
\vbox{\offinterlineskip\tabskip=0pt
\halign{\strut
\vrule#&
~\hfil$#$~&
\vrule$#$& 
~\hfil$#$&  
~\hfil$#$&  
~\hfil$#$&  
~\hfil$#$~& 
\vrule$#$& 
~\hfil$#$~& 
~\hfil$#$~& 
~\hfil$#$~& 
~\hfil$#$~& 
\vrule$#$& 
~\hfil$#$~& 
~\hfil$#$~& 
~\hfil$#$~& 
~\hfil$#$~& 
\vrule$#$& 
~\hfil$#$~& 
~\hfil$#$~& 
~\hfil$#$~& 
~\hfil$#$~& 
\vrule#\cr
\noalign{\hrule}
&&& \multispan4\hfil Bases \hfil
&&  \multispan4\hfil $X_3$-fibrations\hfil
&&  \multispan4\hfil $X_4$-fibrations\hfil
&&  \multispan4\hfil $X_6$-fibrations\hfil
&\cr
\noalign{\hrule}
&$B$ &&\chi^B & h_{11}^B &(c_1^3)_B &(c_1c_2)_B 
&&\chi^X
&(c_2^X)^2
&h_{11}^X
&h_{31}^X
&&\chi^X
&(c_2^X)^2
&h_{11}^X
&h_{31}^X
&&\chi^X
&(c_2^X)^2
&h_{11}^X
&h_{31}^X
&\cr
\noalign{\hrule}
&\IP^3 &&4 &1 &64 &24 
&&4896  &2112 &4(2)& 804    
&&9504  &3648 &3(1)& 1573    
&&23328 &8256 &2 &3878 &\cr
\noalign{\hrule}
&F_0^{(2)} &&6 &2 &54 &24 
&&4176  &1872 & 5(2)&683      
&&8064  &3178 & 4(1) & 1332       
&&19728 &7056 &3 &3277 &\cr
\noalign{\hrule}
&F_1^{(2)} &&6 &2 &56 &24 
&&4320  &1920 & 5(2) & 701    
&&8352  &3264 & 4(1) & 1380      
&&20448 &7296 & 3 &3397 &\cr
\noalign{\hrule}
&F_2^{(2)} &&6 &2 &62 &24 
&&4752  &2064 & 5(2) & 779    
&&9216  &3552 & 4(1) & 1524       
&&22608 &8016 &3 &3757 &\cr
\noalign{\hrule}
&(5) &&6 &2 &54 &24
&&4176  &1728 & 5(2) &  683   
&&8064  &3168 & 4(1) &  1332       
&&19728 &7056 &3 &3277 &\cr
\noalign{\hrule}
&(\IP^1)^3 &&8 &3 &48 &24
&&3744  &1728 & 6(2)&610     
&&7200  &2880 & 5(1) &  1187    
&&17568 &6336 &4 &2916 &\cr
\noalign{\hrule}
&(7) &&8 &3 &52 &24
&&4032  &1824 & 6(2) & 658    
&&7776  &3072 & 5(1) & 1283     
&&19008 &6816 &4 &3156 &\cr
\noalign{\hrule}
&(8) &&8 &3 &44 &24  
&&3456  &1632 & 6(2) &  562   
&&6624  &2688 & 5(1)&  1091   
&&16128 &5856 &4 &2676 &\cr
\noalign{\hrule}
&\IP^1\times F_1^{(1)} &&8 &3 &48 &24
&&3744  &1728 & 6(2) & 610    
&&7200  &2880 & 5(1) &1187
&&17568 &6336 &4 &2916 &\cr
\noalign{\hrule}
&(10) &&8 &3 &50 &24
&&3888  &1776 & 6(2) &  634    
&&7488  &2976 & 5(1) & 1235
&&18288 &6576 &4 &3036 &\cr
\noalign{\hrule}
&(11) &&8 &3 &50 &24
&&3888  &1776 & 6(2) & 634     
&&7488  &2976 & 5(1)& 1235 
&&18288 &6576 &4 &3036 &\cr
\noalign{\hrule}
&(12) &&8 &3 &46 &24
&&3600  &1680 & 6(2) &  586   
&&6912  &2784 & 5(1) &  1139     
&&16848 &6048 &4 &2796 &\cr
\noalign{\hrule}
&\IP^1\times B_2 &&10 &4 &42 &24
&&3312  &1584 & 7(2) & 537     
&&6336  &2592 & 6(1) & 1042       
&&15408 &5616 &5 &2555 &\cr
\noalign{\hrule}
&(14) &&10 &4 &44 &24
&&3456 &1632 & 7(2) & 561     
&&6624 &2688 & 6(1) &  1090   
&&16128 &5856 &5 &2675 &\cr
\noalign{\hrule}
&(15) &&10 &4 &40 &24
&&3168  &1536 & 7(2) &  513   
&&6048  &2496 & 6(1) & 994      
&&14688 &5376 &5 &2435 &\cr
\noalign{\hrule} 
&(16) &&10 &4 &46 &24
&&3600  &1680 & 7(2) & 585     
&&6912  &2784 & 6(1) &  1138    
&&16848 &6096 &  5 &2795 &\cr
\noalign{\hrule}
&\IP^1\times B_3 &&12 &5 &36 &24
&&2880  &1440 & 8(2) & 464     
&&5472  &2302 & 7(1) &  897   
&&13248 &4896 &6 &2194 &\cr
\noalign{\hrule}
&(18) &&12 &5 &36 &24
&&2880   &1440 & 8(2) &464     
&&5472 & 2302  & 7(1) & 897      
&&13248 &4896 &6 &2194 &\cr
\noalign{\hrule}}
\hrule}}
$$ 
\leftskip 1pc\rightskip 1pc
\noindent{\ninepoint\sl \baselineskip=8pt {\bf Tab. 6.1}:
Elliptic fibred fourfolds over toric
Fano bases with fibre $X_3$, $X_4$ and $X_6$. All topological 
numbers\foot{$\int_Y c_{n_1}^{r_1} c_{n_2}^{r_2}$ 
is abbreviated as $ (c_{n_1}^{r_1}c_{n_2}^{r_2})_Y$ 
in the the table.}  are
calculated independently. From \chII~ and $\chi_0=1$ for Fano d-folds,
follows $\int_B c_1 c_2=24$. As further checks serve \ch~a.) and \eul .
$h_{21}^X=0$ for all fibre types and all bases.} 
\endinsert

By construction $\Delta^*$ has a prominent reflexive face $\Theta^*_B$, 
which is the convex hull of $\nu^{ *(i) }$, with 
$\tau=\nu^*_{r+1}=(0,0,0,-2,-3)$ as 
the only interior point. 
$B=V(\tau)\cap X$ gives divisors of type a.), which describes sections of 
the fibration in $X$. The two endpoints of $\Theta_B$ are $\nu^{\pm}=(-1,-1,-1,a^{\pm},b^{\pm})$,
with $(a^+,b^+)=(2,-1),(3,-1),(2,-1)$; $(a^-,b^-)=(-1,2),(-1,1),(-1,-2)$ for
the $X_3,X_4,X_6$ fibres, i.e. $l(\Theta_B)+1=3,2,1$ reflecting the fact
that the fibrations have $3,2,1$ sections.  
The discussion of the other divisors is equally simple. For instance for
the model $\IP^1\times B_3$ $(13)$ we see that all divisors 
$D_i=V(\nu^*_i)\cap X$, (up to $D_{r+1}=B$) with $\nu^*_i$ 
from \dualpoints~are of type d.), with $\chi(D_i,\cO_{D_i})
=1,1,1,0,0,0,0$ for $i=1,\ldots,r$ and $\chi(D_{e_4},\cO_{D_{e_4}})=-109$, 
$\chi(D_{e_5},\cO_{D_{E_5}})=-324$. Especially $D_1,\dots,D_3$ are divisors
which lead to superpotentials while $D_4,\ldots,D_7$ correspond
to embeddings of Calabi-Yau threefolds in $X$.

Let us finally comment on the transitions.  Model (1)-(3) and (5)-(12) 
above are connected by the blow up of a fixed point under the torus 
action. 
We can blow up $\IP^3$ in generic points by adding successively 
the vertices $-e_1$, $-e_2$, $-e_3$ and $e_1+e_2+e_3$ to the $\IP^3$ 
polyhedron. 
If $\hat B$ is obtained from $B$ by blowing up such fixed points then for 
the canonical bundles one has ($n=d-1={\dim}(B)$)
\eqn\excep{\hat K=\pi^* K +(n-1)[E]}
and since $[E]$ ($[E]^2=-1$) does not intersect with classes of 
$B$ one has $c_1^{n}(\hat K)=c_1^n(K)+(n-1)^n [E]^n$ so that 
$\int_{\hat B} c_1^n(\hat B)=\int_B c_1^n(B)-(n-1)^n$. 
In our case the effect of the transition is 
$\chi(\hat B)=\chi(B)+2$, $h^{1,1}(\hat B)=h^{1,1}(B)+1$,
$\int_{\hat B} c_1^3(\hat B)=\int_B c_1^3(B)-8$ and since 
$\int_B c_1 c_2$ is invariant one has $\chi(\hat X )=
\chi(X)-8\cdot 360$ for the $X_6$ fibre 
($360$ has to be replaced by $144,72,36 $ for the other fibres). 
As $h^{1,1}(\hat X)=h^{1,1}(X)+1$ and $h^{2,1}(\hat X)=h^{2,1}(X)$
this means by the index theorem \eul~especially that 
$h^{3,1}(\hat B)=h^{3,1}(B)-471$ for the $X_6$ fibre 
($471$ has to be replaced by $183,87,39$ for the other fibres). 
The seven branes at the discrimante induce a three brane charge
(comp. section 2 and \SVW ) . The contribution from the generic member 
in the class $-12 [\tilde \Delta]$ is for $\IP^3$ $Q(X)=-2300$ and each 
blow up changes this number by $Q(\hat X)=Q(X)+286$. 

For generic moduli values in the above examples we have no codimension 
three enhancements of the elliptic fibre singularities over the 
base. However if we restrict the complex $471$ moduli as we must do 
in order to follow a transition, then enhanced singularities at 
codimension three emerge, which should roughly localize the induced 
negative threebrane charges to points in the base were they annihilate
with the positive threebranes. Figure 1 also shows that the $F$-theory vacua under consideration 
are multiple connected by paths in the moduli space. The associated
fourfold polyhedra are embedded into each other, which implies that 
there are (extremal) transitions among them\foot{Such embeddings are 
expected to connect all fourfolds constructed by reflexive 
polyhedra .}\ref\bkk{P.\ Berglund, S.\ Katz, A.\ Klemm, 
\npb456 (1995) 153}. We will discuss the geometrical aspects of the 
$(5)\leftrightarrow (1)\leftarrow (3)$ transitions in more detail in (9.6).  

One might wonder what are in general the allowed modifications of the three 
dimensional fan $\Sigma_B$ of the base for which the property of the 
elliptic fibration $K_B=-{1\over 12} [\tilde \Delta]$ is kept. From 
the Weierstrass form and $K_B=-\sum_i D_i$ for reflexive polyhedra 
we expect this to be the case when $\Sigma_B$ comes from a reflexive
polyhedron, which would mean that any $K3$ polyhedron can be used in
this construction.

\subsec{The Weierstrass form of $X$}
To study the elliptic fibration and it's possible degeneration 
let us first describe the Weierstrass forms of $X$. Recall that the 
toric variety $\IP^{d+1}_{\Delta^*}$ is defined as follows. We associate
to every integral point $\nu_i\neq (0,\ldots,0)$ in $\Delta^*$ a 
coordinate $x_i$ $i=1,\ldots,q=l(\Delta^*)$ in $\IC^{q}$. 
Next we choose a complete triangulation ${\cal T}$ of $\Delta^*$ in 
$d+1$-dimensional simplices, whose vertices are the $\nu_i$ points. 
The Stanley-Reisner ideal is defined as the common zero of all 
those coordinate sets  $\{x_{i_1},\ldots, x_{i_p}\}$, for which every 
subset $S$ of points $S\subset\{\nu_{i_1},\ldots ,\nu_{i_p}\}$ does 
not lie on a common  $k$ dimensional simplex, we denote these 
zero sets as ${\cal S}^{(k)}_j$.  Linear relations between
points in $\Delta^*$, like $\sum l^{(k)}_i \nu_i=(0,\ldots,0)$ 
define $(q-d-2)$ independent $\IC^*$-actions on the coordinates $x_i$: 
$(x_1,\ldots,x_q) \sim ( \lambda_{(k)}^{l^{(k)}_1}x_1,\ldots,
\lambda_{(k)}^{l^{(k)}_q} x_q)$, with $\lambda_{(k)}\in \IC^*$. The
toric variety is then $\IP^{d+1}_{\Delta^*}=
({\IC^q-\cup_{k,j} {\cal S}^{(k)}_j})/(\IC^*)^{q-d-2}$. For every regular
triangulation ${\cal T}$ there is a canonical choice of $l^{(k)}$ 
such that all $l^{(k)}_i$ are semi-positive. Given such a choice we
can write the hypersurface $p=0$ as the polynomial in the $x_i$ which
scales homogeneously and with the minimal integers $\sum_i l^{(k)}_i$
with respect to all the $k=1,\ldots,q-d-1$ $\IC^*$-actions.
Suppose such $\hat l^{(k)}$ $k=1,\ldots, s$ have been constructed for
a triangulation of the fan of $B$, then there exist always a 
triangulation $\cT$
of $\Delta^*$ such that the following scaling vectors $l^{(k)}$ appear among
the $l^{(k)}$ for $(\Delta^*,{\cal T})$:
\eqn\morigen{
\eqalign{l^{(1)}&=(0,\ldots,0;1,n_1,n_2)\cr 
l^{(k+1)}&=(\ \ \ \hat l^{(k)} \ \ ;0, n_1\sum_I l^{(k)}_i,
n_2 \sum_I l^{(k)}_i)},
\quad (n_1,n_2)= 
\left\{
\matrix{
\!\!\!
&(2,3)\ {\rm for\ the \ } E_8{\rm -fibre}\cr               
\!\!\!
&(1,2)\ {\rm for\ the \ } E_7{\rm -fibre}\cr               
\!\!\!
&(1,1)\ {\rm for\ the \ } E_6{\rm -fibre,}}\right.}
where $k$ runs from  $1$ to $s$.

This implies that $p$ can be written, at least in a certain patch, 
in the following Weierstrass
\foot{Here one omits the first sub-leading terms in $x$ and $y$ to avoid 
redundant deformations of the equation as it is familiar in 
singularity theory. Writing down normal forms compatible with 
\morigen~for the other cases is straightforward:
$E_7:\ y^2=x^4+x^2z^2 e(x_i) + x z^3 f(x_i)+z^4 g(x_i)$ with
$\Delta = 2^8 g^3-2^7 e^2 g^2+2^43^2e g f^2 +2^4e^3 f^2-3^3 f^4$ and 
$E_6:\ y^3+x^3= yz^2 e(x_i) + x z^2 f(x_i) + z g^3(x_i)$ with
$\Delta= 2^4 (f^6+e^6)-2^3 3^3g^2 (e^3 +f^3)-2^5 e^3 f^3+3^6 g^4.$} form 
($y:=x_q$, $x:=x_{q-1}$ and $z:=x_{q-2}$)
\eqn\weierstrass{y^2=x^3 + x z^4 f(x_1,\ldots,x_{q-3})+
z^6 g(x_1,\ldots,x_{q-3}),}
with discriminant 
\eqn\discriminante{\Delta=27 g^2 + 4 f^3.}

As one can see from the table 6.1 the Euler number of $X$ fulfills always
\eulfoursmooth, so one expects that the elliptic fibre does not 
degenerate over codimension one or two in the base. 
This can in fact easily be checked in the Weierstrass models.

\newsec{ Gauge groups in four dimensions and more general 
degeneration of the elliptic fibres }
The degenerations of the fibre are described by Kodaira (table A.1) 
and a practical way to identify or construct such 
degenerations from the functions 
$f$ and $g$ of the Weierstrass form is Tate's alogarithm 
\ref\tate{J. Tate, in Modular Functions in one variable IV,
Lect. Notes in Math., {\bf 476}, Springer Verlag, Berlin,
Heidelberg (1975) 33}. This was used in \bikmsv, 
to analyze the physics associated to the 
degenerations of the elliptic fibre for $F$-theory compactifications 
to six dimensions. 
Here we will be interested in the simplest situation were the
fibre degenerates homogeneously over a codimension one locus $B'$ 
in the base. In this situation the enhancement of the gauge group
in four dimensions can be, at least for $A_n$ singular fibres, 
explained with parallel $7$-branes whose world-volume fills 
$B'\times \IR^4$. We will study situations in which $B$ admits a 
itself a fibration $\IP^1\rightarrow B\rightarrow B'$, such that we get a 
$N=1$ heterotic theory on $\cE'\rightarrow Z\rightarrow B'$.  

Let us consider for this purpose a generalization of the models 
(2)-(3), i.e. we consider as base $B$ the fibration
$\IP({\cal O}_{\IP^r}\otimes {\cal O}_{\IP^r}(n))
\rightarrow B\rightarrow \IP^r$,
which we denote as $F^{(r)}_n$, such that $F^{(1)}_n$ are the ordinary 
Hirzebruch surfaces $F_n$. The fan $\Sigma_B$ for $F^{(r)}_n$ is spanned
by $(e_i,i=1,\ldots, r+1;-e_{r+1},-e_1-\ldots -e_r- n e_{k+1})$. For 
the relevant case $r=2$ we have the following topological properties
of the base
\eqn\topfrn{\chi(F^{(2)}_n)=6,\quad \int_{F^{(2)}_n} c_1 c_2=24,
\quad \int_{F^{(2)}_n}c_1^3=54+2 n^2.}
{}From \morigen~with 
$$\eqalign{\hat l^{(1)}=(1,\ldots,1,n,0)\cr
           \hat l^{(2)}=(1,\ldots,0,1,1)}$$
follows ($y:=x_q$, $x:=x_{q-1}$, $z:=x_{q-2}$) 
$$y^2=x^3+x z^4 \!\!
\sum_{l=0}^{\left[4(n+r+1)\over n\right]}\! \! 
 v^l u^{8-l} 
f_{4(n+r+1)-nl}+\! \!
z^6\sum_{l=0}^{\left[6(n+r+1)\over n\right]} \! \! v^l u^{12-l} 
g_{6(n+r+1)-nl},$$
where  $u=x_{r+3}$, $v=x_{r+2}$ are the coordinates of the  
$\IP^1(\cO_{\IP^r}\otimes \cO(n)_{\IP^r})$ fibre, $\left[a/b \right]$ 
denotes the integer part of $a/b$ and
$f_k$ and $g_k$ are polynomials homogeneous of degree $k$ in the 
coordinates of the $\IP^r$ 
$(x_1,\ldots, x_{r+1})$.
To discuss the simplest degenerations of the the fibres, which lead
to generic gauge groups in space time, we have now just
to look at the leading behavior of the Weierstrass form 
near $(z,u)=(0,0)$.
The basic behavior is determined by the divisibility properties of 
$4(n+r+1)$, $6(n+r+1)$ by $n$; the leading singularity is
\eqn\leadsing{ x f_{(4(n+r+1)\ mod \ n)} u^{8-\left[4(n+r+1)\over n\right]}+
                 g_{(6(n+r+1)\ mod \ n)} u^{12-\left[6(n+r+1)\over n\right]}.}
The general discussion is exactly as  in 
\tate \mvII~for $r=1$ 
apart from the fact that one gets for the four-folds much richer singularity
structure if the functions $f_k$, $g_k$ are not forced to be constant 
over the $\IP^2$ for the leading term in $u$. 
Let us focus on the simple cases  with pure gauge group and no
additional matter. As it is obvious from \leadsing~the pure 
$SO(8)$, $E_6$, $E_7$ and $E_8$ singularities which occur for $r=1$ 
over the base $\IP^1$  in $F^{(1)}_{n^{(1)}}$  for $n^{(1)}=4,6,8,12$, 
will occur in general over the base $\IP^r$ of $F^{(r)}_{(r+1)n^{(1)}/2}$. 
Especially in four dimensions $r=2$ this gives the following examples 
      
\goodbreak\midinsert
$${
\vbox{\offinterlineskip\tabskip=0pt
\halign{\strut
\vrule#&
~\hfil$#$\hfil~&
~\hfil$#$\hfil~& 
\vrule$#$& 
~\hfil$#$~&
~\hfil$#$~\hfil&
~\hfil$#$~&
\vrule$#$&
~\hfil$#$~& 
~\hfil$#$~& 
~\hfil$#$~& 
~\hfil$#$~& 
\vrule#\cr
\noalign{\hrule}
& \multispan2 $B$\hfil
&& \multispan3 $B'$\hfil
&&  \multispan4\hfil $X_6$-fibrations\hfil&\cr
\noalign{\hrule}
&$B$ 
&\int_B c_1(B)^3
&&B'
&\int_{B'}c_1^2(B')
&G 
&&\chi(X)
&h_{11}^X
&h_{21}^X
&h_{31}^X
&\cr
\noalign{\hrule}
&F^{(2)}_{6} &126 &&\IP^2&9&D_4 
&&44136 &7(2) &0& 7341(0) &\cr
\noalign{\hrule}
&F^{(2)}_{9} &216 && \IP^2&9&E_6 
&&69624 &9(2) &0&11587(0) &\cr
\noalign{\hrule}
&F^{(2)}_{12} &342 &&\IP^2&9&E_7
&&101862 &10(0) &0& 16959(0) &\cr
\noalign{\hrule}
&F^{(2)}_{18} &702 && \IP^2&9&E_8 
&&186048 &11(0) &0&30989(0) &\cr
\noalign{\hrule}}
\hrule}}
$$
\leftskip 1pc\rightskip 1pc
\noindent{\ninepoint\sl \baselineskip=8pt {\bf Tab. 6.3}:
Elliptic fibrations over $F_n^{(2)}$, with pure gauge groups.
Note that $\chi=24\cdot 4244 + {1\over 4}$ for the $E_7$ case.} 
\endinsert

The enhancement of the gauge
group can easily studied in detail if we recognize that this cases 
are closely related to the hypersurfaces 
$X_{6(n+3)}(1,1,1,n,2(n+3),3(n+3))$
which in turn are $K_3$ fibrations with generic fibre 
$X_{2(n+3)}(1,n/3,2(n+3)/3,(n+3))$ over $\IP^2$ of the type 
discussed in section (2.3).  That is the intersection form, 
which lead to the gauge symmetry enhancement comes from vanishing
of the corresponding  cycles in the $K_3$.
We can also see this from the embedding of the polyhedra. 
Notice that the points in the $3,4,5$ 
plane cutting $\Delta^*$  
$$\eqalign{\nu^*_1=(0,0,-n/3,-2(n+3)/3,-(n+3)),\nu^*_2=(0,0,1,0,0),
\nu_3^*(0,0,0,1,0),
\nu_4^*(0,0,0,0,1),}$$ 
span the $K_3$ polyhedron. E.g. in the case of the $E_8$ $K3$ ($n=7$)  
one has six points on the edge between $\nu_1^*$ and 
$\nu^*_2$, two between  $\nu_1^*$ and $\nu^*_3$ and one
between $\nu_1^*$ and $\nu^*_4$. Together with the hyperplane class the
$V(\tau)\cap K_3$ the divisors associated to these points make 
up $Pic$ of the $K_3$ in question and have the intersection form 
$E_8\times U$ \ref\hs{B. Hunt and R. Schimmrigk, hep-th/9512138} 
(for the other cases see Kondos's list \ref\kondo{S.\ Kondo, J. Math. 
Soc. Japan {\bf 44} (1992) 75}\hs). 
The nine points on the
edges of the $K_3$ gives rise divisors $D=V(\tau)\cap X$ of the four 
fold of type b.) in addition the point $\nu^*_1$ gives rise
to a divisor of type c.). All of them have $\chi(D,\cO_D)=1$ 
from \chioD. One is horizontal w.r.t. $\pi$ of \ellX~the other are
are $\IP^1$ bundles over $\IP^2$ and vertical w.r.t. $\pi$ and 
but horizontal w.r.t. $\pi''$ of \ellZ, i.e. they will lead to a 
non-perturbative superpotential of the 
heterotic string. In fact we have here a realization of the situation
described in~\kvII\bjpsv~for the $E_8$ 
group. 
   
As a  further simple generalization\foot{The generation of a superpotential  
in ten examples of this kind with generically $I_1$ degeneration 
are discussed in great detail in \pema .} of (7), we chose 
$B$ such that is it is a 
$\IP^1$ bundle $\IP(\cO_{B'}\otimes \cO(b f_1 \otimes c f_2)_{B'})$
over $B'$ with $\IP(\cO\otimes \cO(a)_{\IP^1}) 
\rightarrow B'\rightarrow\IP^1$. 
This base $B$ , say $F^{(2)}_{k,m,n}$ has as fan 
$(-e_1-k e_2-m e_3,-e_2-n e_3,e_1,e_2,e_3,-e_3)$ with coordinates
$(p,s,q,t,v,u)$ and the topological properties 
\eqn\topfabc{\chi(F^{(2)}_{k,m,n})=8,\quad \int_{F^{(2)}_{a,b,c}} c_1 c_2=24,
\quad \int_{F^{(2)}_n}c_1^3=48+4 m n -2 m^2 k.}
In particular if $k=0$ ($B'=\IP^1\times \IP^1$) and $m=n$ the elliptic fibre 
degenerates homogeneously over $\IP^1\times \IP^1$ as can be seen from the
Weierstrass form
\eqn\weirfkmn{y^2=x^3+x z^4 \!\!
\sum_{l=0}^{\left[4(n+2)\over n\right]}\! \! 
 v^l u^{8-l} 
f_{4(n+2)-nl;4(n+2)-nl}^{(s,t;p,q)}+\! \!
z^6\sum_{l=0}^{\left[6(n+2)\over n\right]} \! \! v^l u^{12-l} 
g_{6(n+2)-nl;6(n+2)-nl}^{(s,t;p,q)},} 
such that we get as before the matter free degenerations, but this time 
at $n=3,4,6,8,12$. The case $n=8$ leads however not to reflexive polyhedra 
hence not to a model with a geometrical resolution.

\goodbreak\midinsert
$${
\vbox{\offinterlineskip\tabskip=0pt
\halign{\strut
\vrule#&
~\hfil$#$\hfil~&
~\hfil$#$\hfil~& 
\vrule$#$& 
~\hfil$#$~&
~\hfil$#$~\hfil&
~\hfil$#$~&
\vrule$#$&
~\hfil$#$~& 
~\hfil$#$~& 
~\hfil$#$~& 
~\hfil$#$~& 
\vrule#\cr
\noalign{\hrule}
& \multispan2 $B$\hfil
&& \multispan3 $B'$\hfil
&&  \multispan4\hfil $X_6$-fibrations\hfil&\cr
\noalign{\hrule}
&$B$ 
&\int_Bc_1(B)^3
&&B'
&\int_{B'}c_1^2(B')
&G 
&&\chi(X)
&h_{11}^X
&h_{21}^X
&h_{31}^X
&\cr
\noalign{\hrule}
&F^{(2)}_{0,3,3} &84 &&\IP^1\times \IP^1&8&A_2 
&&30336 &6(1) &0& 5042(0) &\cr
\noalign{\hrule}
&F^{(2)}_{0,4,4} &112 && \IP^1\times \IP^1&8&D_4 
&&39264 &8(2) &0&6528(0) &\cr
\noalign{\hrule}
&F^{(2)}_{0,6,6} &192 &&\IP^1\times \IP^1&8&E_6
&&61920 &10(2) &0& 10302(0) &\cr
\noalign{\hrule}
&F^{(2)}_{0,12,12} &624 && \IP^1\times \IP^1&8&E_8 
&&165498 &12(0) &0&27548(0) &\cr
\noalign{\hrule}}
\hrule}}
$$
\leftskip 1pc\rightskip 1pc
\noindent{\ninepoint\sl \baselineskip=8pt {\bf Tab. 6.3}:
Elliptic fibrations over $\IP^1\times \IP^1$, with pure  gauge groups.} 
\endinsert

Again the $X_{3(n+2)}(1,n/2,(n+2),(n/2+1)3)$ $K_3$ is embedded in 
the $(2,3,4)$ plane and the divisors of $X$ leading to the
enhanced gauge symmetry have very similar properties to
the ones discussed before.   

The general formula for the Euler number for the elliptic fibred
four fold  $X$ for which the $X_6$-fibration degenerates to a
singularity of type $G$ over a codimension one subspace 
$B'$ in the base $B$ is    
\eqn\eulcor{\eqalign{ \chi(X)&=12\int_B c_1(B) c_2(B)+360 \int_B c_1^3(B)-
\delta^{d=4}(B^\prime,G),\qquad {\rm with}\cr
\delta^{d=4}(B^\prime,G)&=r_{(G)} c_{(G)}
\left(c_{(G)} \int_{B^\prime} c_1(B^\prime)^2+(6-\int_{B'}c_2(B'))
\int_{B'}c_2(B') 
\right).}}

For $d=3$ the correction term is
\eqn\eulcorb{\delta^{d=3}=r_{(G)} c_{(G)} \int_{B'} c_1(B'),}
while for $d=5$ we observe for $B'=\IP^3$ 
\eqn\eulcorb{\delta^{d=5}=r_{(G)}\left(
c_{(G)}^3\int_{B'} c_1^3(B')+3 c_{(G)}^2 \int_{B'} c_1(B') c_2(B')+
2 (3 c_{(G)} -c_{(G)}^2) \int_{B'} c_3(B')\right),}
e.g. the elliptic fivefold fibration over the four dimensional base
$F^{(3)}_{18}$ for which the generic fibre $X_6$ degenerates over a 
$\IP^3$ has by \eulel,\eulcorb~the Euler number $\chi=-55556832$. 

If the degeneration of the fibre is not of the same type over a 
subspace of codimension one in the base, but there are 
codimension two loci where the degeneration increases, a positive 
correction to the Euler number \eulel~ is expected. As example we consider
$F^{(2)}_{0,0,n}$. Now the functions $f_{8}'(p,q),g_{12}'(p,q)$ do not
become constants, when we consider the leading singularity around
$(x,u)=(0,0)$ and we get extra singularities when these functions vanish. 
Application of \eulel~gives $\chi_s=17568$, while the actual data are

\goodbreak\midinsert
$${
\vbox{\offinterlineskip\tabskip=0pt
\halign{\strut
\vrule#&
~\hfil$#$\hfil~&
~\hfil$#$\hfil~& 
\vrule$#$& 
~\hfil$#$~&
~\hfil$#$~\hfil&
~\hfil$#$~&
\vrule$#$&
~\hfil$#$~& 
~\hfil$#$~& 
~\hfil$#$~& 
~\hfil$#$~& 
\vrule#\cr
\noalign{\hrule}
& \multispan2 $B$\hfil
&& \multispan3 $B'$\hfil
&&  \multispan4\hfil $X_6$-fibrations\hfil&\cr
\noalign{\hrule}
&$B$ 
&\int_Bc_1(B)^3
&&B'
&\int_{B'}c_1^2(B')
&G 
&&\chi(X)
&h_{11}^X
&h_{21}^X
&h_{31}^X
&\cr
\noalign{\hrule}
&F^{(2)}_{0,0,3} &48 &&\IP^1\times \IP^1&8&A_2 
&&18240 &5(0) &5& 3032(0) &\cr
\noalign{\hrule}
&F^{(2)}_{0,0,4} &48 && \IP^1\times \IP^1&8&D_4 
&&19680&6(0) &10&3276(0) &\cr
\noalign{\hrule}
&F^{(2)}_{0,0,6} &48 &&\IP^1\times \IP^1&8&E_6
&&23328 &8(0) &10& 3882(0) &\cr
\noalign{\hrule}
&F^{(2)}_{0,0,12} &48 && \IP^1\times \IP^1&8&E_8 
&&35808 &24(11) &0&5936(0) &\cr
\noalign{\hrule}}
\hrule}}
$$ 
\leftskip 1pc\rightskip 1pc
\noindent{\ninepoint\sl \baselineskip=8pt {\bf Tab. 6.4}:
Elliptic fibrations over $F_{0,0,n}^{(2)}$.} 
\endinsert

In $F^{(2)}_n$ and the $F^{(2)}_{0,n,n}$ cases we considered a specific 
point $p=(u=0,v=1)$ in the rational fibre over $B'$ and configurations 
such that the degeneration of the elliptic fibre was homogeneous 
over $B'$. $B'$ is of course just one component of the discriminant locus and 
away from $p$ the fibre will degenerate over codimension one to $I_1$, 
but  more complicated in higher codimensions. 
If we allow for special values of the moduli, there will be also 
more complicated degenerations over codimension one surfaces in the base, 
which will lead to non generic gauge group enhancement. In particular one
can design examples with $ADE$ sphere tree's over a $\IP^2$ in the base 
in which non generic gauge groups arise in $M$ theory compactifications to 
three dimensions similarly as in \ref\kkv{S.\ Katz, A.\ Klemm 
and C. \ Vafa, hep-th/9609239}.

Let us discuss in extension of the last examples in table 3.3 
situations where we have a generic $ADE$ fibre over $B'$, 
but additional enhancements over 
lines and points in $B'$. These cases can be designed, 
by ``upgrading'' the corresponding $F_n$ fibrations in 
three dimensions, which were studied in great detail in 
\bikmsv\ref\candfont{P.\ Candelas, A.\ Font, 
{\sl Duality Between the Webs Heterotic and Type II Vacua}, 
hep-th/9603170} to four dimensions. 

These three dimensional Calabi-Yau spaces $Y$ are elliptic fibration 
over $F_n$: $\cE\rightarrow Y\rightarrow F_n$ {\sl and}  
$K_3$ fibrations $K_3\rightarrow Y\rightarrow \IP^1$. Furthermore the
the $K_3$ is  itself a elliptic fibration over the fibre 
$\IP^1$ of $F_n$, i.e. $\cE \rightarrow K_3 \rightarrow \IP^1$. 
These fibration structure\foot{The complete process is the
 generalization of~\rec~with 
$X^{(0)}_{6}(1,2,3)$, $r^{(1)}=2$, $r^{(2)}=2$, and $r^{(3)}=3$ to the 
polyheder description.}  are reflected in the
geometry of the four dimensional polyhedron $\Delta^*$ (cf. \candfont). 
It has the polar polyhedron of the Newton polyhedron of $X_{6}(1,2,3)$ 
in the (say) $(4,5)$ plane, which is augmented to a $K_3$ polyhedron
in the $(3,4,5)$ plane. Now in the threefold polyhedron there are 
two points $p_1=(0,-1,0,2,3)$ and $p_2=(0,1,2n,2,3)$ outside the 
$(3,4,5)$ plane such that a corner of the $K_3$ polyhedron 
$c=(0,0,n,2,3)$ is in the middle of the line $\overline {p_1p_2}$. 
The coordinates associated to $p_1$ and 
$p_2$ are the homogeneous coordinates of the base $\IP^1$. 
It is now very easy to replace the 
base $\IP^1$ by a rational surface $S$. E.g. we can replace it by $\IP^2$
by adding instead of $p_1,p_2$ the points $p_0=(-1,0,0,2,3)$, 
$p_1=(0,-1,0,2,3)$ and $p_2=(1,1,3 n,2,3)$ so that $e$ represents the 
canonical class of $\IP^2$ (or $S$).  
It is important that the only modification in the scaling 
relations~\morigen~from the 
three to the four dimensional case is that the Mori generator with the 
two 1's on the $\IP^1$ coordinates $l=(1,1,n,0,\ldots,0)$ is replaced
by $l=(1,1,1,n,0,\ldots,0)$ with three 1's on the $\IP^2$ coordinates,
all other linear relations between the points $K3$-plane are 
obviously the same. This implies that the Weierstrass form is 
essentially the same but $f$ and $g$ depend now homogeneously
on three coordinates. That is the generic codimension one singularity at
($u=1,v=0$) is as  analysed in \bikmsv\bkkm and indicated in table 
(6.5), while the additional singularities
which give matter in the six dimensional compactification are now at 
codimension  one in the $\IP^2$.  
Let us ``upgrade'' a couple of examples from 
table 3.2 of~\candfont~to four dimensions in order to demonstrate the effect 
of ``unhiggsing'' of the $(u=1,v=0)$ locus in the fibre of $F_n^{(2)}$.

\goodbreak\midinsert
$${
\vbox{\offinterlineskip\tabskip=0pt
\halign{\strut
\vrule#&
~\hfil$#$\hfil~&
\vrule#&
~\hfil$#$~& 
~\hfil$#$~&
~\hfil$#$~&
~\hfil$#$~&
~\hfil$#$~& 
\vrule#\cr
\noalign{\hrule}
&B^0&
&SU(1)
&SU(2)
&SU(3)
&SU(4)
&SU(5)&\cr
\noalign{\hrule}
&F^{(2)}_{3}& 
&(^0 26208;3,1) 
&(^3 17082;4,1) 
&(^0 13032;5,1) 
&(^2 10116;6,1) 
&(^3 7578;7,1)  
&\cr
\noalign{\hrule}
&F^{(2)}_{6}& 
&(^0 44136;7,0)   
&(^3 24642;8,0)   
&(^0 16704;9,0)   
&(^0 11520;10,0)  
&(^0 7416;11,0) 
&\cr
\noalign{\hrule}
&F^{(2)}_{9}& 
&(^0 69624;9,0) 
&(^3 35874;10,0) 
&(^0 22752;11,0) 
&(^2 14652;12,0) 
&(^2 8604;13,0) 
&\cr
\noalign{\hrule}}
\hrule}}
$$
\leftskip 1pc\rightskip 1pc
\noindent{\ninepoint\sl \baselineskip=8pt {\bf Tab. 6.5}:
Topological invariants $(\chi;h^{1,1},h^{1,2})$ of the chains 
of elliptic fibrations over $F_n^{(2)}$. We indicate by the prefix 
$^n$ on the Euler number it's divisibility $6 n=\chi\ mod\ 24$.} 
\endinsert

We will discuss in section (9.6) in detail how 
the aspects of the discussion of the transitions~\bkkm~carries over.

\newsec{Euler number of Elliptic CY manifolds}
For a complex manifold $M$, we denote the tangent bundle, canonical 
bundle, 
 the total Chern 
class of $M$ by $T_M , K_M$ and $c(M)$ respectively.

\noindent
{\bf Lemma 1.} Let $M$ be a $m$-dimensional compact complex manifold, and 
$D$ be an irreducible smooth divisor of $M$ such that ${\cal O}_M(D)$ is the 
$d$-th power of the canonical sheaf of $M$ for some rational number $d$, 
${\cal O}_M(D) = \omega_M^d$. Then
$$
\chi ( D ) = - \sum_{k=1}^m d^k c_1^k c_{m-k},
$$
where $c_j$ is the $j$-th Chern class of $M$. 

\noindent
Let $N$ be the normal bundle of $D$ in $M$. It is known that the Chern class 
of $D$, $1+ c_1(D) + \cdots + c_{m-1}(D)$, is related to $c_1(N)$ and $c_j$'s 
bt he following relations:
$$
c_j(D) + c_1(N) c_{j-1}(D)  = c_{j|D} \ ,  
$$
hence 
$$
c_j(D) = \sum_{k=0}^{j} (-1)^k c_1(N)^kc_{j-k|D}  
$$
for $ 1 \leq j \leq m-1 $. By $c_1(N) = - dc_{1|D}$, the result follows from 
the above relation for $j=m-1$.
$\Box$ 

\noindent
{\bf Lemma 2.} Let $X$ be a $n$-dimensional CY manifold, which is 
a $l$-fold cyclic cover of a manifold $Y$ for $l \geq 2$. Then
$$
{1\over l-1}\chi(X) = {l\over l-1} \chi(Y) + \sum_{k=1}^n 
({-l\over l-1})^k c_1(Y)^kc_{n-k}(Y)
$$

\noindent
{\it Proof.} Let $D$ be the branched locus for the double cover of $X$ 
over $Y$. $D$ is 
a smooth divisor with ${\cal O}_Y(D) = \omega_Y^{{-l\over l-1}}$. The
result follows immediately from Lemma 1.
$\Box$  

\noindent
Let $E$ be a vector bundle over a complex manifold $M$ of rank $r$, and $\IP$ be the associated projective 
bundle,
$$
\pi : \IP = \IP(E) \longrightarrow M \ .
$$
Note that $\IP = \IP(E \otimes L)$ for any line bundle $L$ over $M$. 
We have the exact sequence of vector bundles over $\IP(E)$:
$$
0 \longrightarrow \IP \times \IC \longrightarrow 
\pi^*E  (1) \longrightarrow T_{\IP} \longrightarrow \pi^*( T_M ) \longrightarrow 0 \ .
$$
where  $(\pi^*E )(1)$ is the tensor bundle  
$\pi^*E \otimes {\cal O}(1)$ with ${\cal O}(1)$ the inverse of the tautological bundle over $\IP$ for the 
bundle $E$. Hence 
\eqn\K{
K_{\IP} = \pi^*( K_M \otimes {\rm det}(E^*) ) \otimes {\cal O}(-r)
}
We have the relation
$$
c ( \IP ) = c ( M )  c (\pi^*E (1)) \ .
$$ 
Consider the cohomology ring ${\rm H}^*(M)$  as a subring of 
${\rm H}^*(\IP)$. ${\rm H}^*(\IP)$ is a ${\rm H}^*(M)$-algebra generated 
by the Chern class
$$
\eta = c_1({\cal O}(1)) 
$$ 
with the relation 
\eqn\cd{
c_d ( \pi^*E (1) ) = \sum_{k=0}^r c_k(E) \eta^{r-k} = 0 \ .
}
As $c (\pi^*E (1) )$ is a projective invariant in 
${\rm H}^*(\IP)$,( i.e. an invariant under changing $E$ to $E \otimes L$), 
one can in principal derive all the projective invariants of $E$ in 
${\rm H}^*(M)$. 
For later purpose, let us work out the cases for $r=2, 3$. For $r=2$, 
we have  
$$
c_1(\pi^*E (1) ) = c_1(E) + 2 \eta \ .
$$
Using \cd~to eliminate $\eta$, we have the well-known projective 
invariant $E$ in 
${\rm H}^*(M)$:
\eqn\idii{
i(E) : = c_1(E)^2 - c_2(E) = c_1( \pi^*E (1) )^2 \in {\rm H}^*(M) \ .
}
For $r=3$, by 
\eqn\ciEi{
c_1( \pi^*E (1)) = c_1(E) + 3 \eta \ , \ \ \ 
c_2( \pi^*E (1) ) = c_2(E) + 2 c_1(E) \eta + 3 \eta^2 \ ,
} 
we obtain the the projective invariant of $E$: 
\eqn\idiii{\eqalign{
i_2(E) : =& c_1(E)^2 - 3 c_2(E)   = c_1(\pi^*E (1))^2 - 3 
c_2(\pi^*E (1))  \ , \cr
i_3(E) :  =& 2c_1(E)^3 - 9 c_1(E)c_2(E) + 27c_3(E)   = 
2c_1(\pi^*E (1))^3 - 9 c_1(\pi^*E (1))
c_2(\pi^*E (1)).}}
One can always express the Chern numbers of $\IP$ in terms of 
those of $M$ and projective invariants of $E$. We are going to derive 
the relations for $r=2, 3$.
For $r=2$, we have  
$$
c_i( \IP ) = c_i(M) + c_{i-1}(M) ( c_1(E) + 2 \eta ) \ ,
$$
which implies $\chi ( \IP ) = 2 \chi(M) $ for $ i = m+1$. 
Using \idii, we have
$$
c_k(\IP)c_{m+1-k} (\IP)  = 2 c_k(M)c_{m-k}(M) 
+ 2c_{k-1}(M)c_{m+1-k}(M)  \ \ {\rm for} \ 1 \leq k \leq m \ .
$$
All the relations of Chern numbers for $r= 2, m=2,3$ are given by
\eqn\chdii{\eqalign{
m=2 :& \cases{ 
c_2(\IP)c_1 (\IP) =& $2 c_2 (M) + 2 c_1(M)^2$ \cr
c_1^3(\IP) =& $6c_1(M)^2 + 2 i(E)$ ; } \cr
m=3 :& \cases{ 
c_3(\IP)c_1 (\IP)   = & $2 c_3(M)
+ 2c_2(M)c_1(M)$ \cr
c_2(\IP)^2  = & $4 c_2(M)c_1(M)$ \cr
c_2(\IP)c_1(\IP)^2 =& $4 c_2(M)c_1(M) + 2 c_1(M)^3 + 2c_1(M)i(E)$\cr
c_1(\IP)^4 =& $8 c_1(M)^3 + 8 c_1(M)i(E)$ } }}
For $r=3$, we have  
$$
c_i( \IP ) = c_i(M) + c_{i-1}(M) c_1(\pi^*E (1) ) + 
c_{i-2}(M)c_2((\pi^*E )(1) ) \ , 
$$
which implies $\chi ( \IP ) = 3 \chi(M) $ for $ i = m+1$. By \idiii~we 
have
$$
\eqalign{
c_k(\IP)c_{m+2-k} (\IP)  =& 
 3  c_k(M) c_{m-k}(M) + 3 c_{k-2}(M)
 c_{m+2-k}(M)
+ 9  c_{k-1}(M) c_{m+1-k}(M)  \cr & 
+  c_{k-2}(M) c_{m-k}(M)i_2(E)} 
$$
The relations of Chern numbers for $r= 3, m=2,3$ are given as 
follows:
\eqn\chdiii{\eqalign{
m=2 : &
\cases{ 
c_3(\IP)c_1 (\IP)  = & $9c_2(M) + 3 c_1(M)^2$ \ ,  \cr
c_2(\IP)^2 = & $6c_2(M) + 9c_1(M)^2 + i_2(E)$ \ , \cr
c_2(\IP)c_1(\IP)^2 = & $ 9c_2(M) + 21 c_1(M)^2 + 6i_2(E)$ \ , \cr
c_1(\IP)^4 = & $54 c_1(M)^2 + 27 i_2(E)$ \ ; }
\cr
m=3 :&
\cases{
c_4(\IP)c_1 (\IP)  =& $ 9 c_3(M)+ 3 c_2(M)c_1(M)$\ , \cr
c_3(\IP)c_2(\IP)   =& $ 9c_3(M) + 12 c_2(M)c_1(M) + c_1(M)i_2(E)$ \ , \cr
c_3(\IP) c_1(\IP)^2=& $ 9c_3(M) + 18 c_2(M)c_1(M) + 3c_1(M)^3 
+ 6c_1(M)i_2(E)$ \ , \cr
c_2(\IP)^2c_1(\IP) =& $ 9 c_1(M)^3 + 24 c_2(M)c_1(M) + 13 c_1(M)i_2(E)
- i_3(E)$ \ ,  \cr
c_2(\IP)c_1(\IP)^3 =& $ 27 c_2(M)c_1(M) + 30c_1(M)^3 + 
45c_1(M)i_2(E) - 3i_3(E)$ \ ,  \cr
c_1(\IP)^5 =& $ 90 c_1(M)^3 + 135 c_1(M)i_2(E) - 9i_3(E)$ \ . }
}}

We now discuss the $n$-dimensional CY manifolds $X$ which 
is either a hypersurface or a cyclic branched cover of a projective 
bundle $\IP (E) $ over a complex manifold $M$. Such $X$ is always an elliptic 
fibration over $M$. By Lemma 1 and 2, the Euler number $\chi(X)$ 
can be expressed by the Chern numbers of $M$ and the projective 
invariants of $E$. By \chdii~and \chdiii , we have the 
following results for $n=3,4$:     

\noindent
{\bf Proposition 1.} Let $X$ be a $n$-dimensional CY manifold. 

\noindent
(I) If $X$ is a double cover of a projective bundle $\IP$ associated to 
a rank 2 bundle 
$E$ over a $(n-1)$-dimensional complex manifold $M$ for $n=3,4$, then
$$
\chi(X) = \left\{\matrix{ 
 - 28 c_1(M)^2 - 8 i(E) & {\rm for} \ n = 3 \ , \cr
 12c_2(M)c_1(M) + 72c_1(M)^3 + 72c_1(M)i(E) & {\rm for} \ n = 4 \ . 
}\right.
$$ 

(II) If $X$ is a  
hypersurface of a projective bundle $\IP$ associated to a rank 3 bundle 
$E$ over a $(n-1)$-dimensional complex manifold $M$ for $n=3,4$, then
$$
\chi(X) = \left\{\matrix{
 -  18 c_1(M)^2 - 6 i_2(E)  & {\rm for} \ n = 3 \ , \cr
   12 c_2(M)c_1(M) 
+ 27c_1(M)^3 +39 c_1(M)i_2(E) - 3i_3(E) & {\rm for} \ n = 4 \ \Box }\right. $$ 

\noindent
{\bf Remarks.} 
\noindent
(1) For $n=4$ in (I), by $12 | c_2(M)c_1(M)$, 
 we have 
$$
72 | \chi(X) \ .
$$
When $E = K_M^{-2} \oplus 1$, one obtains the formula (2.12) in 
\SVW.

\noindent
(2) 
For $n=4$ and $E =$ the trivial bundle in (II), we have the 
following criterion for the integral property of ${\chi(X)\over 24}$:
$$
24  | \chi(X) \ \Longleftrightarrow \ 8 | c_1(M)^3 \ .
$$
Note that above condition do not hold for $M = \IP^1 \times \IP^2 $, in which 
case, $c_1(M)^3 = 54 $ and $X$ is an elliptic CY 4-fold in 
$\IP^1 \times \IP^2 \times \IP^2$ with $\chi(X) = 1746 $  $\Box$

\newsec{Elliptic CY manifolds with sections}
In this section we consider the structure of elliptic CY $n$-fold
$\pi : X \longrightarrow M $ 
with an involution $\sigma$, and a (holomorphic) 
section $s: M \longrightarrow X$. Here the involution $\sigma$ means an 
order 2 automorphism of $X$ commuting with $\pi$ having the non-empty 
fixed points on the general fiber of $\pi$, and the section $s$ will 
always assume its image $s(M)$ lying outside the critical points of $\pi$.

For a line bundle $L$ over $M$, we shall denote $\overline{L}$ the 
$\IP^1$-bundle $\IP(L \oplus 1)$ over $M$,
$$
\pi_0: \overline{L} \longrightarrow M \ ,
$$
 $M_0$ the zero-divisor $\IP(0\oplus 1)$ 
and $M_\infty$ the infinity-divisor $\IP(L\oplus 0)$ in $\overline{L}$. 
As ${\cal O}(1)$ and $\pi_0^* L^{-1}$ are the line bundles associated to the divisor 
$M_\infty$ and $-M_0 + M_\infty$ respectively, by \K~we have
$$
K_{\overline{L}} = \pi_0^*K_M \otimes {\cal O}( - M_0 - M_\infty ) \ .
$$
Now set $L = K_M^{-2}$, and consider a smooth divisor $D$ contained in $L$ such 
that the restriction of $\pi_0$ defines a 3-fold branched covering over 
$M$. Then $D$ is defined by the equation:
\eqn\eqDii{
\xi^3 + a_1 \xi^2 + a_2 \xi + a_3 = 0 \ , \ \ \xi \in L \ , \ \ 
a_i \in \Gamma ( M , L^i ) \ \ {\rm for} \ i =1,2,3.
} 
Hence $D$ is 
linearly equivalent to $3M_0$ in $\overline{L}$ and 
we have
$$
K_{\overline{L}}^{-2} = {\cal O}(D + M_\infty ) \ .
$$ 
The double cover of $\overline{L}$ branched at $D + M_\infty$ becomes 
an elliptic CY $n$-fold over $M$, denoted by $Z(2)$, with the involution 
$\sigma$ and the projection 
given by the following diagram:
$$
\eqalign{
Z(2) & \longrightarrow  \overline{L} = Z(2)/<\sigma> \cr
\pi \downarrow & \quad \quad \downarrow \pi_0 \cr
M & \ \ \ \ \ = \ \ \ \ \  M  \ .}$$
The infinity-section of $\overline{L}$ over $M$ induces a section of 
the fibration $Z(2)$ over $M$ fixed by $\sigma$.

\noindent
{\bf Proposition 2.} The Euler number of $Z(2)$ is given by 
$$
\chi(Z(2)) =    2 
 \sum_{k=1}^{n-1} (-1)^{k-1}6^k c_1(M)^k c_{n-1-k}(M) 
$$

\noindent
{\it Proof.} We have 
$$
\chi(Z(2)) = 2 \chi( \overline{ K_M^{-2} } ) - \chi(M_\infty) - 
\chi(D) = 3 \chi(M) - \chi(D) \ .
$$
We may assume $a_1= a_2=0, a_3 \neq 0$ in the equation \eqDii~of 
$D$. Hence $D$ is a 3-fold cyclic cover of $M$ branched at the zeros of 
$a_3$, which is a divisor in $M$ for $K_M^{-6}$. By Lemma 1,
$$
\chi(D) = 3 \chi(M) + 2 
 \sum_{k=1}^{n-1} (-6)^k c_1(M)^k c_{n-1-k}(M) \ ,
$$ 
hence the result follows immediately $\Box$ 

\noindent
For $L = K_M^{-1}$, and a smooth divisor $D$  contained in 
$L$ with the restriction of $\pi_0$ defining a 4-fold branched covering over 
$M$. Then $D$ is linearly equivalent to $4M_0$ in $\overline{L}$, and 
$$
K_{\overline{L}}^{-2} = {\cal O}(D ) \ .
$$
Denote $Z(1)$ the double cover of $\overline{L}$ branched at $D$, 
and $\sigma$ the involution.
$Z(1)$ is an elliptic CY $n$-fold over $M$ with the projection 
$$
\pi : Z(1) \longrightarrow M 
$$
induced by $\pi_0$. Since the infinity-section $M_\infty$ of 
$\overline{L}$ does not intersect $D$, it 
gives rise to two disjoint sections of $Z(1)$ over $M$ permuted by 
$\sigma$. With the similar argument in Propostion 2, we have 
the following result: 

\noindent
{\bf Proposition 3.} The Euler number of $Z(1)$ is given by 
$$
\chi(Z(1)) =  3 
 \sum_{k=1}^{n-1} (-1)^{k-1}4^k c_1(M)^k c_{n-1-k}(M) \ \Box
$$

\noindent
{\bf Remarks} 

(1) The formulas of $\chi(Z(i))$ for small $N$ are as follows: 
$$
\chi(Z(2)) = \cases{
-60c_1^2(M) & {\rm for} \ n = 3 \ , \cr
 12c_1(M)c_2(M)  + 360 c_1^3(M)  & {\rm for} \ n = 4 \ ,\cr
12 c_1(M)c_3(M)- 72c_1(M)^2c_2(M)  - 2160c_1^4(M)
& {\rm for} \ n = 5 \ .} 
$$
and
$$
\chi(Z(1)) =\cases{
-36c_1^2(M) & {\rm for} \ n = 3 \ , \cr
 12c_1(M)c_2(M)  + 144 c_1^3(M)  & {\rm for} \ n = 4 \ ,\cr
12 c_1(M)c_3(M)- 48c_1(M)^2c_2(M)  - 576c_1^4(M)
& {\rm for} \ n = 5 \ . }
$$ 
The above $\chi(Z(2)$ for $n=4$ is the formula in \SVW .  

\noindent
(2) When $M= \IP^{n-1}$, $Z(2) , Z(1)$ are the CY manifolds for the 
hypersurface $X_{6n}(\underbrace{1, \cdots, 1}_{n}, 2n , 3n)$ and 
 $X_{4n}(\underbrace{1, \cdots, 1}_{n}, n , 2n)$ respectively. 
In general,  
a CY $n$-fold $X_{6k}(w_1, \cdots, w_{n+2})$ 
with $\sum_{j=1}^n w_j = k$ and $w_{n+1}= 2k, w_{n+2}= 3k$ has the 
above $Z(2)$ structure with $M$ as a non-singular toric variety dominating 
$\IP(w_1, \cdots, w_n)$. Similarly the  CY $n$-fold 
$X_{4k}(w_1, \cdots, w_{n+2})$ with 
$\sum_{j=1}^n w_j = k$ and $w_{n+1}= k, w_{n+2}= 2k$ for $Z(1)$.
However the construction 
of $Z(2)$ can also be applied to a non-toric variety $M$, e.g. a del Pezzo 
surface.   

\noindent
The above elliptic CY fibration $Z(i)$ 
has the following characterization: 

\noindent
{\bf Proposition 4.} Let X be an 
elliptic CY fibration over a complex manifold $M$ with 
an involution $\sigma$ such that all the 
fibers are irreducible. 

\noindent
(I) If there is a section of $X$ over $M$ fixed by $\sigma$, and 
$H^1(M, K_M^{-2}) = 0$, then
$X$ isomorphic to $Z(2)$ over $M$.

\noindent
(II) If there exist two disjoint sections of $X$ over $M$ permuted by 
$\sigma$, and 
$H^1(M, K_M^{-1}) = 0$, then $X$ isomorphic to $Z(1)$ over $M$.  

\noindent
{\it Proof.} Let $\pi$ be the projection of $X$ onto $M$, and $s$ a 
section fixed by $\sigma$. The image $s(M)$ 
is a smooth divisor of $X$ isomorphic to $M$. 
By the irreducibility of fibers of $\pi$,   $\pi_*{\cal O}(2s(M))$ is a rank 
2 vector bundle $M$, and denote its dual bundle by $E$. The section of 
$\pi_*{\cal O}(2s(M))$ 
determined by $2s(M)$ gives rise the trivial line sub-bundle of 
$\pi_*{\cal O}(2s(M))$, hence one has 
the extension
$$
0 \longrightarrow L \longrightarrow E \longrightarrow 1 
\longrightarrow 0 
$$
where $L$ is a line bundle over $M$. The ratio of values of 
local sections of $\pi_*{\cal O}(2s(M))$ induces a double cover of $X$ over 
$\IP(E)$, in which $\IP(L)$ lies as a component of the branched locus 
corresponding to $s(M)$. As the normal bundle of $s(M)$ in $X$ is 
equal to $s^*K_M $, one obtains  $L = K_M^{-2}$. By 
$H^1(M, K_M^{-2}) = 0$, the above extension of $E$ splits and we have 
$E= K_M^{-2} \oplus 1$, hence (I) follows immediately. 
By the same argument one obtains (II)$\Box$ 
\noindent

\newsec{Topological correlation functions and mirror symmetry}

The $A$ and the $B$ models are topological $N=2$ supersymmetric 
$\sigma$-models with a Calabi-Yau d-fold $X$ as their target space. 
They correspond to two possibilities to twist the 
$N=2$, $c=3\cdot$d superconformal $\sigma$-model on the world-sheet 
\ref\wab{E. Witten, {\sl Mirror symmetry and topological field 
theory}, Essays on Mirror Manifolds (ed. S.-T. Yau), Int. Press 
(1992) Hong Kong}. The algebra of observable (BRST invariants) 
of the $A$ model is identified with the quantum deformation of 
the classical intersection algebra on 
$\cA=\oplus_{p=0}^n H^p(X,\wedge^p T^*)$. 
More precisely the corresponding cubic forms has the form 
\eqn\dumb{Q(a,b,c)=\int_X a\wedge b\wedge c+\sum N_d(a,b,c) {q^d\over 1-q^d}}
where $N_d(a,b,c)$ can be defined as certain intersection numbers on a moduli
space of mappings. Here $q^d=q_1^{d_1}\cdots q_m^{d_m}$, $m=h^{1,1}(X)$,
where $q_1,..,q_m$ are some local coordinates on 
the complexified K\"ahler cone of $X$. The series above is expected
to converge for small $|q_i|$.
The algebra of observables of the $B$ model is identified with an algebra 
on\foot{The equivalence is due to the unique
holomorphic $(d,0)$-form $\Omega$ present on every Calabi-Yau $d$-fold.}  
$\cB=\oplus_{p=0}^n H^p(X,\wedge^p T)\sim 
\oplus_{p=0}^n H^p(X,\wedge^{d-p} T^*)$, whose structure constants
can be analyzed using Griffith's transversality of the Gauss-Manin
connection on the middle dimensional cohomology of $X$. 
Especially the marginal operators of the $A$ and $B$ model 
are identified with elements of $H^{1,1}(X)$ and $H^{d-1,1}(X)$ 
respectively. 
All correlations functions of the topological theories can be 
obtained from these structure constants or equivalently from the 
2- and 3-point correlators. The 2-point correlator in a topological
field theory is purely topological: in the present cases it is simply
the Poincar\'e pairing on $\cA$ or $\cB$ respectively. 
Relative to any given base, we denote the matrix value of this
inner product $\bra,\ket:H^p(X)\otimes H^{d-p}(X)\ra\C$ by $\eta_{(p)}^{\alpha\beta}$.
It's inverse is denoted $\eta^{(p)}_{\alpha\beta}$. By the identification
of the marginal operators, the 3-point correlators will depend on 
$h^{1,1}(X)$ complexified K\"ahler moduli in the $A$ model and 
$h^{d-1,1}(X)$ complex structure moduli in the $B$ model. 
Our goal here is to show {\it all 3-point correlators 
of the $B$ model can be written  explicitly in terms of the periods 
for the middle dimensional cohomology of the Calabi-Yau $d$-fold $X$.} 
We give explicit expression for the periods and 3-point correlators 
containing two marginal operators, which are a direct generalisation 
of the formulas in~\hktyII. Using further properties of the Frobenius algebra 
one can derive explicit expressions for all correlators of the B-model 
on $X$ from them. By mirror symmetry the formalism can therefore 
be used to obtain the $A$-model correlation functions on $X$, after
suitable identification, from the $B$-model correlation functions 
on the mirror manifold $X^*$. This is in fact the main application 
we have in mind.  For one moduli Calabi-Yau of arbitrary 
dimensions this was discussed in~\ref\gmp{B. Greene, D. R. Morrison, 
R. Plesser, \cmp173(1995)559}. Some aspects of the generalization to 
multimoduli cases can be found in~\ref\jin{M. Jinzenij and M. Nagura \ijmpa11(1996) 455}
\ref\nag{M. Nagura, \mpl 10 (1995) 1667}\ref\sug{K. Sugiyama, 
hep-th 9504114; hep-th/9504115},\pema . Here we generalize the 
$d=4$ case treated in~\pema~to $d$-folds.

\subsec{The B-model algebra}

Let $\pi:{\cal X}\ra S$ be a family whose generic fiber 
is a Calabi-Yau $n$-fold $X_z$. One writes now 
the 3-point correlators as a cubic form on the groups  
$H^p(X_z,\wedge^p T)$. Put $\cB_z=\oplus H^p(X_z,\wedge^pT)$.
The cubic forms  are defined by 
\eqn\dumb{C(a,b,c)=\int\Omega(a\wedge b\wedge c)\wedge\Omega}
where $\Omega(a\wedge b\wedge c)$ is the contraction along the tangent direction
producing an $n$-form on $X_z$.

Mirror symmetry provides a vector space isomorphism $\phi_z:\cB_z\ra\cA$,
a mapping $z\mapsto z(q)$ and a normalization ${1\over f}$ such that
near the large radius limit $q=0$, we have
\eqn\dumb{{1\over f} C(\phi_z a,\phi_z b,\phi_z c)=Q(a,b,c)(q(z)),}
where $Q$ is the quantum corrected cubic form on $\cA$.
It's clear that $Q$ should be independent of the choice of $\Omega$. 
But $C$ depends
on $\Omega$ quadratically. Thus we expect that ${1\over f}$ must
be a holomorphic function near $q=0$ which cancels this dependence.
Near the large radius limit, there is a unique holomorphic period $\omega_0(z)=
\int_\gamma\Omega(z)$. The choice ${1\over f}={1\over\omega_0^2}$ therefore
provides a natural resolution to this cancellation problem. Equivalently
we can replace $\Omega$ by ${1\over \omega_0}\Omega$ and set $f=1$. 
This is what we shall do.
We shall first fix a base point $0\in S$, a topological
base of homology cycles and the dual base $\gamma_a^{(p)}$
on $H^n(X_0)$ with the property that 
$\bra\gamma_a^{(p)},\gamma_b^{(q)}\ket=0$ 
for $p+q\leq n$. For fixed $p$, the label $a$ in $\gamma^{(p)}_a$
takes $h^{n-p,p}(X_0)$ different values.
Due to mirror symmetry such a base will be the image of a 
base on $\cA$ under $\phi_0$. In fact in practice, there is usually a 
canonical choice of such a base on the A-model side.

There is a filtration of holomorphic
vector bundles  over $S$: $F_{(0)}\subset F_{(1)}\subset\cdots\subset F_{(n)}$,
where the fiber over $z\in S$ of $F_{(k)}$ is the vector space
$\oplus_{p=0}^{k} H^p(X_z,\wedge^p T)$. 
We now provide a set of frames for the these bundles. We shall
express these frames as linear combinations in the base $\gamma_a^{(p)}$
with holomorphically varying coefficients.
We shall see that these coefficients completely determine the
cubic form $C$. For each $k$, let $\{\alpha^{(0)}:=
\Omega,\alpha^{(1)}_a,..,\alpha^{(k)}_b\}$
be a frame of $F_{(k)}$ having the following upper-triangular
property with respect to the $\gamma^{(p)}_a$:
\eqn\alphaEQ{
\alpha^{(k)}_a=\gamma^{(k)}_a+\sum_{p>k}g^{(p)c}_a\gamma^{(p)}_c.}
(The $g^{(p)}$ actually depends on $k$, which we have suppressed
in the notation above.)
These frames can be obtained by row reduction on a given arbitrary base
of sections. (See \gmp.) Note that for $k=0$
the coefficients $g^{(p)}$ are exactly the periods 
of the above given homology cycles.
These periods are solutions to the Picard-Fuchs equations (in an
appropriate gauge). We will give explicit formulas later for these
periods for Calabi-Yau complete intersections in a toric variety. Note that
in $\alpha^{(0)}$ the coefficients $t_a:=g^{(1)}_a$ are regarded
as local coordinates on $S$. These are the so-called flat coordinates.
In these coordinates the Gauss Manin connection $\nabla_{a}$ becomes $\partial_{t_a}$, 
and the cubic form of type $(1,k,d-k-1)$ is given by 
\eqn\CubicForm{C^{(1,k,d-k-1)}_{a,b,c}
=\int_X \alpha^{(d-k-1)}_a\wedge \partial_{t_a} \alpha^{(k)}_b =:
\bra \partial_{t_a}\alpha^{(k)}_b,\alpha^{(n-k-1)}_c\ket.}

Using the upper-triangular property of the $\alpha^{(k)}_a$ and
the topological basis $\gamma^{(k)}$, it
is easy to show that
\eqn\InnerProduct{\eta_{ab}^{(k)}:=\bra\alpha^{(k)}_a,
\alpha^{(d-k)}_b\ket=
\bra\gamma^{(k)}_a,\gamma^{(n-k)}_b\ket.}
In particular these matrix coefficients are independent of $t$. 
Furthermore we claim that
\eqn\Transversality{\partial_{t_a}\alpha^{(k)}_b=
C^{(1,k,d-k-1)}_{a,b,c}\eta_{(d-k-1)}^{cd}\ \alpha^{(k+1)}_d.}
By Griffith's transversality, we have
$\partial_{t_a}\alpha^{(k)}_b\in F_{(k+1)}=Span\{\alpha^{(0)},..,
\alpha^{(k+1)}_a\}$. But because of the 
upper triangular form of $\alpha^{(k)}_b$, 
$\partial_{t_a}\alpha^{(k)}_b$ has
zero component along $\gamma^{(0)},..,\gamma^{(k)}_a$. Thus
it can be expressed as a linear combination (with holomorphically
varying coefficients) of the $\alpha^{(k+1)}_b$. To determine the
coefficients, we take its inner product with $\alpha^{(n-k-1)}_c$
and apply eqns ~\CubicForm,~\InnerProduct. The claim above then follows.

To summarize, our strategy for computing the A-model cubic form $Q$ on $X$
by mirror symmetry is as follows. Actually we will only do it for
a Frobenius subalgebra $\cA$ (see below) of the A-model algebra. First
we fix a topological basis on $\cA$ (In the case
of toric hypersurfaces, this basis
will come from toric geometry). We define
our isomorphism $\phi_z$ so that it sends this basis to the
holomorphically varying basis $\alpha^{(k)}_a$ of the B-model
with $1\mapsto\alpha^{(0)}$.
Then  we shall use eqns \CubicForm,\InnerProduct~ and
\Transversality~ as our crucial ingredients for computing the
B-model cubic forms $C$ explicitly. For this we shall need some elementary
theory of Frobenius algebras which we now discuss.

\subsec{Frobenius algebras}

In this section, all vector spaces are finite dimensional.
A Frobenius algebra is a commutative graded algebra
$A=\oplus_{i=0}^nA_{(i)}$, generated by $A_{(1)}$, has $A_{(0)}=\C\cdot 1$,
and a nondegenerate degree $n$ bilinear symmetric invariant pairing
$\bra,\ket:A\times A\ra\C$. Note that because we require
generation by $A_{(1)}$, this notion is slightly stronger than
the usual notion of a Frobenius algebra. We give some well-known
examples from geometry. Let $\P$ be a complete toric variety,
and $A^*(\P)$ be its Chow ring. Then $A^*(\P)\otimes\C$ is
a Frobenius algebra. The pairing here is the Poincar\'e pairing.
If $X$ is a hypersurface in $\P$, then it can be shown that the
ring
\eqn\dumb{\tilde A^*(X):=Im(A^*(\P)\ra A^*(X))=A^*(\P)/Ann([X])}
tensored with $\C$
is a Frobenius algebra. More generally, if $A$ is a Frobenius algebra,
and $x\in A_{(1)}$ is a nonzero element, then
$\tilde A:=A/Ann(x)$ is a Frobenius algebra with the
induced pairing $\bra a+Ann(x),b+Ann(x)\ket:=\bra a,b\cdot x\ket$
having degree $n-1$.

Let $V_1,V_2,V_3$ be vector spaces, and $C:V_1\otimes V_2\otimes V_3\ra\C$
be a cubic form. It is call $V_1$-nondegenerate if that $C_{(a,b,c)}=0$
for all $b,c$ implies that $a=0$. Similar notion of $V_i$-nondegeneracy
applies. We call the form nondegenerate if it is $V_i$-nondegenerate for all
$i$. Now suppose $C$ is $V_3$-nondegenerate. Then we have the following
invertibility property. Let $D:V_3^*\otimes V_4\ra\C$ be any bilinear
form. Then the knowledge of the 3-form
$E_{(a,b,d)}:=C_{(a,b,c_i)}D_{(\gamma^i,d)}$ ($\{c_i\},\{\gamma^i\}$ being dual bases),
allows us to determine $D$ completely. In fact, there exists (in general
not unique) a 3-form $F$ such that
$D_{(\gamma,d)}=F_{(\gamma,\alpha^i,\beta^j)}E_{(a_i,b_j,d)}$. 
This is just the statement that the $V_3$-nondegenerate cubic form $C$
defines an onto map $V_1\otimes V_2\ra V_3^*$, hence choosing
a section gives us a left inverse $F$ to this map.

We now return to a Frobenius algebra $A$.
it determines a collection of cubic forms
$C^{(ijk)}: A_{(i)}\otimes A_{(j)}\otimes A_{(k)}\ra\C$ with
$i,j,k\geq0, i+j+k=n$.
These cubic forms are $A_{(i)}$-nondegenerate
whenever either $j=1$  or $k=1$ because $A_{(1)}\cdot A_{(i)}=A_{(i+1)}$.

\subsec{Reconstruction}

Let $A=\oplus_{i=0}^n A_{(i)}$ be a graded space with $A_{(0)}=\C$
and equipped with a degree $n$ nondegenerate symmetric bilinear form
$\eta$. Suppose we are given cubic forms:
$C^{(ijk)}:A_{(i)}\otimes A_{(j)}\otimes A_{(k)}\ra\C$,
$i,j,k\geq0$ with the following properties:
\item{(a)} (Degree) $C^{(ijk)}=0$ unless $i+j+k=d$.
\item{(b)} (Unit) $C^{(0ij)}_{(1,b,c)}=\eta^{(i)}_{b,c}$.
\item{(c)} (Nondegeneracy) $C^{(1ij)}$ is nondegenerate in the second slot.
\item{(d)} (Symmetry) For any permutation $\sigma$ of 3 letters,
$C^{(ijk)}_{(a,b,c)}=C^{\sigma(ijk)}_{\sigma(a,b,c)}$.
\item{(e)} (Associativity)
$$C^{(i,j,n-i-j)}_{(a,b,c_p)}\eta_{(n-i-j)}^{pq}C^{(i+j,k,n-i-j-k)}_
{(d_q,e,f)}
=C^{(i,k,n-i-k)}_{(a,e,c_p')}\eta_{(n-i-k)}^{pq}C^{(i+k,j,n-i-j-k)}_
{(d_j',b,f)}$$
where the $c$ and the $d$ are bases of the appropriate spaces.

Then $A$ is a Frobenius algebra with the product
\eqn\dumb{a\cdot b=C_{(a,b,c_p)}\eta^{pq}d_q.}
The rules above are known as fusion rules.
One can also build a $k$-form by fusing together 2- and 3-forms.
The associativity law says that there will often be many ways
to build a given $k$-form. Similarly the 3-forms are not
independent. We claim that the forms of type $(i,j,n-i-j)$ for
$i,j>1$ are determined by the those of type $(1,r,n-r-1)$.
To see this without loss of generality, we can assume $1<n-i-j\leq i,j$.
Now by the associativity law above with $k=n-i-j-1$ and the invertibility
property of $C^{(i+j,k,n-i-j-k)}=C^{(i+j,k,1)}$, it follows that
$C^{(i,j,n-i-j)}$ are determined  in terms of forms of type
$(i,n-i-j-1,j+1)$ and $(i+k,j,1)$. By the symmetry property,
$(i,n-i-j-1,j+1)$ is equivalent to $(i,j+1,n-i-j-1)$.
Thus we have reduced the value of $n-i-j$ by 1. 
By induction, we see that all $(i,j,n-i-j)$ can be expressed
in terms of those of type $(1,r,n-r-1)$. In terms of the algebra
$A$ itself, an alternative way to state the result is that
all the products $A_{(i)}\otimes A_{(j)}\ra A_{(i+j)}$ is
determined by those of the form $A_{(1)}\otimes A_{(r)}\ra
A_{(r+1)}$ because $A$ is generated by $A_{(1)}$ and that
\eqn\dumb{(a_1\cdots a_i)(a_{i+1}\cdots a_{i+j})
=a_1(a_2\cdots a_{i+j}).}

\subsec{Application}

Let $X$ be a Calabi-Yau $n$-fold, and let $\cA$ be a Frobenius subalgebra
of $\oplus_{p=0}^n H^p(X,\wedge^pT^*)$. Suppose mirror symmetry holds:
there is a mirror family $X^*$ whose B-model algebra coincides with
the A-model algebra of $X$. We shall now compute the Frobenius
subalgebra $\cB$ of the B-model algebra corresponding to $\cA$.
{}From our general discussion of Frobenius algebras, it is enough
to compute the cubic forms $C$ of types $(1,r,n-r-1)$ which come
with $\cB$. Once we have a period expansion in the topological base \alphaEQ~these can be 
easily obtained using eqns ~\CubicForm,~\InnerProduct and ~\Transversality. 
To obtain the coefficients in \alphaEQ~we will use the fact\ref\hkty{S. Hosono, A. Klemm, S. Theisen and  
S.T. Yau, \npb433(1995)501-554}\hly
 that the universal structure of the solution of 
the Picard-Fuchs equation on $X^*$ at the large radius point mirrors the 
primitive part of the vertical cohomology of $X$ and the 
leading structure of logarithm enables us to associate  
this solutions with the expansion of the periods in a 
topological base. This leads to a direct generalisation of the formulas of 
\hkty\ to some correlation functions on $d$-folds. 

More precisely there are $h^{r,r}_{prim}(X)$ solutions $0<r<d$ 
with leading degree $r$ in the $\log(z_i)$, which have the form 
\eqn\sol{\tilde \Pi^{(r)}_k=\sum_{\Pi} 
{^0C^{d-r,1\ldots 1}_{k,i_1,\ldots,i_r}} 
\bigl({1\over r!} l_{i_1} \ldots l_{i_r} S_0 + 
{1\over (r-1)!} l_{i_1} \ldots l_{i_{r-1}} S_{i_r} +
\ldots + S_{i_1,\ldots,i_r}\bigr),}
here we defined $l_i:=\log(z_i)$ and the $S_{i_1,\ldots i_r}$ are 
holomorphic series in the $z_i$, whose explicit form are given 
below. The map to an specific element of the cohomology $H^{d-r,d-r}$ of 
$X$ can be made precise by noting that the 
$^0C^{d-r,1\ldots 1}_{k,i_1,\ldots i_r}$ are given by the 
classical intersection of that specific element with the intersection of
divisors $J_{i_1}\cdot\ldots\cdot J_{i_r}$. We discuss the primitive 
part of the (co)homology generated by $J_1\ldots J_{h^{1,1}}$ only and by
Poincare duality, this data fix the element in $H^{d-r,d-r}$ completely.

As mentioned above the covariant derivative $\nabla_{a}$ in \gmp\ 
becomes the ordinary derivative in the flat complexified K\"ahler structure coordinates $t_k$.
The coordinate change from the natural complex structure coordinates 
$z_a$ to the $t_k$ variables is given by the mirror map 
$t_k={\tilde \Pi^{(1)}_k(z_i) \over \tilde \Pi^{(0)}(z_i)}=\log(z_k)+{S_k\over S_0}$. 
If we substitute this coordinate transformation in the normalized periods 
$\Pi_i^{(r)}={\tilde \Pi_i^{(r)}\over \tilde \Pi^{(0)}}$ 
some simplifications occur as the first subleading terms in the 
$t_i$ cancel out:       
\eqn\soldr{\Pi^{(r)}_k=\sum_{\Pi} 
{^0C^{d-r,1\ldots 1}_{k,i_1,\ldots,i_r}} 
\bigl({1\over r!} t_{i_1} \ldots t_{i_r}   +  
{1\over (r-2)!}  t_{i_1} \ldots t_{i_{r-2}} \hat S_{i_{r-1}} \hat S_{i_r}+
\ldots + \hat S_{i_1,\ldots,i_r}\bigr).}
Now we notice from the monodromy around $z_i=0$ ($t_i\rightarrow t_i+1$) 
that the periods $\Pi_k^{(r)}$ correspond to a expansion 
of $\alpha^{(0)}=\Omega$ in terms of the topological 
basis\foot{This is actually only true up to the addition of 
solutions with subleading logarithms, which however does not
affect the holomorphic couplings discussed below. It will 
affect however the non-holomorphic Weil-Peterson metric.} 
$\gamma_{(r)}^k$ of \alphaEQ~ $\alpha^{(0)}=\sum_{k,r} \Pi^{(r)}_k {\gamma}_{(r)}^k$.

The coupling $C^{(1,1,d-2)}_{a,b,c}:H^{1,1}\times 
H^{1,1}\times H^{d-2,d-2}\rightarrow \IC$ is especially 
simple to obtain. Applying \Transversality~ in the case $k=0$ we have 
$\partial_{t_a}\alpha^{(0)}=
\alpha^{(1)}_a$. This determines $\alpha^{(1)}_a$, hence all its coefficients.
Now using \CubicForm~ for $k=1$, \alphaEQ~  for $k=1,d-2$,
and the fact that
$\bra\gamma^{(k)}_a,\gamma^{(l)}_b\ket=0$ for $k+l>d$, we see
that
\eqn\IIdmII{C^{(1,1,d-2)}_{a,b,c}=\partial_{t_a}g^{(2)d}_b\eta^{(2)}_{dc}=
\partial_{t_b}\partial_{t_b}\Pi^{(2)}_c,} 
where the $g^{(2)}$ are the coefficients of the $\gamma^{(2)}$ in the 
$\alpha^{(1)}$. Note that the last equation follows from the fact that 
$\Pi^{(r)}_a$ is an expansion in the dual base $\gamma^a_{(r)}$ 
and that the associativity of the classical parts in \IIdmII~is manifest.
Eqs.~\soldr\IIdmII~are direct generalizations of eqs. (4.9) and (4.18)
to the $d$-fold case. For $d=4$ an equivalent description has been 
given in~\pema. For $H^{1,1}$ we have always a canonical choice 
of the basis say $J_1\ldots J_{h^{1,1}}$, as there is a canonical basis 
for the tangent space of the moduli space corresponding to elements 
$H^{d-1,1}(X^*)$, which is mapped by the monomial divisor mirror map 
to $H^{1,1}(X)$ and \IIdmII\ reduces for $d=3$ to the 
expressions given in \hkty. 
For $d>3$ there is a priori no canonical choice for the 
basis of $H^{d-2,d-2}$. However toric geometry can be used as in~\hly~to 
show that the graded ring 
$${\cal R}=\IC[\theta_1,\ldots ,\theta_{h^{1,1}}]/{\cal J},$$ 
where ${\cal J}$ is the ideal generated by the leading $\theta$-terms of 
Picard-Fuchs equations, gives, by the identification 
$\theta_i\rightarrow J_i$, a presentation of the primitive part of 
$H^{*,*}$. Because of Poincare duality it is of course sufficient to pick
a basis of half of $H^{*,*}$ and as mentioned above the choice 
of the basis in $H^{1,1}$ is canonical. It was shown in 
\hkty\hly\ that any element of $\cal R$ can be mapped to a 
solution \sol , i.e.  the $^0C^{d-r,1\ldots 1}_{i_1,\ldots,i_r}$ are 
determined by the principal part of the Picard-Fuchs equation. 
This can be viewed as a proof of mirror symmetry at the level
of the classical intersections, which readily generalizes to $d$-folds.

Now proceed by induction. Suppose we know (the coefficients of)
the $\alpha_{(i)}$ 
and the cubic forms of types $(1,i,n-i-1)$
for $i=0,1,..,k$. Then by the invertibility property of
a cubic form of type $(1,k,n-k-1)$ in a Frobenius algebra,
we can solve for the $\alpha_{(k+1)}$ using \Transversality.
Thus the $\alpha^{(k+1)}$ are determined.
By \alphaEQ, we can write $\partial_{t_a}\alpha^{(k+1)}_b=
\partial_{t_a}g^{(k+2)d}_b\gamma^{(k+2)}_d+\cdots$ (which is
now known),
arguing as before using \CubicForm~ with $k$ replaced by $k+1$, 
and using the inner product property of the $\gamma$, we find that
$C^{(1,k+1,n-k-2)}_{abc}=
\partial_{t_a} g^{(k+2)d}_b\eta^{(k+2,n-k-2)}_{dc}$.
Thus the cubic form of type $(1,k+1,n-k-2)$ is also determined.
This shows that all cubic forms of type $(1,k,n-k-1)$ for $k=1,2,..,n-1$
can be expressed in terms of the coefficients of $\alpha_{(0)}$ alone.

\subsec{Explicit expressions for periods and instanton sums for 
complete intersections in toric varieties}   

Following \hkty\ we can determine the holomorphic series  
$S_{i_1,\ldots,i_r}$ from the generators of the Mori cone.
Consider a Calabi-Yau $d$-fold defined as complete 
intersection with $p$ polynomial constraints in a toric 
variety of dimension $d+p$. The generators of the Mori cone will be
of the form
$$l^{(i)}=(\hat l^{(i)}_0,\ldots,\hat l^{(i)}_{p-1};
l^{(i)}_1,\ldots,l^{(i)}_q),$$ 
where $q=d+p+h^{d-1,1}$. The series $S_{i_1,\ldots,i_r}$ are
obtained by the Frobenius method from the coefficients of the 
holomorphic function $\omega(\vec z,\vec \rho)$
$$\eqalign{\omega(z,\vec \rho)&=\sum c(\vec n,\vec \rho)
\prod_{j=1}^{h^{1,D-1}} z_j^{n_j+\rho_j}\cr
c(\vec n,\vec \rho)&={ \prod_{k=1}^p \Gamma(1-\sum_{i=1}^{h^{1,D-1}} 
\hat l^{(i)}_k(n_i+\rho_i))\over
 \prod_{k=1}^q \Gamma(1-\sum_{i=1}^{h^{1,D-1}} 
 l^{(i)}_k (n_i+\rho_i))}\cr 
S_{i_1,\ldots,i_r}&=\partial_{\rho_{i_1}}\ldots\partial_{\rho_{i_r}}
\omega(\vec z,\vec \rho)|_{\vec \rho=\vec 0}}$$     
Notably with leading behavior 
$S_0=1+\ldots$, $S_i=z_i+\ldots$.

This gives the explicit expansion of $C^{(d-2,1,1)}_{A,b,c}
=^0C^{(d-2,1,1)}_{A,b,c}+\cO(q_i)$, with $q_i=e^{t_i}$. The latter has 
a conjectural interpretation as being the counting  function 
for invariants of maps from the two sphere into $X$. These maps are 
defined such that two fixed points $P_b$, $P_c$ are mapped to 
the divisors $\cD_b$, $\cD_c$, while one point $P_A$ is mapped to the 
codimension $r$ subvariety $A$ in a class of $H^{r,r}(X)$. And the 
invariant is the Euler class of the moduli space of that curve, 
weighted by $(-1)^{{\rm dim \ } \cM}$. From the 
definition of the degree a generic rational curves of degree $d_l$ 
will pass through the divisor $\cD_l$ in $d_l$ points, 
but a generic curve does not pass through the submanifold $A$ 
of higher codimension then one. If we require the latter this 
imposes a restriction and the invariants of that specific 
curves will be labeled by the class of $A$. Moreover in the path 
integral definition of $C^{(d-2,1,1)}_{A,b,c}$ 
one integrates over the points $P_i$ and has accordingly to divide 
by a combinatorial factor of $d_b d_c$ in order to extract the 
invariant for the elementary rational curves $n^{(A)}_{\vec d}$ from 
the three-point function. 
By a similar reasoning as in 
\ref\am{P. Aspinwall, D. Morrison, \cmp151(1993)245}
is was described in \gmp\ how to subtract the multiple 
wrapping contributions from the lower degree curves in order 
to get the invariants of the elementary curves at given 
multidegree $\vec d$. Taking both effects into account the expansion of the 
three-point function in terms of invariants $n_{\vec d}$ is as 
follows\foot{For all toric varieties  these invariants can be calculated 
with a updated version of the program INSTANTON (which is available on 
request) from the Mori generators and the classical intersections.}     
\eqn\inst{C^{(d-2,1,1)}_{A,b,c}=^0C^{(d-2,1,1)}_{A,b,c}+
\sum_{\vec d} {d_a d_b n^{(A)}_{\vec d}\over 1-
\prod_{i=1}^{h^{1,1}} q_i^{d_i} } \prod_{i=1}^{h^{1,1}} q_i^{d_i}.}

\subsec{Examples of the quantum cohomology rings and transitions}

Let us discuss as the simplest example case (1) of chapter 5,
the elliptic fibration with $X_6(1,2,3)$ fibre over $\IP^3$ and 
its transition by the blow up at an equivariant fix point in 
$\IP^3$ to model (3) and along the irreducible subvariety to 
model (5). Evaluation of the explicit quantum cohomology 
in other cases can be found in \pema .

The toric representation of the mirror of (1) is defined 
by (4.1) were $\Delta^*$, is given by (5.1) as the convex hull 
of the following points 
\eqn\cpoly{\eqalign{
\nu_0^*&=(\phm 0,\phm 0,\phm 0,\phm 0,\phm 0)\cr 
\nu^*_1&=(\phm 1,\phm 0,\phm 0,\phm 0, \phm 0)\cr      
\nu^*_2&=(\phm 0,\phm 1,\phm 0,\phm 0, \phm 0)\cr      
\nu^*_3&=(\phm 0,\phm 0,\phm 1,\phm 0, \phm 0)\cr      
\nu^*_4&=(-1,-1,-1,-8,-12)\cr      
\nu^*_6&=(\phm 0, \phm 0,\phm 0,\phm 1 , \phm 0)\cr      
\nu^*_7&=(\phm 0,\phm 0,\phm 0,\phm 0, \phm 1 )\cr      
\nu^*_8&=(\phm 0,\phm 0,\phm 0, -2 , -3).}}
The manifold itself can be described by considering the vanishing
of the Newtonpolynom of the polar polyhedron $\Delta$ in $P_\Delta^*$. 
It turns out to be a degree $24$ Fermat hypersurface in a weighted 
projective space $X_{24}(1,1,1,1,8,12)$. 

There is a unique triangulation of the polyhedron 
$\Delta^*$ from its origin $\nu_0^*=(0,0,0,0,0)$. Note that the points 
$\nu^*_1,\nu_2^*,\nu_3^*,\nu_4^*,\nu_7^*$ all lie on 
a codim $2$ face of $\Delta^*$, with $\nu_7$ the interior point of
that face, while the points $\nu_5^*,\nu_6^*,\nu_7^*$ and $\nu_0^*$ lie on 
a codim $3$ plane, which cuts the polyhedron. The two linear relation 
implied by this lead to the two generators of the Mori cone. 

$$\eqalign{l^{(1)}&=(\phm  0;1,1,1,1,0,0,-4)\cr
           l^{(2)}&=(  -6  ;0,0,0,0,2,3,\phm 1)\cr}$$

The two K\"ahler classes $J_1,J_2$ dual to this Mori generators measure 
classically the volume of the base $\IP^3$ and  the size of the fiber 
respectively. While the the divisor $D_1$ associated to the 
first Mori cone represents the section and is horizontally, $D_2$ is a vertical divisor, which 
intersects the base $\IP^3$ in codim $2$. Since three planes do not 
intersect generically in  $\IP^3$ the classical 4-point coupling
${\cal D}_1\cdot {\cal D}_1 \cdot {\cal D}_1\cdot 
{\cal D}_2=\int J_1^3 J_2$ is zero. The other classical 
4-point couplings 
$\int J_i J_k J_l J_m$ and the evaluation 
$\int c_2 J_i J_k$, and $\int c_3 J_i$ 
are summarized by the coefficients in the following formal polynomials    
$$\eqalign{
{\cal C}_0&=J_2 J_1^3+ 4 J_2^2 J_1^2 + 16 J_2^3 J_1+ 64 J_2^4\cr
{\cal C}_2&= 48 J_1^2 +182 J_1 J_2 + 728 J_2^2\cr
{\cal C}_3&= -960 J_1 - 3860 J_2}$$

The Picard-Fuchs equations for the mirror manifold are

$$\eqalign{{\cal L}_1&=  
\theta_1^4  - (4\theta_1 - \theta_2-4)
( 4\theta_1 - \theta_2-3 )(4\theta_1 - \theta_2-2)
(4\theta_1 - \theta_2-1)z_1\cr
{\cal L}_2&=  \theta_2(\theta_2-4 \theta_1 ) - 
12(6\theta_2-5)(6\theta_2-1) z_2},$$

have the following discriminant

$$\eqalign{
\Delta_1&=(1-256 z_1)\cr
\Delta_2&=(1-432 z_2)^4-z_1 z_2^4.}$$

The mirror map $z_2(q_1=0,q_2)=P(J(t_2))$ is defined by the 
ratio of two periods of holomorphic $1$-form on the elliptic 
curve $X_{6}(1,2,3)$, while mirror map $z_1(q_1,q_2=0)$ is described 
by the ratio of periods over a  meromorphic differential on the 
$K_3$ surface $X_4(1,1,1,1)$.    

The basis of $H^{1,1}$ are denoted by $J_1,\ldots, J_r$.
We choose then a basis of $H^{2,2}$ 
$$\eqalign{b^{(2)}_1&=J_1^2\cr
           b^{(2)}_2&=J_1 J_2+ 4 J_2^2.}$$
The intersection matrix between elements of $H^{2,2}$ 
in this basis is 
$$\eta_{(2,2)} =\left(\matrix{0& 17\cr 17& 1156}\right).$$
If we determine the basis of $H^{(3,3)}$ by the requirement 
that Poincar\`e bilinear pairing takes the simplest form 
$\eta_{(1,3)}^{i,j}= \delta^{i,h^{1,1}-i+1}$ with the canonical basis 
of $H^{1,1}$, then we get 
$$\eqalign{b^{(3)}_1&=J_1^3\cr 
           b^{(3)}_2&={1\over 273}(J_1^2 J_2+
4 J_1 J_2^2+16 J_2^3)-4 J_1^3.}$$
The basis for $H^{4,4}$ is fixed up to a volume normalization of the
d-fold, which we choose so that $\eta_{0,d}^{1,1}=1$. In our 
case above $b^{(4)}={1\over 75} \cC_0$. 

The leading order logarithms in the periods are according 
to \sol
$$\eqalign{\Pi^{(2)}_1&=S_0(l_1 l_2+2 l_1^2)+\cO(l)\cr
           \Pi^{(2)}_2&=S_0({17\over 2} l_1^2+68 l_1 l_2+
                        136 l_2^2)+ \cO(l).}$$ 

The invariants for the genus zero curves from the normalized 
three-point functions listed in the two tables below 

\noindent $b_1^{(2)}=J_1^2$, ${1\over 20} C^{(2,1,1)}_{1,i,j}$:
$$
\fivepoint{\vbox{\offinterlineskip\tabskip=0pt
\halign{\strutf
\vrule#&
{}~\hfil$#$~&
\vrule#&
{}~\hfil$#$~&
{}~\hfil$#$~&
{}~\hfil$#$~&
{}~\hfil$#$~&
{}~\hfil$#$~&
\vrule$#$\cr
\noalign{\hrule}
& m &&n^{(1)}_{0,m} &  n^{(1)}_{1,m}&n^{(1)}_{2,m}
&n^{(1)}1_{3,m}&n^{(1)}_{4,m} & \cr
\noalign{\hrule}
&0 && 0     &  0&   0     &         0&             0&\cr
&1 &&-1     &   384&-90000   &  13919744& 31152804996 &\cr
&2 &&-41    &24576&-7990080 &1785169920& -301991420880&\cr
&3 &&-3403  &2812800 & -1230118560 &369021660288& -84154079407488&\cr
&4 &&-374322& 397171200 &-219729224832 & 83117668597760& -23932769831261760&\cr
&5 &&-48251945&62575303680&-41951914533360&19174105171468800&-6670224866876828160
&\cr
\noalign{\hrule}
}
\hrule}}$$

\noindent $b_2^{(2)}=J_1 J_2+4J_2^2$, ${1\over 16320} C^{(2,1,1)}_{2,i,j}$: 
$$
\fivepoint{\vbox{\offinterlineskip\tabskip=0pt
\halign{\strutf
\vrule#&
{}~\hfil$#$~&
\vrule#&
{}~\hfil$#$~&
{}~\hfil$#$~&
{}~\hfil$#$~&
{}~\hfil$#$~&
{}~\hfil$#$~&
\vrule$#$\cr
\noalign{\hrule}
& m &&n^2_{0,m} &  n^2_{1,m}&n^2_{2,m}&n^2_{3,m}&n^2_{4,m} & \cr
\noalign{\hrule}
&0 && 0     &  1&   2     &         3&             4&\cr
&1 &&0&6&-1893&439256&2661669198&\cr
&2 &&0&189& -102750&31221300&-6618229812&\cr
&3 && 0&14366& -11162250&4632513522&-1326773710832&\cr
&4 && 0 & 1518750& -1537867338&816075268892&-297124091742240 &\cr
&5 && 0& 191238192&-238866784083&154724059936392&-68479975849390752 &\cr
\noalign{\hrule}
}
\hrule}}$$

Adding of the point $\nu_5^*=(0,-1,-1,-6,-9)$ correspond
to an blow up of $\IP^3$ along an $\IP^1$ and leads to model (5).
This transition has a close similarity to the transition by 
shrinking (blowing) a Del Pezzo surface studied in~\mvII\kvm~as in
the fourfold a {\sl six-cycle} shrinks along the $E_8$ Del 
Pezzo\foot{Similarly one can observe the shrinking of $E_7$, $E_6$, ($D_5$) Del Pezzo
surface in the corresponding  fibrations types. } surface to $T$ 
invariant orbit in the base. In fact we will see the $E_8$ partition function 
$$\hat \Lambda_{E_8}={1 \over 2} \sum_{\alpha=even} {\theta_\alpha^8(\tau)
\over \eta(\tau)^{12}}=1+252q+5130q^2+\ldots $$
appearing as counting functional of the instantons in the appropriate
normalized threepoint functions, marked by the $*$ in the table below 
(as well as the higher degree invariants of the shrinking Del Pezzo, 
marked with the $\diamond$). This model has two phases and 
in the first the Stanley Reisner ideal is given by 
${\cS}=\{x_2 x_5, x_1 x_3, x_1 x_3 x_4, x_6 x_7 x_8 \}.$
The Mori generators below correspond to the classes of the curve 
in the $\IP(\cO\oplus \cO_{\IP^1}\oplus_{\IP^1}(1))$ bundle (2), 
a section of the $\IP^1$ base in this bundle (1) and the class 
of the elliptic fibre over $B$ (3): 
\eqn\mcIV{\eqalign{
l^{(1)}&=(\phm  0;0, 1, 0,  -1, 1, 0, 0, -1),\cr 
l^{(2)}&=(\phm  0;1, 0, 1, \phm 1, 0, 0, 0, -3),\cr 
l^{(3)}&=(     -6;0, 0, 0, \phm 0, 0, 2, 3,  \phm 1),}} 
\noindent
The classical couplings are 
\eqn\classIV{\eqalign{\cC_0&=
J_1J_2^2J_3 + J_2^3J_3 + 3J_1J_2J_3^2 + 4J_2^2J_3^2 + \cr & 
9J_1J_3^3 + 15J_2J_3^3 + 54J_3^4\cr
\cC_2&=36J_1J_2+102J_1J_3+48J_2^2+172J_2J_3+618J_3^2\cr
\cC_3&=-540J_1-900J_2-3258J_3.}}
Analogous as in~\mvII~one has to flop the $\IP^1$ in $B$ first.
As such flops were not discussed in the fourfolds context 
let us give the data of this transition to the second phase
whose Mori generators are 
$l^{\prime (1)} =-l{(1)}$, 
$l^{\prime (2)} =l^{(1)}+l^{(2)}$ and 
$l^{\prime (3)} =l^{(1)}+l^{(3)}$. The Stanley Reisner Ideal
changes to $\cS=\{x_4 x_8,x_1 x_3 x_4, x_6 x_7 x_8, 
x_1 x_2 x_3 x_5, x_2 x_5 x_6 x_7\}$ while the classical couplings
become 

\eqn\PHASEII{
\eqalign{
\cC_0&=
J'_3 J^{\prime 3}_2+4 J^{\prime 2}_2 J^{\prime 2}_3+
15J^{\prime 3}_3J'_2+54J^{\prime 4}_3+J'_1J^{\prime 3}_2+
4J'_1J^{\prime 2}_2 J'_3
+16J'_1J^{\prime 2}_3J'_2\cr & 
+60J'_1J^{\prime 3}_3+4J^{\prime 2}_1
J^{\prime 2}_2+16J'_2J'_3J^{\prime 2}_1+64J^{\prime 2}_1
J^{\prime 2}_3+16J^{\prime 3}_1 J'_2+
64J^{\prime 3}_1 J'_3+64J^{\prime 4}_1\cr
\cC_2&=48J^{\prime 2}_2+172J'_2J'_3+182J'_2J'_1+
618J^{\prime 2}_3+688J'_3J'_1+728J^{\prime 2}_1\cr
\cC_3 &=-900J'_2-3258J'_3-3620J_1'.}}

The positive scaling relations on the variables $x_1,\dots,x_8$ are 
\eqn\mcIV{\eqalign{
&(-18;1, 0, 1, 1, 0, 6, 9,0),\cr 
&(-24;1, 1, 1, 0, 1, 8, 12,0),\cr 
&(-6 ;0, 0, 0, 0, 0, 2, 3,1),}}
and the Weierstrass form  
$$x_7^2=x_6^3+x_6 x_8^4\sum_{\mu,\nu,\rho}x_1^\mu x_3^\rho 
x_2^\nu x_5^{16-\mu-\nu-\rho} x_4^{12-\mu-\rho}+
x_8^6\sum_{\mu,\nu,\rho}x_1^\mu x_3^\rho 
x_2^\nu x_5^{24-\mu-\nu-\rho} x_4^{18-\mu-\rho}.$$
The singularity at $D_4$, near $x_2=x_5=0$ and along $(x_1,x_3)$ 
is recognized as the {\sl canonical singularity with crepant blowup }
which signals the collapse of the $E_8$ Del Pezzo surface~\mvII~and
is smoothed to a generic member of the family $X_{24}(1,1,1,1,8,12)$ 
by perturbing with those terms, which were forbidden by the first 
scaling relation. This completes the transition to the fibration over 
$\IP^3$.

\noindent
With the choice of basis   
\eqn\sbasIV{\eqalign{
b_1^{(2)} = J_1J_2, \ \ \ b_2^{(2)} = J_1J_3 + J_3^2
, \ \ \
b_3^{(2)} = J_2^2, \ \ \ b_4^{(2)} = J_2J_3 + 3J_3^2,}}
we have the following  data for the quantum cohomology ring
$$\eta_{2,2}=\left(\matrix{0& 3& 0& 10\cr 3& 72& 5& 207\cr
             0& 5& 0& 13&\cr 10& 207& 13& 580}\right)$$

\noindent $b_1^{(2)}=J_1 J_2$, $C^{(2,1,1)}_{1,i,j}$:
$$
\fivepoint{\vbox{\offinterlineskip\tabskip=0pt
\halign{\strutf
\vrule#&
{}~\hfil$#$~&
\vrule#&
{}~\hfil$#$~&
{}~\hfil$#$~&
{}~\hfil$#$~&
{}~\hfil$#$~&
{}~\hfil$#$~&
{}~\hfil$#$~&
{}~\hfil$#$~&
{}~\hfil$#$~&
{}~\hfil$#$~&
{}~\hfil$#$~&
{}~\hfil$#$~&
\vrule$#$\cr
\noalign{\hrule}
& m &&
n^{(1)}_{m,0,0} &  n^{(1)}_{m,0,1}&n^{(1)}_{m,0,2}&
n^{(1)}_{m,0,3} &  n^{(1)}_{m,0,4}&
n^{(1)}_{m,1,0} &  n^{(1)}_{m,1,1}&n^{(1)}_{m,1,2}&
n^{(1)}1_{m,2,0} &  n^{(1)}_{m,2,1}&n^{(1)}_{m,2,2}& \cr
\noalign{\hrule}
&0 && 0&0&0&0&0&3&-1080&143370&-12&5400&-1149120 &\cr
&1 && 1^*&252^*&5130^*&54760^*&419895^*&-19&6840&-1578960&344&-
182520&5206830&\cr
&2 &&0  &0&-2\cdot 9252^\diamond&-2\cdot 673760^\diamond&
-2\cdot 20534040^\diamond & 1&-360&156060&-798&447480&-140472720&\cr
\noalign{\hrule}
}
\hrule}}$$

\noindent $b_2^{(2)}=J_1 J_3+J_3^2$, ${1 \over 12 } C^{(2,1,1)}_{(2,i,j)}$:
$$
\fivepoint{\vbox{\offinterlineskip\tabskip=0pt
\halign{\strutf
\vrule#&
{}~\hfil$#$~&
\vrule#&
{}~\hfil$#$~&
{}~\hfil$#$~&
{}~\hfil$#$~&
{}~\hfil$#$~&
{}~\hfil$#$~&
{}~\hfil$#$~&
{}~\hfil$#$~&
{}~\hfil$#$~&
{}~\hfil$#$~&
\vrule$#$\cr
\noalign{\hrule}
& m &&
n^{(2)}_{m,0,0} &  n^{(2)}_{m,0,1}&n^{(2)}_{m,0,2}&
n^{(2)}_{m,1,0} &  n^{(2)}_{m,1,1}&n^{(2)}_{m,1,2}&
n^{(2)}_{m,2,0} &  n^{(2)}_{m,2,1}&n^{(2)}_{m,2,2}& \cr
\noalign{\hrule}
&0 &&315&630&945&0&-630&167265&0&1575&-670320&\cr
&1 && 0&249&9495&0&1890&-577485&0&34020&16320375&\cr
&2 && 0&0&-17268& 0& 0& 56970&0& 59535& -31350510&\cr
\noalign{\hrule}
}\hrule}}$$

\noindent $b_3^{(2)}=J_2^2$, ${1\over 2} C^{(2,1,1)}_{3,i,j}$: 
$$
\fivepoint{\vbox{\offinterlineskip\tabskip=0pt
\halign{\strutf
\vrule#&
{}~\hfil$#$~&
\vrule#&
{}~\hfil$#$~&
{}~\hfil$#$~&
{}~\hfil$#$~&
{}~\hfil$#$~&
{}~\hfil$#$~&
{}~\hfil$#$~&
{}~\hfil$#$~&
{}~\hfil$#$~&
{}~\hfil$#$~&
\vrule$#$\cr
\noalign{\hrule}
& m &&
n^{(3)}_{m,0,0} &  n^{(3)}_{m,0,1}&n^{(3)}_{m,0,2}&
n^{(3)}_{m,1,0} &  n^{(3)}_{m,1,1}&n^{(3)}_{m,1,2}&
n^{(3)}_{m,2,0} &  n^3_{m,2,1}&n^{(3)}_{m,2,2}& \cr
\noalign{\hrule}
&0&&0&0&0&4&-1260&236520&-19&7920&-1624950&\cr
&1&& 0&0&0&-10&3600&-831600&256&-133560&38111040&\cr
&2&&0&0&0&0&0&20520&-410&230400&-72511020&\cr
\noalign{\hrule}}
\hrule}}$$

\noindent $b_4^{(2)}=J_2 J_3+3 J_3^2$, ${1\over 12} C^{(2,1,1)}_{4,i,j}$:
$$
\fivepoint{\vbox{\offinterlineskip\tabskip=0pt
\halign{\strutf
\vrule#&
{}~\hfil$#$~&
\vrule#&
{}~\hfil$#$~&
{}~\hfil$#$~&
{}~\hfil$#$~&
{}~\hfil$#$~&
{}~\hfil$#$~&
{}~\hfil$#$~&
{}~\hfil$#$~&
{}~\hfil$#$~&
{}~\hfil$#$~&
\vrule$#$\cr
\noalign{\hrule}
& m &&
n^{(4)}_{m,0,0} &  n^{(4)}_{m,0,1}&n^{(4)}_{m,0,2}&
n^{(4)}_{m,1,0} &  n^{(4)}_{m,1,1}&n^{(4)}_{m,1,2}&
n^{(4)}_{m,2,0} &  n^{(4)}_{m,2,1}&n^{(4)}_{m,2,2}& \cr
\noalign{\hrule}
&0&& 0&885&1770&0&-1770&469935&0&4425&-1883280&\cr
&1&& 0&489&18945&0&-5310&-1606995&0&-95580&45813825&\cr
&2&&0&0&-34383&0&0&113670&0&167265&-87245010&\cr
\noalign{\hrule}}
\hrule}}$$

The blow up to (3), which is the $\IP^1$-bundle $\IP(\cO)_{\IP^2}\otimes 
\cO(1)_{\IP^2}$ over $\IP^2$ is described torically by adding the point 
$\nu_5^*=(0,-1,0,-4,-6)$ to~\cpoly . For this case we have the 
Mori generators:
\eqn\mcIII{\eqalign{
l^{(1)}&=(\phm  0;1, 0, 1,-1, 1, 0, 0, -2),\cr 
l^{(2)}&=(\phm  0;0, 1, 0, \phm 1, 0, 0, 0, -2),\cr 
l^{(3)}&=(      -6;0, 0, 0, \phm 0, 0, 2, 3,  \phm 1),}}
The associated K\"ahler classes control the volume of the $\IP^2$, 
the volume of the $\IP^1$ fibre and the volume of the elliptic fibre.  
The  Picard-Fuchs equations are :
\eqn\pfIII{\eqalign{
\cL_1 &= -\theta_1^3 - (-1 + \theta_1 - \theta_2)
(-2 + 2\theta_1 + 2\theta_2 - \theta_3)
    (-1 + 2\theta_1 + 2\theta_2 - \theta_3)z_1\cr
\cL_2 &= \theta_2(-\theta_1 + \theta_2) - 
(-2 + 2\theta_1 + 2\theta_2 - \theta_3)
    (-1 + 2\theta_1 + 2\theta_2 - \theta_3)z_2\cr
\cL_3 &= \theta_3(-2\theta_1 - 2\theta_2 + \theta_3) - 
12(-5 + 6\theta_3)(-1 + 6\theta_3)z_3}}
 
\noindent
The classical couplings 
\eqn\classIII{\eqalign{\cC_0&=
J_3J_1^2J_2+J_3J_2^2J_1+J_3J_2^3+2J_3^2J_1^2+4J_2J_1J_3^2\cr &
+4J_3^2J_2^2+12J_3^3J_1+16J_3^3J_2+56J_3^4\cr
\cC_2&=24J_1^2+48J_1J_2+138J_1J_3+48J_2^2+182J_2J_3+640J_3^2\cr
\cC_3&=-720J_1-960J_2-3378J_3.}}
show that there is also a $K_3$ fibration over the $\IP^2$. 
\noindent
Basis of $H^{2,2}$:  
\eqn\sbasII{\eqalign{
b_1^{(2)} = J_1^2, \ \ \ b_2^{(2)} = J_1J_2+J_2^2, \ \ \
b_3^{(2)} = J_1J_3 + 2J_3^2, \ \ \ b_4^{(2)} = J_2J_3 + 2J_3^2,}}
with 
$$\eta_{2,2}=\left(\matrix{0& 0& 4& 5\cr 0& 0& 18& 18\cr
             4& 18& 274& 284&\cr 5& 18& 284& 292}\right)$$

\noindent
The following invariants are read off from the normalized
threepoint functions  

\noindent $b_1^{(2)}=J_1^2$, $C^{(2,1,1)}_{1,i,j}$:
$$
\fivepoint{\vbox{\offinterlineskip\tabskip=0pt
\halign{\strutf
\vrule#&
{}~\hfil$#$~&
\vrule#&
{}~\hfil$#$~&
{}~\hfil$#$~&
{}~\hfil$#$~&
{}~\hfil$#$~&
{}~\hfil$#$~&
{}~\hfil$#$~&
{}~\hfil$#$~&
{}~\hfil$#$~&
{}~\hfil$#$~&
\vrule$#$\cr
\noalign{\hrule}
& m &&
n^{(1)}_{m,0,0} &  n^{(1)}_{m,0,1}&n^{(1)}_{m,0,2}&
n^{(1)}_{m,1,0} &  n^{(1)}_{m,1,1}&n^{(1)}_{m,1,2}&
n^{(1)}1_{m,2,0} &  n^{(1)}_{m,2,1}&n^{(1)}_{m,2,2}& \cr
\noalign{\hrule}
&0 && 0&0&0&0&0&0&0&0&0 &\cr
&1 && -1&240&141444&-14&5040&-1096200&51&22800&-5263920&\cr
&2 &&1  &-240&28200&-6&2640&-703800& -616&356160&-110457000&\cr
\noalign{\hrule}
}
\hrule}}$$

\noindent $b_2^{(2)}=J_1 J_2+J_2^2$, ${1 \over 2 } C^{(2,1,1)}_{(2,i,j)}$:
$$
\fivepoint{\vbox{\offinterlineskip\tabskip=0pt
\halign{\strutf
\vrule#&
{}~\hfil$#$~&
\vrule#&
{}~\hfil$#$~&
{}~\hfil$#$~&
{}~\hfil$#$~&
{}~\hfil$#$~&
{}~\hfil$#$~&
{}~\hfil$#$~&
{}~\hfil$#$~&
{}~\hfil$#$~&
{}~\hfil$#$~&
\vrule$#$\cr
\noalign{\hrule}
& m &&
n^{(2)}_{m,0,0} &  n^{(2)}_{m,0,1}&n^{(2)}_{m,0,2}&
n^{(2)}_{m,1,0} &  n^{(2)}_{m,1,1}&n^{(2)}_{m,1,2}&
n^{(2)}_{m,2,0} &  n^{(2)}_{m,2,1}&n^{(2)}_{m,2,2}& \cr
\noalign{\hrule}
&0 && 0&0&0&-1& 720&424332&0&0&1440&\cr
&1 && 0&0&0&-20& 7680&-1716840&-138&-62400&-15292440&\cr
&2 && 0&0&0& 0& 0& 0&-820& 491520& -155976240&\cr
\noalign{\hrule}
}
\hrule}}$$

\noindent $b_3^{(2)}=J_1 J_3+2 J_3^2$, ${1\over 24} C^{(2,1,1)}_{3,i,j}$: 
$$
\fivepoint{\vbox{\offinterlineskip\tabskip=0pt
\halign{\strutf
\vrule#&
{}~\hfil$#$~&
\vrule#&
{}~\hfil$#$~&
{}~\hfil$#$~&
{}~\hfil$#$~&
{}~\hfil$#$~&
{}~\hfil$#$~&
{}~\hfil$#$~&
{}~\hfil$#$~&
{}~\hfil$#$~&
{}~\hfil$#$~&
\vrule$#$\cr
\noalign{\hrule}
& m &&
n^{(3)}_{m,0,0} &  n^{(3)}_{m,0,1}&n^{(3)}_{m,0,2}&
n^{(3)}_{m,1,0} &  n^{(3)}_{m,1,1}&n^{(3)}_{m,1,2}&
n^{(3)}_{m,2,0} &  n^3_{m,2,1}&n^{(3)}_{m,2,2}& \cr
\noalign{\hrule}
&0&&0&310&620&0&310&501273&0&0&620&\cr
&1&& 0&0&64710&0&1860&-586830&0& 9300&-3818580&\cr
&2&&0&0&0&0&0&0&0&58590&-31852500&\cr
\noalign{\hrule}
}
\hrule}}$$

\noindent $b_4^{(2)}=J_2 J_3+2 J_3^2$, ${1\over 24} C^{(2,1,1)}_{4,i,j}$:
$$
\fivepoint{\vbox{\offinterlineskip\tabskip=0pt
\halign{\strutf
\vrule#&
{}~\hfil$#$~&
\vrule#&
{}~\hfil$#$~&
{}~\hfil$#$~&
{}~\hfil$#$~&
{}~\hfil$#$~&
{}~\hfil$#$~&
{}~\hfil$#$~&
{}~\hfil$#$~&
{}~\hfil$#$~&
{}~\hfil$#$~&
\vrule$#$\cr
\noalign{\hrule}
& m &&
n^{(4)}_{m,0,0} &  n^{(4)}_{m,0,1}&n^{(4)}_{m,0,2}&
n^{(4)}_{m,1,0} &  n^{(4)}_{m,1,1}&n^{(4)}_{m,1,2}&
n^{(4)}_{m,2,0} &  n^{(4)}_{m,2,1}&n^{(4)}_{m,2,2}& \cr
\noalign{\hrule}
&0&& 0&130&260&0&130&235266&0&0&260&\cr
&1&& 0&-260&69030&0&-2080&761670&0&-7020&3118050&\cr
&2&&0&320&640&0&320&547029&0&0&640&\cr
\noalign{\hrule}}}}$$

Let us finally discuss the transition between the first two models
in table (6.5). The four parameter model has as polyhedron the convex
hull of 
\eqn\polytrans{\eqalign{
\nu_1^*&=(-1,0,0,2,3),\,\ 
\nu_2^*=(0,-1,0,2,3),\,\
\nu_3^*=(0,0,0,0,-1),\ 
\nu_4^*=(0,0,0,-1,0) \cr
\nu_5^*&=(0,0,0,2,3),\,\ 
\nu_6^*=(0,0,1,2,3),\,\
\nu_7^*=(1,1,3,2,3),\,\ 
\nu_8^*=(0,0,-1,2,3), \cr
\nu_9^*&=(0,0,-1,1,2)}}

\eqn\mcII{\eqalign{
l^{(1)}&=(-2      ;0, 0, 1, 0,\phm  1, \phm 0, 0, -2, \phm 2),\cr 
l^{(2)}&=(\phm   0;1, 1, 0, 0,\phm  0, -3, 1,  \phm   0, \phm 0),\cr 
l^{(3)}&=(\phm   0;0, 0, 0, 0, -2,  1, \phm 0, \phm 1,\phm  0),\cr 
l^{(4)}&=(-2      ;0, 0, 1, 1,\phm  0, \phm  0,  0,\phm 1,-1).}}

\eqn\classtII{\eqalign{\cC_0&=
12J_2J_1^3+6J_2^2J_4^2+18J_1^2J_3^2+324J_1J_4^3+9J_1J_3^3+18J_4J_3^3+\cr &
54J_4^2J_3^2+162J_3J_4^3+72J_1^4+54J_2J_4^3+216J_1^2J_4^2+36J_3J_1^3+\cr &
144J_4J_1^3+2J_2^2J_1^2+6J_2J_4J_3^2+3J_2J_1J_3^2+36J_2J_1J_4^2+6J_2J_3J_1^2+
\cr & 24J_2J_4J_1^2+18J_2J_3J_4^2+108J_1J_3J_4^2+2J_2^2J_3J_4+J_2^2J_3J_1+
4J_2^2J_1J_4+\cr &
36J_1J_4J_3^2+72J_4J_3J_1^2+486J_4^4+12J_2J_4J_3J_1\cr
\cC_2&= 216J_3^2+582J_3J_4+408J_3J_1+72J_3J_2+1746J_4^2
\cr & +1164J_4J_1+198J_4J_2+816J_1^2+138J_1J_2+24J_2^2\cr
\cC_3&=-1674J_3-5076J_4-3366J_1-558J_2.}}

The transition to the three parameter model is described by the
omission of the point $\nu^*_9$ from the polyhedron \polytrans . 
The Mori generators of the three parameter model 
are $l^{(1')}=2l^{(4)}+l^{(1)}$, $l^{(2')}=l^{(2)}$ 
and $l^{(3')}=l^{(3)}$. We have adapted our notation to 
\bkkm , so that the indices of $x_i$ are shifted by one to make 
place for the additional coordinate of the $\IP^2$ (instead of 
$\IP^1$) at $x_1$. The elliptic fibre has  again type 
$(1,0,0,2)$. The conic bundle at $D_9=0$ is 
$x_3^2 f_8+ x^2_4 + x_8^2f_{20}+x_3 x_4 f_4+ x_3 x_8 f_{14} + 
x_4 x_8 f_10$ over $\IP^2$ with $x_1,x_2,x_6$ 
coordinates degenerates over a curve of genus $351$. The 
contraction of the conic bundle to a singular form of the
parameter model is given by the map 
$(x_1,\ldots, x_9)\mapsto (x_1,x_2,x_3 x_9,x_4 x_9,x_5,
x_6 x_7, x_8 x_9)$. 

The classical couplings of the three 
parameter model are essentially obtained by restricting~\classtII~to 
$J_4=0$, only $\cC_3$ changes to $\cC_3=-4338J_1-720J_2-2160J_3$.

\vfill
\eject
\noindent 
{{\bf Appendix A:} {\sl Kodaira's classification of elliptic fibre singularities.} }
\medskip

$${
\vbox{\offinterlineskip\tabskip=0pt
\halign{\strut
\vrule#&
~\hfil$#$~& 
\vrule$#$& 
~\hfil$#$~& 
\vrule$#$& 
~\hfil$#$~& 
\vrule$#$& 
~\hfil$#$~& 
\vrule$#$& 
~\hfil$#$~& 
\vrule$#$& 
~\hfil$#$~& 
\vrule#\cr
\noalign{\hrule}
&   {\rm ord}(f) 
&&  {\rm ord}(g)
&&  {\rm ord}(\Delta)
&&  {\rm fibre}
&&  {\rm singularity}
&&  a_i&\cr
\noalign{\hrule}
&\ge 0&&\ge 0&&0&&smooth&& none&&-&\cr
& 0   &&    0&&n&&I_n   && A_{n-1}&&{n\over 12}&\cr
&\ge 1&&    1&&2&&II    && none   &&{1\over 6}&\cr
&\ge 1&&\ge 2&&3&&III   && A_1    &&{1\over 4}&\cr
&\ge 2&&    2&&4&&IV    && A_2    &&{1\over 3}&\cr
&    2&&\ge 3&&n+6&&I^*_n&&D_{n+4}&&{1\over 2}+{n\over 12}&\cr
&\ge 2&&    3&&n+6&&I^*_n&&D_{n+4}&&{1\over 2}+{n\over 12}&\cr
&\ge 3&&    4&&8&&IV^*   && E_6   &&{5\over 6}&\cr
&    3&&\ge 5&&9&&III^*   && E_7&&{3\over 4}&\cr
&\ge 4&&    5&&10&&II^*    && E_8   &&{2\over 3}&\cr
\noalign{\hrule}}
\hrule}}$$ 
\noindent {\bf Tab. 1} {\sl Classification of the singular fibres 
occurring in an non-singular elliptic surface with section
\ref\kodaira{K.\ Kodaira, Annals of Math. {\bf 77} (1963) 563; 
{\bf 78} (1963) 1}\mvI. The last entry is the Euler number of the 
singular fibre  divided by $12$. For ${\rm ord}(\Delta>10)$ 
there exist no resolution with trivial canonical bundle.}
\medskip
\vfill
\eject

\noindent {{\bf Appendix B:} {\sl Tables of Calabi-Yau manifolds} }
$$\eqalign{ & {\rm{\bf Table\ B.1 \ 
CY-Fourfolds\ with\ negative\ Euler\ number }}\cr 
&\phantom{\rm{\bf Table\ B.1\ }} ^* {\rm \ indicates\ that \ no\ 
reflexive\ polyhedron\ exists.}\cr
& \fivepoint{\vbox{\offinterlineskip\tabskip=0pt
\halign{\strutf
\vrule#&
~\hfil$#$~~~&
\vrule$#$& 
~\hfil$#$&  
~\hfil$#$&  
~\hfil$#$&  
~\hfil$#$&  
~\hfil$#$~& 
\vrule$#$& 
~\hfil$#$~& 
~\hfil$#$~& 
~\hfil$#$~& 
~\hfil$#$~& 
~\hfil$#$~& 
~\hfil$#$~& 
\vrule$#$&
~\hfil$#$~& 
\vrule#\cr
\noalign{\hrule}
\tabsp
&N^o &&\chi & h_{11} &h_{21}&h_{22}&h_{31}
&&w_1&w_2&w_3&w_4&w_5&w_6&&m&\cr
\tabsp 
\noalign{\hrule} 
\tabsp
&   1&&-240&  54& 228& 308& 126&&   9&  9& 70& 72& 80&120&&  360&\cr &   
   2&&-198&  22& 272& 424& 209&&   5&  5&  5& 18& 24& 33&&   90&\cr & 
   3&&-198&  30& 101&  82&  30&&  21& 24& 25& 25& 25& 30&&  150&\cr & 
   4&&-192&  24& 147& 178&  83&&   9&  9& 18& 28& 32& 48&&  144&\cr & 
   5&&-192&  47& 195& 274& 108&&   7& 13& 27& 27& 27& 88&&  189&\cr & 
   6&&-192&  47& 195& 274& 108&&  13& 14& 54& 54& 54&189&&  378&\cr & 
   7&&-168&  81& 144& 188&  27&&  21& 24& 52& 97& 97& 97&&  388&\cr & 
   8&&-144&  22& 165& 246& 111&&   7&  7&  7& 12& 18& 33&&   84&\cr & 
   9&&-144&  23& 242& 400& 187&&   5&  5& 10& 24& 32& 44&&  120&\cr & 
  10&&-144&  24& 141& 198&  85&&   9& 11& 11& 11& 15& 42&&   99&\cr & 
  11&&-144&  24& 141& 198&  85&&  15& 18& 22& 22& 22& 99&&  198&\cr & 
  12&&-144&  28&  91&  98&  31&&  15& 18& 19& 19& 19& 24&&  114&\cr & 
  13&&-144&  29& 147& 210&  86&&   8& 14& 15& 15& 15& 53&&  120&\cr & 
  14&&-144&  29& 102& 120&  41&&  12& 15& 22& 22& 22& 39&&  132&\cr & 
  15&&-144&  31&  91&  98&  28&&  25& 25& 28& 32& 40& 50&&  200&\cr & 
  16&&-144&  33& 273& 462& 208&&   5&  5& 20& 36& 48& 66&&  180&\cr & 
  17&&-144&  41& 102& 120&  29&&  25& 25& 42& 48& 60&100&&  300&\cr & 
  18&&-144&  44& 322& 560& 246&&   5&  5& 30& 48& 64& 88&&  240&\cr & 
  19&&-138&  22& 144& 208&  91&&   9&  9&  9& 10& 16& 37&&   90&\cr & 
  20&&-138&  76& 462& 844& 355&&   5&  5& 55& 78&104&143&&  390&\cr & 
  21&&-120&  26& 135& 202&  81&&   9&  9&  9& 19& 23& 30&&   99&\cr & 
  22&&-120&  26& 135& 202&  81&&  12& 12& 12& 23& 33& 40&&  132&\cr & 
  23&&-120&  58& 198& 328& 112&&   9&  9& 61& 63& 71&102&&  315&\cr & 
  24&&-120&  81& 135& 202&  26&&  30& 33& 37&100&100&100&&  400&\cr & 
  25&& -96&  23&  90& 128&  43&&  11& 13& 13& 13& 16& 25&&   91&\cr & 
  26&& -96&  23&  90& 128&  43&&  11& 15& 15& 15& 21& 28&&  105&\cr & 
  27&& -96&  23& 147& 242& 100&&   7&  7& 14& 16& 24& 44&&  112&\cr & 
  28&& -96&  24& 240& 428& 192&&   5&  5&  5& 16& 23& 31&&   85&\cr & 
  29&& -96&  24& 240& 428& 192&&   6&  6&  6& 17& 32& 35&&  102&\cr & 
  30&& -96&  24& 240& 428& 192&&  10& 10& 10& 23& 32& 85&&  170&\cr & 
  31&& -96&  25& 126& 200&  77&&  11& 11& 12& 20& 22& 56&&  132&\cr & 
  32\tm&& -96&  27&  90& 128&  39&&  20& 20& 20& 21& 24& 35&&  140&\cr & 
  33&& -96&  29&  82& 112&  29&&  19& 19& 20& 24& 32& 38&&  152&\cr & 
  34&& -96&  30& 132& 212&  78&&  11& 11& 15& 25& 33& 70&&  165&\cr & 
  35&& -96&  30&  92& 132&  38&&  20& 24& 33& 33& 66& 88&&  264&\cr & 
  36&& -96&  30& 132& 212&  78&&  22& 22& 25& 30& 66&165&&  330&\cr & 
  37&& -96&  32& 166& 280& 110&&   7&  7& 24& 28& 36& 66&&  168&\cr & 
  38&& -96&  32& 146& 240&  90&&   9&  9& 20& 32& 36& 74&&  180&\cr & 
  39&& -96&  36& 144& 236&  84&&  10& 18& 19& 19& 38& 86&&  190&\cr & 
  40&& -96&  38&  92& 132&  30&&  19& 19& 30& 36& 48& 76&&  228&\cr & 
  41&& -96&  39&  90& 128&  27&&  24& 25& 35& 42& 42& 42&&  210&\cr & 
  42&& -96&  42& 150& 248&  84&&  15& 15& 16& 28& 60&106&&  240&\cr & 
  43&& -96&  43&  90& 128&  23&&  31& 35& 36& 36& 42& 72&&  252&\cr & 
  44&& -96&  45& 180& 308& 111&&   6& 13& 25& 25& 25& 81&&  175&\cr & 
  45&& -96&  45& 180& 308& 111&&   7& 12& 26& 26& 26& 85&&  182&\cr & 
  46&& -96&  45& 168& 284&  99&&  11& 11& 24& 40& 66&112&&  264&\cr & 
  47&& -96&  45& 180& 308& 111&&  12& 13& 50& 50& 50&175&&  350&\cr & 
  48&& -96&  50& 165& 278&  91&&  12& 22& 23& 23& 69&127&&  276&\cr & 
  49&& -96&  52& 228& 404& 152&&   7&  7& 40& 56& 60&110&&  280&\cr & 
  50\tm&& -96&  59& 150& 248&  67&&   8& 15& 54& 54& 54& 85&&  270&\cr & 
  51&& -96&  61& 228& 404& 143&&   9&  9& 40& 64& 90&148&&  360&\cr & 
  52&& -96&  68& 204& 356& 112&&   9& 17& 35& 35& 70&149&&  315&\cr & 
  53&& -96&  78& 450& 848& 348&&   5&  5& 55& 76&103&141&&  385&\cr & 
  54&& -90&  26&  81& 114&  32&&  12& 17& 17& 17& 18& 21&&  102&\cr 
\tabsp
\noalign{\hrule}}
\hrule}
\quad
\vbox{\offinterlineskip\tabskip=0pt
\halign{\strutf
\vrule#&
~\hfil$#$~~~&
\vrule$#$& 
~\hfil$#$&  
~\hfil$#$&  
~\hfil$#$&  
~\hfil$#$&  
~\hfil$#$~& 
\vrule$#$& 
~\hfil$#$~& 
~\hfil$#$~& 
~\hfil$#$~& 
~\hfil$#$~& 
~\hfil$#$~& 
~\hfil$#$~& 
\vrule$#$& 
~\hfil$#$~& 
\vrule#\cr
\noalign{\hrule}
\tabsp
&N^o &&\chi & h_{11} &h_{21}&h_{22}&h_{31}
&&w_1&w_2&w_3&w_4&w_5&w_6&&m&\cr 
\tabsp
\noalign{\hrule} 
\tabsp
&  55&& -90&  27&  92& 136&  42&&   9& 15& 19& 19& 19& 33&&  114&\cr & 
  56&& -84&  30& 126& 208&  74&&   9&  9& 18& 26& 31& 42&&  135&\cr & 
  57&& -84&  30& 126& 208&  74&&  12& 12& 24& 31& 45& 56&&  180&\cr & 
  58\tm&& -72&  27&  84& 132&  37&&  16& 16& 18& 21& 32& 41&&  144&\cr & 
  59\tm&& -72&  34&  90& 144&  36&&  16& 16& 31& 37& 44& 48&&  192&\cr & 
  60\tm&& -72&  34&  90& 144&  36&&  20& 20& 37& 48& 55& 60&&  240&\cr & 
  61\tm&& -72&  36&  90& 144&  34&&  12& 17& 33& 33& 33& 37&&  165&\cr & 
  62\tm&& -72&  36&  90& 144&  34&&  15& 16& 36& 36& 36& 41&&  180&\cr & 
  63\tm&& -72&  36&  90& 144&  34&&  16& 17& 44& 44& 44& 55&&  220&\cr & 
  64\tm&& -72&  37&  84& 132&  27&&  22& 25& 25& 29& 49& 75&&  225&\cr & 
  65\tm&& -72&  37&  84& 132&  27&&  27& 29& 32& 32& 72& 96&&  288&\cr & 
  66&& -72&  56& 144& 252&  68&&   7& 15& 51& 51& 51& 80&&  255&\cr & 
  67&& -72&  68& 144& 252&  56&&  16& 16& 65& 75&100&128&&  400&\cr & 
  68&& -72&  78& 135& 234&  37&&  17& 18& 61& 96& 96& 96&&  384&\cr & 
  69&& -66&  27& 220& 408& 174&&   5&  5& 10& 22& 31& 42&&  115&\cr &
  70&& -66&  27& 220& 408& 174&&   6&  6& 12& 23& 44& 47&&  138&\cr & 
  71&& -66&  27& 220& 408& 174&&  10& 10& 20& 31& 44&115&&  230&\cr & 
  72&& -60&  50&  96& 164&  28&&  20& 24& 49& 49& 54& 98&&  294&\cr & 
  73&& -48&  27&  73& 126&  30&&  16& 17& 17& 24& 28& 34&&  136&\cr & 
  74&& -48&  28&  83& 146&  39&&  12& 19& 19& 20& 38& 44&&  152&\cr & 
  75&& -48&  28&  83& 146&  39&&  16& 20& 27& 27& 54& 72&&  216&\cr & 
  76&& -48&  35&  82& 144&  31&&  17& 17& 24& 36& 42& 68&&  204&\cr & 
  77&& -48&  37&  94& 168&  41&&  18& 19& 19& 30& 66& 76&&  228&\cr & 
  78&& -48&  39&  83& 146&  28&&  19& 19& 25& 30& 40& 57&&  190&\cr & 
  79&& -48&  41&  94& 168&  37&&  20& 22& 22& 25& 65& 66&&  220&\cr & 
  80&& -48&  41&  94& 168&  37&&  25& 30& 33& 33& 99&110&&  330&\cr &  
  81\tm&& -48&  45& 114& 208&  53&&  13& 13& 33& 48& 75& 91&&  273&\cr & 
  82&& -48&  46& 153& 286&  91&&  11& 11& 21& 35& 55& 98&&  231&\cr & 
  83&& -48&  46& 280& 540& 218&&   6&  6& 30& 41& 80& 83&&  246&\cr & 
  84\tm&& -48&  53& 210& 400& 141&&   7&  7& 36& 49& 54& 99&&  252&\cr & 
  85\tm&& -48&  53& 114& 208&  45&&  14& 19& 40& 40& 80& 87&&  280&\cr & 
  86&& -48&  61& 120& 220&  43&&  14& 22& 61& 61& 86&122&&  366&\cr & 
  87&& -48&  63& 108& 196&  29&&  12& 28& 33& 73& 73& 73&&  292&\cr & 
  88&& -36&  24&  80& 148&  42&&  11& 11& 16& 20& 22& 30&&  110&\cr & 
  89&& -36&  25&  82& 152&  43&&   9& 12& 17& 17& 17& 30&&  102&\cr & 
  90&& -36&  25& 182& 352& 143&&   5& 10& 15& 36& 48& 66&&  180&\cr & 
  91\tm&& -36&  26& 120& 228&  80&&  10& 10& 13& 20& 24& 53&&  130&\cr & 
  92&& -36&  33&  71& 130&  24&&  25& 42& 48& 50& 60& 75&&  300&\cr & 
  93&& -36&  62& 108& 204&  32&&  12& 21& 40& 73& 73& 73&&  292&\cr & 
  94\tm&& -30&  31& 120& 232&  76&&  10& 13& 13& 19& 26& 62&&  143&\cr & 
  95\tm&& -30&  31& 120& 232&  76&&  11& 14& 14& 20& 28& 67&&  154&\cr & 
  96\tm&& -30&  31& 120& 232&  76&&  19& 20& 26& 26& 52&143&&  286&\cr & 
  97&& -24&  30&  78& 152&  36&&  10& 17& 36& 36& 36& 45&&  180&\cr & 
  98&& -24&  30&  72& 140&  30&&  16& 32& 39& 45& 48& 60&&  240&\cr & 
  99&& -24&  39& 117& 230&  66&&   9& 18& 45& 56& 64& 96&&  288&\cr & 
 100&& -24&  66& 192& 380& 114&&   8& 17& 33& 33& 66&140&&  297&\cr & 
 101&& -24&  66& 192& 380& 114&&   9& 16& 34& 34& 68&145&&  306&\cr & 
 102&& -24&  66& 117& 230&  39&&  12& 17& 52& 81& 81& 81&&  324&\cr & 
 103&& -24&  69& 108& 212&  27&&  24& 28& 39& 91& 91& 91&&  364&\cr & 
 104&& -12&  33& 108& 220&  65&&   9& 18& 18& 35& 40& 60&&  180&\cr & 
 105&& -12&  65& 108& 220&  33&&  18& 29& 34& 81& 81& 81&&  324&\cr & 
 106&&  -6&  36&  74& 156&  29&&  17& 17& 20& 30& 35& 51&&  170&\cr & 
 107&&  -6&  38&  85& 178&  38&&  15& 19& 19& 25& 55& 57&&  190&\cr & 
 108&&  -6&  38&  85& 178&  38&&  20& 25& 27& 27& 81& 90&&  270&\cr  
\tabsp
\noalign{\hrule}}
\hrule}}}
$$

$$\eqalign{ & {\rm{\bf Table\ B.2 \ 
CY-Fourfolds\ with\ vanishing\ Euler\ number }}\cr
&\phantom{\rm{\bf Table\ A.1\ }} ^* {\rm \ indicates\ that \ no\ 
reflexive\ polyhedron\ exists.}\cr
& \fivepoint{\vbox{\baselineskip=6pt\offinterlineskip\tabskip=0pt
\halign{\strutf
\vrule#&
~\hfil$#$~~~&
\vrule$#$& 
~\hfil$#$&  
~\hfil$#$&  
~\hfil$#$&  
~\hfil$#$&  
~\hfil$#$~& 
\vrule$#$& 
~\hfil$#$~& 
~\hfil$#$~& 
~\hfil$#$~& 
~\hfil$#$~& 
~\hfil$#$~& 
~\hfil$#$~& 
\vrule$#$&
~\hfil$#$~& 
\vrule#\cr
\noalign{\hrule}
\tabsp
&N^o &&\chi & h_{11} &h_{21}&h_{22}&h_{31}
&&w_1&w_2&w_3&w_4&w_5&w_6&&d&\cr
\tabsp 
\noalign{\hrule} 
\tabsp
& 109&&   0&  21&  75& 162&  46&&   9& 11& 11& 11& 14& 21&&   77&\cr & 
 110&&   0&  23&  84& 180&  53&&  12& 12& 12& 14& 21& 25&&   96&\cr & 
 111&&   0&  24& 126& 264&  94&&   7&  7& 13& 14& 23& 41&&  105&\cr & 
 112\tm&&   0&  24&  80& 172&  48&&  10& 10& 18& 20& 23& 29&&  110&\cr & 
 113&&   0&  24& 126& 264&  94&&   8&  8& 13& 16& 30& 45&&  120&\cr & 
 114&&   0&  24& 126& 264&  94&&   8&  8& 15& 16& 26& 47&&  120&\cr & 
 115\tm&&   0&  24&  80& 172&  48&&  12& 12& 22& 24& 27& 35&&  132&\cr & 
 116\tm&&   0&  24&  80& 172&  48&&  15& 15& 23& 27& 30& 55&&  165&\cr & 
 117&&   0&  24& 126& 264&  94&&  14& 14& 23& 26& 28&105&&  210&\cr & 
 118\tm&&   0&  25&  72& 156&  39&&  12& 13& 13& 21& 26& 32&&  117&\cr & 
 119&&   0&  25& 111& 234&  78&&   7& 14& 21& 24& 36& 66&&  168&\cr & 
 120&&   0&  26&  74& 160&  40&&  12& 16& 17& 17& 34& 40&&  136&\cr &
 121&&   0&  27&  84& 180&  49&&  15& 15& 15& 16& 24& 35&&  120&\cr & 
 122&&   0&  27&  96& 204&  61&&  11& 18& 22& 30& 33& 84&&  198&\cr & 
 123\tm&&   0&  30&  72& 156&  34&&  16& 16& 23& 28& 29& 32&&  144&\cr & 
 124\tm&&   0&  30&  72& 156&  34&&  20& 20& 27& 28& 40& 45&&  180&\cr 
\tabsp
\noalign{\hrule}}
\hrule}
\quad
\vbox{\offinterlineskip\tabskip=0pt
\halign{\strutf
\vrule#&
~\hfil$#$~~~&
\vrule$#$& 
~\hfil$#$&  
~\hfil$#$&  
~\hfil$#$&  
~\hfil$#$&  
~\hfil$#$~& 
\vrule$#$& 
~\hfil$#$~& 
~\hfil$#$~& 
~\hfil$#$~& 
~\hfil$#$~& 
~\hfil$#$~& 
~\hfil$#$~& 
\vrule$#$& 
~\hfil$#$~& 
\vrule#\cr
\noalign{\hrule}
\tabsp
&N^o &&\chi & h_{11} &h_{21}&h_{22}&h_{31}
&&w_1&w_2&w_3&w_4&w_5&w_6&&d&\cr 
\tabsp
\noalign{\hrule} 
\tabsp
& 125\tm&&   0&  30&  72& 156&  34&&  20& 20& 29& 35& 36& 40&&  180&\cr & 
 126&&   0&  31&  64& 140&  25&&  19& 30& 36& 38& 48& 57&&  228&\cr & 
 127&&   0&  33& 108& 228&  67&&  12& 12& 24& 29& 39& 52&&  168&\cr & 
 128&&   0&  34&  84& 180&  42&&  17& 17& 18& 24& 60& 68&&  204&\cr & 
 129&&   0&  37& 108& 228&  63&&   9& 18& 36& 49& 56& 84&&  252&\cr & 
 130&&   0&  40& 120& 252&  72&&  12& 12& 36& 37& 51& 68&&  216&\cr & 
 131&&   0&  41& 108& 228&  59&&  12& 12& 38& 57& 60& 61&&  240&\cr & 
 132&&   0&  49&  84& 180&  27&&  28& 32& 38& 49& 49& 98&&  294&\cr & 
 133&&   0&  53&  84& 180&  23&&  35& 37& 40& 56& 56&112&&  336&\cr & 
 134&&   0&  56& 108& 228&  44&&  14& 18& 55& 55& 78&110&&  330&\cr & 
 135&&   0&  62& 168& 348&  98&&  12& 12& 61& 72& 87&116&&  360&\cr & 
 136&&   0&  63& 108& 228&  37&&  12& 25& 41& 78& 78& 78&&  312&\cr & 
 137&&   0&  63& 108& 228&  37&&  14& 25& 52& 91& 91& 91&&  364&\cr & 
 138&&   0&  64& 114& 240&  42&&   9& 20& 48& 77& 77& 77&&  308&\cr 
\tabspI
\tabsp
\noalign{\hrule}}
\hrule}}}
$$

$$\eqalign{
& {\rm{\bf Table\ B.3\  \ Elliptic \  fibred \ K3 }}\cr 
& \fivepoint{\vbox{\offinterlineskip\tabskip=0pt
\halign{\strut
\vrule#&
\hfil$\, # \, $&
\vrule$\, # \, $& 
\hfil$\, # \, $&
\hfil$\, # \, $&
\hfil$\, # \, $&
\hfil$\, # \, $&
\vrule$\, # \, $& 
\hfil$\, # \, $\hfil&
\vrule$\, # \, $&
\hfil$\, # \, $&
\vrule#\cr
\noalign{\hrule}
&d &&w_1&w_2&w_3&w_4&&\hfil P \hfil &&{\cal E}&\cr
\noalign{\hrule}
&   6&&    1&    1&    2&    2&&
x_1^{ 6}
+x_2^{ 6}
+x_3^{ 3}
+x_4^{ 3}
 &&
 E_6
 &\cr
&   9&&    1&    2&    3&    3&&
x_1^{ 9}
+x_2^{ 4}x_1
+x_3^{ 3}
+x_4^{ 3}
 &&
 E_6
 &\cr
&  12&&    1&    3&    4&    4&&
x_1^{12}
+x_2^{ 4}
+x_3^{ 3}
+x_4^{ 3}
 &&
 E_6
 &\cr
&  15&&    2&    3&    5&    5&&
x_1^{ 6}x_2
+x_2^{ 5}
+x_3^{ 3}
+x_4^{ 3}
 &&
 E_6
 &\cr
\noalign{\hrule}
&   8&&    1&    1&    2&    4&&
x_1^{ 8}
+x_2^{ 8}
+x_3^{ 4}
+x_4^{ 2}
 &&
 E_7
 &\cr
&  12&&    1&    2&    3&    6&&
x_1^{12}
+x_2^{ 6}
+x_3^{ 4}
+x_4^{ 2}
 &&
 E_8
 E_7
 &\cr
&  16&&    1&    3&    4&    8&&
x_1^{16}
+x_2^{ 5}x_1
+x_3^{ 4}
+x_4^{ 2}
 &&
 E_7
 &\cr
&  20&&    2&    3&    5&   10&&
x_1^{10}
+x_2^{ 6}x_1
+x_3^{ 4}
+x_4^{ 2}
 &&
 E_7
 &\cr
&  20&&    1&    4&    5&   10&&
x_1^{20}
+x_2^{ 5}
+x_3^{ 4}
+x_4^{ 2}
 &&
 E_7
 &\cr
&  28&&    3&    4&    7&   14&&
x_1^{ 8}x_2
+x_2^{ 7}
+x_3^{ 4}
+x_4^{ 2}
 &&
 E_7
 &\cr
\noalign{\hrule}
&   9&&    1&    1&    3&    4&&
x_1^{ 9}
+x_2^{ 9}
+x_3^{ 3}
+x_4^{ 2}x_1
 &&
 E_8'
 &\cr
&  15&&    1&    2&    5&    7&&
x_1^{15}
+x_2^{ 4}x_4
+x_3^{ 3}
+x_4^{ 2}x_1
 &&
 E_8'
 &\cr
&  21&&    1&    3&    7&   10&&
x_1^{21}
+x_2^{ 7}
+x_3^{ 3}
+x_4^{ 2}x_1
 &&
 E_8'
 &\cr
\noalign{\hrule}
&  10&&    1&    1&    3&    5&&
x_1^{10}
+x_2^{10}
+x_3^{ 3}x_1
+x_4^{ 2}
 &&
 E_8''
 &\cr
&  16&&    1&    2&    5&    8&&
x_1^{16}
+x_2^{ 8}
+x_3^{ 3}x_1
+x_4^{ 2}
 &&
 E_8''
 &\cr
&  18&&    1&    3&    5&    9&&
x_1^{18}
+x_2^{ 6}
+x_3^{ 3}x_2
+x_4^{ 2}
 &&
 E_8''
 &\cr
&  22&&    1&    3&    7&   11&&
x_1^{22}
+x_2^{ 5}x_3
+x_3^{ 3}x_1
+x_4^{ 2}
 &&
 E_8''
 &\cr
&  28&&    1&    4&    9&   14&&
x_1^{28}
+x_2^{ 7}
+x_3^{ 3}x_1
+x_4^{ 2}
 &&
 E_8''
 &\cr
\noalign{\hrule}}
\hrule}
\quad
\vbox{\offinterlineskip\tabskip=0pt
\halign{\strut
\vrule#&
\hfil$\, # \, $&
\vrule$\, # \, $& 
\hfil$\, # \, $&
\hfil$\, # \, $&
\hfil$\, # \, $&
\hfil$\, # \, $&
\vrule$\, # \, $& 
\hfil$\, # \, $\hfil&
\vrule$\, # \, $&
\hfil$\, # \, $&
\vrule#\cr
\noalign{\hrule}
&d &&w_1&w_2&w_3&w_4&&\hfil P \hfil &&{\cal E}&\cr
\noalign{\hrule}
&  12&&    1&    1&    4&    6&&
x_1^{12}
+x_2^{12}
+x_3^{ 3}
+x_4^{ 2}
 &&
 E_8
 &\cr
&  18&&    2&    3&    4&    9&&
x_1^{ 9}
+x_2^{ 6}
+x_3^{ 4}x_1
+x_4^{ 2}
 &&
 E_8
 &\cr
&  18&&    1&    2&    6&    9&&
x_1^{18}
+x_2^{ 9}
+x_3^{ 3}
+x_4^{ 2}
 &&
 E_8
 &\cr
&  24&&    1&    3&    8&   12&&
x_1^{24}
+x_2^{ 8}
+x_3^{ 3}
+x_4^{ 2}
 &&
 E_8
 &\cr
&  30&&    4&    5&    6&   15&&
x_1^{ 6}x_3
+x_2^{ 6}
+x_3^{ 5}
+x_4^{ 2}
 &&
 E_8
 &\cr
&  30&&    1&    4&   10&   15&&
x_1^{30}
+x_2^{ 5}x_3
+x_3^{ 3}
+x_4^{ 2}
 &&
 E_8
 &\cr
&  36&&    1&    5&   12&   18&&
x_1^{36}
+x_2^{ 7}x_1
+x_3^{ 3}
+x_4^{ 2}
 &&
 E_8
&\cr
&  42&&    3&    4&   14&   21&&
x_1^{14}
+x_2^{ 7}x_3
+x_3^{ 3}
+x_4^{ 2}
 &&
 E_8
 &\cr
&  42&&    2&    5&   14&   21&&
x_1^{21}
+x_2^{ 8}x_1
+x_3^{ 3}
+x_4^{ 2}
 &&
 E_8
 &\cr
&  42&&    1&    6&   14&   21&&
x_1^{42}
+x_2^{ 7}
+x_3^{ 3}
+x_4^{ 2}
 &&
 E_8
  &\cr
&  48&&    3&    5&   16&   24&&
x_1^{16}
+x_2^{ 9}x_1
+x_3^{ 3}
+x_4^{ 2}
 &&
 E_8
 &\cr
&  54&&    4&    5&   18&   27&&
x_1^{ 9}x_3
+x_2^{10}x_1
+x_3^{ 3}
+x_4^{ 2}
 &&
 E_8
 &\cr
&  66&&    5&    6&   22&   33&&
x_1^{12}x_2
+x_2^{11}
+x_3^{ 3}
+x_4^{ 2}
 &&
 E_8
&\cr
\noalign{\hrule}}
\hrule}}}
$$ 

\vfill\eject

$$\eqalign{ & {\rm{\bf Table\ B.4 \ 
Elliptic\ fibred\ CY-Fourfolds\ with\ small\ Picard\ number }}\cr 
&\phantom{\rm{\bf Table\ A.1\ }} ^* {\rm \ indicates\ that \ \chi \ 
not\ divisible\ by\ 24.}\cr
& \fivepoint{\vbox{\offinterlineskip\tabskip=0pt
\halign{\strutf
\vrule#&
\hfil$\, # \, $~&
\hfil$\, # \, $~&
\vrule$\, # \, $&
\hfil$\, # \, $&  
\hfil$ #\,  $&  
\hfil$ #\,  $&  
\hfil$ #\,  $&  
\hfil$ #\,  $& 
\vrule$\, # \, $& 
\hfil$\, #  $& 
\hfil$ #  $& 
\hfil$ #  $& 
\hfil$ #  $& 
\hfil$ #  $& 
\hfil$ # \, $& 
\vrule$\, # \, $&
\hfil$\, # \, $& 
\vrule#\cr
\noalign{\hrule}
& N^o&F &&\chi & h_{11} &h_{21}&h_{22}&h_{31}
&&w_1&w_2&w_3&w_4&w_5&w_6&&d&\cr 
\noalign{\hrule} &  
1&{E8}&&^* 13362&     2&     0&  8920&  2217&&   1&  3&  4&  5& 26& 39&&   78&\cr &
2&{E8}&& 16176&     2&     0& 10796&  2686&&   1&  2&  3&  5& 22& 33&&   66&\cr &
2&{E8}&& 20832&     2&     0& 13900&  3462&&   1&  1&  2&  3& 14& 21&&   42&\cr &
4&{E8}&& 22776&     2&     0& 15196&  3786&&   1&  1&  1&  2& 10& 15&&   30&\cr &
5&{E8}&& 23328&     2&     0& 15564&  3878&&   1&  1&  1&  1&  8& 12&&   24&\cr &
6&{E8}&&  8424&     3&     0&  5628&  1393&&   2&  3&  4&  5& 28& 42&&   84&\cr &
7&{E7}&&^*  8484&     3&     0&  5668&  1403&&   1&  1&  2&  3&  7& 14&&   28&\cr &
8&{E7}&&^*  9276&     3&     0&  6196&  1535&&   1&  1&  1&  2&  5& 10&&   20&\cr &
9&{E7}&&  9504&     3&     0&  6348&  1573&&   1&  1&  1&  1&  4&  8&&   16&\cr &
10&{E{8'}}&& 10992&     3&     0&  7340&  1821&&   1&  2&  3&  5& 17& 28&&   56&\cr &
11&{E8}&&^* 12810&     3&     0&  8552&  2124&&   1&  4&  5&  7& 34& 51&&  102&\cr &
12&{E8}&& 14232&     3&     0&  9500&  2361&&   1&  3&  5&  7& 32& 48&&   96&\cr &
13&{E{8'}}&&^* 14484&     3&     0&  9668&  2403&&   1&  1&  2&  3& 11& 18&&   36&\cr &
14&{E8}&& 15240&     3&     0& 10172&  2529&&   1&  2&  3&  4& 20& 30&&   60&\cr &
15&{E8}&& 15624&     3&     0& 10428&  2593&&   1&  2&  2&  3& 16& 24&&   48&\cr &
16& {E{8'}}&&^* 16242&     3&     0& 10840&  2696&&   1&  1&  1&  2&  8& 13&&   26&\cr &
17&{E{8'}}&& 18528&     3&     0& 12364&  3077&&   1&  1&  2&  3& 13& 20&&   40&\cr &
18&{E{8'}}&&^* 18954&     3&     0& 12648&  3148&&   1&  1&  1&  1&  7& 11&&   22&\cr &
19&{E8}&& 19056&     3&     0& 12716&  3165&&   1&  2&  3&  7& 26& 39&&   78&\cr &
20&{E{8'}}&&^* 19308&     3&     0& 12884&  3207&&   1&  1&  1&  2&  9& 14&&   28&\cr &
21&{E8}  && 19728&     3&     0& 13164&  3277&&   1&  1&  2&  2& 12& 18&&   36&\cr &
22&{E8}  &&^* 22122&     3&     1& 14762&  3677&&   1&  1&  3&  4& 18& 27&&   54&\cr &
23&{E8} && 26208&     3&     1& 17486&  4358&&   1&  1&  1&  3& 12& 18&&   36&\cr &
24&{E6}  &&  4368&     4&     0&  2924&   716&&   1&  1&  2&  3&  7&  7&&   21&\cr &
25&{E8}  &&  4704&     4&     0&  3148&   772&&   4&  5&  6&  7& 44& 66&&  132&\cr &
26&{E6}  &&  4776&     4&     0&  3196&   784&&   1&  1&  1&  2&  5&  5&&   15&\cr &
27&{E6}  &&  4896&     4&     0&  3276&   804&&   1&  1&  1&  1&  4&  4&&   12&\cr &
28&{E{8'}} &&^*  6228&     4&     0&  4164&  1026&&   2&  3&  4&  5& 23& 37&&   74&\cr &
29&{E7} &&  6240&     4&     0&  4172&  1028&&   1&  2&  3&  4& 10& 20&&   40&\cr &
30&{E7}  &&  6408&     4&     0&  4284&  1056&&   1&  2&  2&  3&  8& 16&&   32&\cr &
31&{E{8'}}  &&^*  6708&     4&     0&  4484&  1106&&   2&  3&  5&  7& 29& 46&&   92&\cr &
32&{E7}  &&  8064&     4&     0&  5388&  1332&&   1&  1&  2&  2&  6& 12&&   24&\cr &
33&{E{8''}} &&  8640&     4&     0&  5772&  1428&&   1&  2&  3&  5& 17& 23&&   51&\cr &
34&{E8}  &&  8640&     4&     0&  5772&  1428&&   2&  4&  5&  6& 34& 51&&  102&\cr &
35&{E8}  &&  8856&     4&    12&  5940&  1476&&   2&  3&  3&  5& 26& 39&&   78&\cr &
36&{E8} &&^*  8916&     4&     1&  5958&  1475&&   2&  3&  5&  8& 36& 54&&  108&\cr &
37&{E7} &&  9000&     4&     1&  6014&  1489&&   1&  1&  3&  4&  9& 18&&   36&\cr &
38&{E8} && 10248&     4&     1&  6846&  1697&&   2&  3&  5& 11& 42& 63&&  126&\cr &
39&{E{8'}} && 10344&     4&     0&  6908&  1712&&   1&  3&  4&  5& 22& 35&&   70&\cr &
40&{E8}  && 10608&     4&     1&  7086&  1757&&   2&  2&  3&  5& 24& 36&&   72&\cr &
41&{E7}  && 10656&     4&     1&  7118&  1765&&   1&  1&  1&  3&  6& 12&&   24&\cr &
42&{E{8'}} &&^* 10884&     4&     0&  7268&  1802&&   1&  2&  3&  4& 16& 26&&   52&\cr &
43&{E{8'}} && 11376&     4&     0&  7596&  1884&&   1&  2&  2&  3& 13& 21&&   42&\cr &
44&{E{8''}} && 11568&     4&     0&  7724&  1916&&   1&  1&  2&  3& 11& 15&&   33&\cr &
45&{E8}  && 11568&     4&     0&  7724&  1916&&   2&  2&  3&  4& 22& 33&&   66&\cr &
46&{E{8'}}  &&^* 11850&     4&     0&  7912&  1963&&   1&  2&  3&  4& 17& 27&&   54&\cr &
47&{E{8'}}  && 12000&     4&     0&  8012&  1988&&   1&  2&  3&  7& 19& 32&&   64&\cr &
48&{E8}  && 12480&     4&     1&  8334&  2069&&   1&  4&  5&  6& 32& 48&&   96&\cr &
49&{E{8'}}  &&^* 12708&     4&     0&  8484&  2106&&   1&  2&  2&  3& 14& 22&&   44&\cr &
50&{E{8'}}  &&^* 12900&     4&     0&  8612&  2138&&   1&  2&  3&  5& 19& 30&&   60&\cr &
51&{E{8''}} && 13200&     4&     0&  8812&  2188&&   1&  1&  1&  2&  8& 11&&   24&\cr &
52&{E8}  &&^* 13434&     4&    10&  8988&  2237&&   1&  3&  3&  4& 22& 33&&   66&\cr &
53&{E{8'}} &&^* 14028&     4&     0&  9364&  2326&&   1&  2&  3&  4& 19& 29&&   58&\cr &
54&{E8}  && 14856&     4&     0&  9916&  2464&&   1&  2&  3&  3& 18& 27&&   54&\cr &
55&{E{8'}} && 14928&     4&     0&  9964&  2476&&   1&  1&  2&  2& 10& 16&&   32&\cr &
56&{E{8'}}  && 15168&     4&     0& 10124&  2516&&   1&  1&  3&  4& 14& 23&&   46&\cr &
57&{E8}  && 15792&     4&     1& 10542&  2621&&   1&  2&  4&  5& 24& 36&&   72&\cr &
58&{E{8''}}  && 16776&     4&     0& 11196&  2784&&   1&  1&  1&  1&  7& 10&&   21&\cr &
59&{E8}  && 16776&     4&     0& 11196&  2784&&   1&  2&  2&  2& 14& 21&&   42&\cr &
60&{E{8'}}  &&^* 17082&     4&     1& 11402&  2836&&   1&  1&  1&  3&  9& 15&&   30&\cr 
\noalign{\hrule}}
\hrule}
\ \ 
\vbox{\offinterlineskip\tabskip=0pt
\halign{\strutf
\vrule#&
\hfil$\, # \, $&
\hfil$\, # \, $&
\vrule$\, # \, $& 
\hfil$\, # \, $&  
\hfil$ #\,  $&  
\hfil$ #\,  $&  
\hfil$ #\,  $&  
\hfil$ #\,  $& 
\vrule$\, # \, $& 
\hfil$\, #  $& 
\hfil$ #  $& 
\hfil$ #  $& 
\hfil$ #  $& 
\hfil$ #  $& 
\hfil$ #  $& 
\vrule$\, # \, $& 
\hfil$\, # \, $& 
\vrule#\cr
\noalign{\hrule}
&N^o&F &&\chi & h_{11} &h_{21}&h_{22}&h_{31}
&&w_1&w_2&w_3&w_4&w_5&w_6&&d&\cr
\noalign{\hrule} &  
61&{E{8'}}  && 17184&     4&     0& 11468&  2852&&   1&  1&  2&  2& 11& 17&&   34&\cr &
62&{E{8''}}  && 17328&     4&     0& 11564&  2876&&   1&  1&  2&  3& 13& 19&&   39&\cr &
63&{E8}  && 17328&     4&     0& 11564&  2876&&   1&  2&  4&  6& 26& 39&&   78&\cr &
64&{E{8''}} && 17544&     4&     0& 11708&  2912&&   1&  1&  1&  2&  9& 13&&   27&\cr &
65&{E8}  && 17544&     4&     0& 11708&  2912&&   1&  2&  2&  4& 18& 27&&   54&\cr &
66&{E{8'}} && 20208&     4&     0& 13484&  3356&&   1&  1&  3&  4& 17& 26&&   52&\cr &
67&{E8}  && 20688&     4&     7& 13818&  3443&&   1&  1&  3&  3& 16& 24&&   48&\cr &
68&{E{8'}} &&^* 22854&     4&     0& 15248&  3797&&   1&  1&  1&  3& 11& 17&&   34&\cr &
69&{E8}  && 23328&     4&     1& 15566&  3877&&   1&  1&  2&  4& 16& 24&&   48&\cr &
70&{E8}  &&^* 24234&     4&     0& 16168&  4027&&   1&  1&  4&  5& 22& 33&&   66&\cr &
71&{E8}  && 24264&     4&     2& 16192&  4034&&   1&  1&  3&  5& 20& 30&&   60&\cr &
72&{E8}  &&^* 31194&     4&     0& 20808&  5187&&   1&  1&  1&  4& 14& 21&&   42&\cr &
73&{E6}  &&  3240&     5&     0&  2172&   527&&   1&  2&  3&  4& 10& 10&&   30&\cr &
74&{E6}  &&  3336&     5&     0&  2236&   543&&   1&  2&  2&  3&  8&  8&&   24&\cr &
75&{E6}  &&  3408&     5&     0&  2284&   555&&   1&  2&  3&  5& 11& 11&&   33&\cr &
76&{E7} &&^*  3516&     5&     0&  2356&   573&&   2&  3&  5&  7& 17& 34&&   68&\cr &
77&{E6} &&  4176&     5&     0&  2796&   683&&   1&  1&  2&  2&  6&  6&&   18&\cr &
78&{E8} &&^*  4938&     5&     0&  3304&   810&&   4&  5&  7& 13& 58& 87&&  174&\cr &
79&{E6} &&  5472&     5&     1&  3662&   900&&   1&  1&  1&  3&  6&  6&&   18&\cr &
80&{E7} &&  5976&     5&     1&  3998&   984&&   1&  3&  4&  7& 15& 30&&   60&\cr &
81&{E7} &&^*  6108&     5&     0&  4084&  1005&&   1&  2&  3&  3&  9& 18&&   36&\cr &
82&{E{8'}} &&  7416&     5&     0&  4956&  1223&&   2&  3&  5&  7& 31& 48&&   96&\cr &
83&{E{8'}} &&  7440&     5&     0&  4972&  1227&&   2&  2&  3&  5& 19& 31&&   62&\cr &
84&{E{8'}} &&  7512&     5&     0&  5020&  1239&&   2&  3&  4&  5& 26& 40&&   80&\cr &
85&{E7}  &&^*  7764&     5&     0&  5188&  1281&&   1&  2&  3&  7& 13& 26&&   52&\cr &
86&{E8}  &&  8328&     5&     7&  5578&  1382&&   2&  3&  5&  6& 32& 48&&   96&\cr &
87&{E8}  &&  8760&     5&     0&  5852&  1447&&   2&  3&  7&  8& 40& 60&&  120&\cr &
88&{E{8''}} &&  8856&     5&     0&  5916&  1463&&   1&  2&  3&  4& 16& 22&&   48&\cr &
89&{E{8''}} &&^*  8874&     5&     0&  5928&  1466&&   1&  3&  4&  5& 22& 31&&   66&\cr &
90&{E{8''}} &&  8928&     5&     0&  5964&  1475&&   1&  2&  3&  7& 19& 25&&   57&\cr &
91&{E8} &&  8928&     5&     0&  5964&  1475&&   2&  4&  6&  7& 38& 57&&  114&\cr &
92&{E7} &&  9504&     5&     1&  6350&  1572&&   1&  1&  2&  4&  8& 16&&   32&\cr &
93&{E7} &&  9864&     5&     2&  6592&  1633&&   1&  1&  3&  5& 10& 20&&   40&\cr &
94&{E{8'}} && 10032&     5&     0&  6700&  1659&&   1&  4&  5&  7& 29& 46&&   92&\cr &
95&{E8}  && 10176&     5&     9&  6814&  1692&&   2&  2&  3&  3& 20& 30&&   60&\cr &
96&{E8}  && 10464&     5&     0&  6988&  1731&&   2&  3&  7& 13& 50& 75&&  150&\cr &
97&{E{8'}} && 10704&     5&     9&  7166&  1780&&   1&  3&  3&  4& 19& 30&&   60&\cr &
98&{E7} &&^* 10788&     5&     0&  7204&  1785&&   1&  1&  2&  5&  9& 18&&   36&\cr &
99&{E{8'}}  &&^* 11094&     5&     0&  7408&  1836&&   1&  2&  4&  5& 19& 31&&   62&\cr &
100&{E{8''}} && 11256&     5&     0&  7516&  1863&&   1&  2&  2&  3& 14& 20&&   42&\cr &
101&{E{8'}} && 11256&     5&     0&  7516&  1863&&   1&  2&  3&  3& 15& 24&&   48&\cr &
102&{E{8''}} && 11280&     5&     0&  7532&  1867&&   1&  2&  3&  5& 19& 27&&   57&\cr &
103&{E8} && 11280&     5&     0&  7532&  1867&&   2&  3&  4& 10& 38& 57&&  114&\cr &
104&{E{8'}} &&^* 11748&     5&     0&  7844&  1945&&   1&  3&  4&  7& 26& 41&&   82&\cr &
105&{E{8''}} &&^* 11994&     5&     0&  8008&  1986&&   1&  1&  3&  4& 14& 19&&   42&\cr &
106&{E{8'}} && 12144&     5&     0&  8108&  2011&&   1&  2&  2&  4& 14& 23&&   46&\cr &
107&{E8} && 12432&     5&    19&  8338&  2078&&   1&  4&  4&  5& 28& 42&&   84&\cr &
108&{E{8''}}&& 12624&     5&     0&  8428&  2091&&   1&  1&  2&  2& 10& 14&&   30&\cr &
109&{E7} && 12672&     5&     0&  8460&  2099&&   1&  1&  1&  4&  7& 14&&   28&\cr &
110&{E{8''}}&& 13032&     5&     1&  8702&  2160&&   1&  1&  1&  3&  9& 12&&   27&\cr &
111&{E8} && 13032&     5&     1&  8702&  2160&&   2&  2&  2&  3& 18& 27&&   54&\cr &
112&{E8} && 13248&     5&    12&  8868&  2207&&   1&  4&  5&  8& 36& 54&&  108&\cr &
113&{E{8'}} && 13320&     5&     0&  8892&  2207&&   1&  2&  2&  2& 12& 19&&   38&\cr &
114&{E8} && 13896&     5&     6&  9288&  2309&&   1&  3&  4&  6& 28& 42&&   84&\cr &
115&{E{8''}} && 14256&     5&     0&  9516&  2363&&   1&  1&  2&  3& 12& 17&&   36&\cr &
116&{E{8'}} &&^* 15078&     5&     6& 10076&  2506&&   1&  1&  3&  3& 13& 21&&   42&\cr &
117&{E{8'}} && 15216&     5&     1& 10158&  2524&&   1&  1&  2&  4& 12& 20&&   40&\cr &
118&{E8} &&^* 15354&     5&     0& 10248&  2546&&   1&  4&  5& 11& 42& 63&&  126&\cr &
119&{E{8'}} && 15744&     5&     0& 10508&  2611&&   1&  2&  3&  7& 23& 36&&   72&\cr &
120&{E8} &&^* 16554&     5&     5& 11058&  2751&&   1&  3&  4&  9& 34& 51&&  102&\cr 
\noalign{\hrule}}
\hrule}}}
$$

$$\eqalign{ & {\rm{\bf Table\ B.4\ (continued) \ \ 
CY-Fourfolds\ with\ small\ Picard\ number }}\cr 
& \fivepoint{\vbox{\offinterlineskip\tabskip=0pt
\halign{\strutf
\vrule#&
\hfil$\, # \, $&
\hfil$\, # \, $&
\vrule$\, # \, $& 
\hfil$\, # \, $&  
\hfil$   # \,  $&  
\hfil$   # \,  $&  
\hfil$   # \,  $&  
\hfil$   # \, $& 
\vrule$\, # \, $& 
\hfil$\, #  $& 
\hfil$ #  $  & 
\hfil$ #  $ & 
\hfil$ #  $ & 
\hfil$ #  $ & 
\hfil$ #  $& 
\vrule$\, # \, $&
\hfil$\, # \, $& 
\vrule#\cr
\noalign{\hrule}
&N^o&F &&\chi & h_{11} &h_{21}&h_{22}&h_{31}
&&w_1&w_2&w_3&w_4&w_5&w_6&&d&\cr 
\noalign{\hrule} &  
121&{E{8'}} && 16728&     5&     1& 11166&  2776&&   1&  1&  3&  4& 15& 24&&   48&\cr &
122&{E8} && 18288&     5&     2& 12208&  3037&&   1&  2&  2&  5& 20& 30&&   60&\cr &
123&{E8} &&^* 18930&     5&     0& 12632&  3142&&   1&  3&  4& 11& 38& 57&&  114&\cr &
124&{E{8'}} && 18960&     5&     0& 12652&  3147&&   1&  1&  2&  4& 14& 22&&   44&\cr &
125&{E{8'}} &&^* 19044&     5&     0& 12708&  3161&&   1&  1&  1&  4& 10& 17&&   34&\cr &
126&{E8} &&^* 20844&     5&     0& 13908&  3461&&   1&  2&  3&  8& 28& 42&&   84&\cr &
127&{E{8'}} &&^* 21054&     5&     0& 14048&  3496&&   1&  1&  2&  4& 15& 23&&   46&\cr &
128&{E{8''}} && 21120&     5&     0& 14092&  3507&&   1&  1&  1&  3& 11& 16&&   33&\cr &
129&{E8} && 21120&     5&     0& 14092&  3507&&   1&  2&  2&  6& 22& 33&&   66&\cr &
130&{E{8'}} &&^* 22374&     5&     0& 14928&  3716&&   1&  1&  3&  5& 19& 29&&   58&\cr &
131&{E8} && 22704&     5&    13& 15174&  3784&&   1&  1&  4&  4& 20& 30&&   60&\cr &
132&{E{8'}} &&^* 24228&     5&     0& 16164&  4025&&   1&  1&  2&  5& 17& 26&&   52&\cr &
133&{E8} && 26880&     5&     5& 17942&  4472&&   1&  1&  3&  6& 22& 33&&   66&\cr &
134&{E8}&& 27072&     5&     0& 18060&  4499&&   1&  2&  3& 11& 34& 51&&  102&\cr &
135&{E8} &&^* 28554&     5&     0& 19048&  4746&&   1&  1&  4&  7& 26& 39&&   78&\cr &
136&{E6} &&  1848&     6&     0&  1244&   294&&   2&  3&  4&  5& 14& 14&&   42&\cr &
137&{E6} &&  2832&     6&     0&  1900&   458&&   1&  3&  4&  5& 13& 13&&   39&\cr &
138&{E6} &&  3192&     6&     0&  2140&   518&&   1&  2&  3&  3&  9&  9&&   27&\cr &
139&{E7} &&  3672&     6&     1&  2462&   599&&   2&  3&  5&  8& 18& 36&&   72&\cr &
140&{E7} &&  4392&     6&     1&  2942&   719&&   2&  2&  3&  5& 12& 24&&   48&\cr &
141&{E7} &&  4440&     6&     0&  2972&   726&&   2&  2&  3&  4& 11& 22&&   44&\cr &
142&{E6} &&  4632&     6&     1&  3102&   759&&   1&  1&  3&  4&  9&  9&&   27&\cr &
143&{E6} &&  4896&     6&     1&  3278&   803&&   1&  1&  2&  4&  8&  8&&   24&\cr &
144&{E6} &&  5064&     6&     2&  3392&   832&&   1&  1&  3&  5& 10& 10&&   30&\cr &
145&{E7} &&  5112&     6&     1&  3422&   839&&   1&  4&  5&  6& 16& 32&&   64&\cr &
146&{E{8''}} &&  5808&     6&     0&  3884&   954&&   2&  3&  5&  7& 29& 41&&   87&\cr &
147&{E8} &&  5808&     6&     0&  3884&   954&&   4&  5&  6& 14& 58& 87&&  174&\cr &
148&{E8} &&  6048&     6&     1&  4046&   995&&   3&  4&  5&  6& 36& 54&&  108&\cr &
149&{E8} &&  6048&     6&     2&  4048&   996&&   3&  4&  7& 10& 48& 72&&  144&\cr &
150&{E{8'}} &&  6144&     6&     0&  4108&  1010&&   2&  3&  5&  8& 28& 46&&   92&\cr &
151&{E8} &&^*  6210&     6&     0&  4152&  1021&&   3&  4&  7& 11& 50& 75&&  150&\cr &
152&{E7} &&^*  6612&     6&     0&  4420&  1088&&   1&  2&  2&  2&  7& 14&&   28&\cr &
153&{E7} &&  6840&     6&     0&  4572&  1126&&   1&  2&  5&  6& 14& 28&&   56&\cr &
154&{E7} &&  6960&     6&     0&  4652&  1146&&   1&  2&  2&  4&  9& 18&&   36&\cr &
155&{E8} &&^*  7002&     6&     2&  4684&  1155&&   3&  3&  4&  5& 30& 45&&   90&\cr &
156&{E{8''}} &&  7032&     6&     0&  4700&  1158&&   2&  3&  4&  5& 26& 38&&   78&\cr &
157&{E{8'}} &&  7080&     6&     0&  4732&  1166&&   2&  3&  5&  8& 31& 49&&   98&\cr &
158&{E8} &&  7272&     6&     3&  4866&  1201&&   2&  5&  6&  9& 44& 66&&  132&\cr &
159&{E{8'}} &&  7320&     6&    11&  4914&  1217&&   2&  3&  3&  5& 23& 36&&   72&\cr &
160&{E8} &&^*  7380&     6&     0&  4932&  1216&&   2&  5&  8& 11& 52& 78&&  156&\cr &
161&{E7} &&  7488&     6&     2&  5008&  1236&&   1&  2&  2&  5& 10& 20&&   40&\cr &
162&{E{8'}} &&^*  7794&     6&     0&  5208&  1285&&   2&  3&  5&  7& 32& 49&&   98&\cr &
163&{E{8'}} &&  7896&     6&     8&  5292&  1310&&   2&  2&  3&  3& 17& 27&&   54&\cr &
164&{E{8''}} &&  7944&     6&     0&  5308&  1310&&   1&  3&  4&  5& 21& 29&&   63&\cr &
165&{E8} &&  7944&     6&     0&  5308&  1310&&   2&  5&  6&  8& 42& 63&&  126&\cr &
166&{E{8'}} &&  8160&     6&     0&  5452&  1346&&   2&  3&  5&  8& 34& 52&&  104&\cr &
167&{E7} &&  8496&     6&     0&  5676&  1402&&   1&  2&  3&  8& 14& 28&&   56&\cr &
168&{E{8'}} &&  8592&     6&     1&  5742&  1419&&   2&  2&  3&  5& 21& 33&&   66&\cr &
169&{E{8'}} &&  8664&     6&     0&  5788&  1430&&   2&  2&  3&  4& 19& 30&&   60&\cr &
170&{E8} &&^*  8754&     6&     0&  5848&  1445&&   3&  4&  5& 17& 58& 87&&  174&\cr &
171&{E8} &&  8928&     6&     1&  5966&  1475&&   2&  3&  3&  4& 24& 36&&   72&\cr &
172&{E{8'}} &&  9120&     6&     0&  6092&  1506&&   1&  4&  5&  6& 26& 42&&   84&\cr &
173&{E{8'}} &&^*  9174&     6&     0&  6128&  1515&&   1&  5&  6&  7& 32& 51&&  102&\cr &
174&{E{8'}} &&^*  9276&     6&     0&  6196&  1532&&   2&  2&  3&  5& 22& 34&&   68&\cr &
175&{E{8''}} &&^*  9366&     6&     0&  6256&  1547&&   1&  2&  2&  3& 13& 18&&   39&\cr &
176&{E8} &&^*  9366&     6&     0&  6256&  1547&&   2&  3&  4&  4& 26& 39&&   78&\cr &
177&{E{8''}} &&  9528&     6&     0&  6364&  1574&&   1&  2&  3&  3& 15& 21&&   45&\cr &
178&{E8} &&  9528&     6&     0&  6364&  1574&&   2&  3&  4&  6& 30& 45&&   90&\cr &
179&{E{8'}} &&  9672&     6&     5&  6470&  1603&&   1&  3&  4&  6& 22& 36&&   72&\cr &
180&{E7} &&  9864&     6&     0&  6588&  1630&&   1&  1&  4&  5& 11& 22&&   44&\cr &
181&{E{8''}} &&  9888&     6&     0&  6604&  1634&&   1&  3&  4&  5& 23& 33&&   69&\cr
\noalign{\hrule}}
\hrule}
\quad
\vbox{\offinterlineskip\tabskip=0pt
\halign{\strutf
\vrule#&
\hfil$\, # \, $&
\hfil$\, # \, $&
\vrule$\, # \, $& 
\hfil$\, # \, $&  
\hfil$ #\,  $&  
\hfil$ # \, $&  
\hfil$ #\,  $&  
\hfil$ # \,  $& 
\vrule$\, # \, $& 
\hfil$\, #  $& 
\hfil$ #  $& 
\hfil$ #  $& 
\hfil$ #  $& 
\hfil$ #  $& 
\hfil$ #  $& 
\vrule$\, # \, $& 
\hfil$\, # \, $& 
\vrule#\cr
\noalign{\hrule}
&N^o&F &&\chi & h_{11} &h_{21}&h_{22}&h_{31}
&&w_1&w_2&w_3&w_4&w_5&w_6&&d&\cr
\noalign{\hrule} &  
182&{E8} &&  9888&     6&     0&  6604&  1634&&   2&  3&  8& 10& 46& 69&&  138&\cr &
183&{E{8'}} &&^* 10050&     6&     0&  6712&  1661&&   2&  2&  3&  4& 20& 31&&   62&\cr &
184&{E8} && 10128&     6&     0&  6764&  1674&&   2&  3&  3&  3& 22& 33&&   66&\cr &
185&{E{8'}} && 10152&     6&     1&  6782&  1679&&   1&  4&  5&  6& 28& 44&&   88&\cr &
186&{E{8'}} && 10512&     6&     4&  7028&  1742&&   1&  3&  4&  9& 25& 42&&   84&\cr &
187&{E8} && 10800&     6&     5&  7222&  1791&&   2&  3&  5& 12& 44& 66&&  132&\cr &
188&{E{8'}} && 10992&     6&     0&  7340&  1818&&   2&  2&  2&  3& 16& 25&&   50&\cr &
189&{E{8''}} && 11616&     6&     1&  7758&  1923&&   1&  1&  2&  4& 12& 16&&   36&\cr &
190&{E{8'}} &&^* 11628&     6&     6&  7776&  1930&&   1&  3&  4&  6& 25& 39&&   78&\cr &
191&{E{8'}} && 11640&     6&     0&  7772&  1926&&   1&  2&  5&  6& 22& 36&&   72&\cr &
192&{E{8''}} && 11760&     6&     0&  7852&  1946&&   1&  2&  3&  4& 18& 26&&   54&\cr &
193&{E{8'}} && 11784&     6&     0&  7868&  1950&&   1&  2&  4&  6& 20& 33&&   66&\cr &
194&{E{8'}} && 11904&     6&     2&  7952&  1972&&   1&  2&  2&  5& 15& 25&&   50&\cr &
195&{E8} && 12024&     6&     3&  8034&  1993&&   2&  2&  3&  7& 28& 42&&   84&\cr &
196&{E8} && 12192&     6&     2&  8144&  2020&&   1&  5&  6&  8& 40& 60&&  120&\cr &
197&{E{8''}} && 12432&     6&     6&  8312&  2064&&   1&  1&  3&  3& 13& 18&&   39&\cr &
198&{E8} && 12432&     6&     6&  8312&  2064&&   2&  2&  3&  6& 26& 39&&   78&\cr &
199&{E{8'}} && 12432&     6&     1&  8302&  2059&&   1&  3&  4&  7& 27& 42&&   84&\cr &
200&{E{8'}} && 12744&     6&     0&  8508&  2110&&   1&  2&  3&  8& 20& 34&&   68&\cr &
201&{E{8''}} && 12792&     6&     0&  8540&  2118&&   1&  2&  3&  5& 20& 29&&   60&\cr &
202&{E8} && 13248&     6&     1&  8846&  2195&&   1&  3&  4&  4& 24& 36&&   72&\cr &
203&{E{8'}} && 13344&     6&     0&  8908&  2210&&   1&  2&  2&  6& 16& 27&&   54&\cr &
204&{E8} && 13344&     6&     6&  8920&  2216&&   1&  5&  8& 12& 52& 78&&  156&\cr &
205&{E8} &&^* 13410&     6&     3&  8958&  2224&&   1&  4&  7&  9& 42& 63&&  126&\cr &
206&{E{8'}} && 13464&     6&     0&  8988&  2230&&   1&  2&  4&  6& 22& 35&&   70&\cr &
207&{E8} && 13752&     6&     2&  9184&  2280&&   1&  3&  5&  6& 30& 45&&   90&\cr &
208&{E{8''}} &&^* 13842&     6&     0&  9240&  2293&&   1&  1&  1&  4& 10& 13&&   30&\cr &
209&{E{8'}} && 13896&     6&     0&  9276&  2302&&   1&  3&  4&  7& 29& 44&&   88&\cr &
210&{E{8''}} && 14136&     6&     1&  9438&  2343&&   1&  1&  3&  4& 15& 21&&   45&\cr &
211&{E8} && 14136&     6&     1&  9438&  2343&&   2&  2&  3&  8& 30& 45&&   90&\cr &
212&{E{8'}} && 14424&     6&     0&  9628&  2390&&   1&  2&  2&  2& 13& 20&&   40&\cr &
213&{E8} && 14424&     6&     0&  9628&  2390&&   1&  3&  3&  3& 20& 30&&   60&\cr &
214&{E{8''}} && 14424&     6&     0&  9628&  2390&&   1&  2&  3&  5& 21& 31&&   63&\cr &
215&{E8} && 14424&     6&     0&  9628&  2390&&   1&  4&  6& 10& 42& 63&&  126&\cr &
216&{E{8'}} && 14664&     6&     0&  9788&  2430&&   1&  2&  2&  4& 16& 25&&   50&\cr &
217&{E{8'}} &&^* 14754&     6&     0&  9848&  2445&&   1&  2&  4&  5& 23& 35&&   70&\cr &
218&{E{8'}}  && 15408&     6&     0& 10284&  2554&&   1&  2&  3& 11& 23& 40&&   80&\cr &
219&{E8}  && 15408&     6&     6& 10296&  2560&&   1&  3&  7&  9& 40& 60&&  120&\cr &
220&{E{8'}} && 15528&     6&     0& 10364&  2574&&   1&  2&  2&  5& 18& 28&&   56&\cr &
221&{E{8'}} && 15600&     6&     0& 10412&  2586&&   1&  2&  2&  4& 17& 26&&   52&\cr &
222&{E8} && 15600&     6&     0& 10412&  2586&&   1&  3&  3&  6& 26& 39&&   78&\cr &
223&{E{8''}} && 15864&     6&     0& 10588&  2630&&   1&  1&  2&  2& 11& 16&&   33&\cr &
224&{E8} && 15864&     6&     0& 10588&  2630&&   1&  2&  4&  4& 22& 33&&   66&\cr &
225&{E8} && 16128&     6&    10& 10784&  2684&&   1&  4&  5& 12& 44& 66&&  132&\cr &
226&{E{8'}} && 16176&     6&    12& 10820&  2694&&   1&  1&  4&  4& 16& 26&&   52&\cr &
227&{E{8'}} && 16464&     6&     0& 10988&  2730&&   1&  1&  4&  5& 17& 28&&   56&\cr &
228&{E{8'}} &&^* 16806&     6&     4& 11224&  2791&&   1&  1&  3&  6& 16& 27&&   54&\cr &
229&{E8} && 17568&     6&     1& 11726&  2915&&   1&  2&  3&  6& 24& 36&&   72&\cr &
230&{E{8'}} &&^* 17844&     6&     0& 11908&  2960&&   1&  1&  4&  5& 18& 29&&   58&\cr &
231&{E8} && 17904&     6&     3& 11954&  2973&&   1&  2&  4&  7& 28& 42&&   84&\cr &
232&{E{8'}} &&^* 17994&     6&     0& 12008&  2985&&   1&  1&  5&  6& 20& 33&&   66&\cr &
233&{E8} && 18672&     6&     3& 12466&  3101&&   1&  2&  5&  8& 32& 48&&   96&\cr &
234&{E{8''}} && 19200&     6&     0& 12812&  3186&&   1&  1&  3&  4& 17& 25&&   51&\cr &
235&{E8} && 19200&     6&     0& 12812&  3186&&   1&  2&  6&  8& 34& 51&&  102&\cr &
236&{E{8'}} &&^* 22074&     6&     0& 14728&  3665&&   1&  1&  2&  5& 16& 25&&   50&\cr &
237&{E8} && 22560&     6&     0& 15052&  3746&&   1&  2&  2&  7& 24& 36&&   72&\cr &
238&{E8} && 26208&     6&     3& 17490&  4357&&   1&  1&  4&  6& 24& 36&&   72&\cr &
239&{E{8'}} && 27744&     6&     0& 18508&  4610&&   1&  1&  1&  4& 13& 20&&   40&\cr &
240&{E8} && 30336&     6&     0& 20236&  5042&&   1&  1&  2&  6& 20& 30&&   60&\cr &
241&{E{8'}} &&^* 33594&     6&     0& 22408&  5585&&   1&  1&  1&  5& 15& 23&&   46&\cr
\noalign{\hrule}}
\hrule}}}
$$

\listrefs
\bye